\documentclass[preprint,authoryear]{elsarticle}
\usepackage[left=1in,right=1in,top=1in,bottom=1in]{geometry}
\usepackage{lineno}
\usepackage[colorlinks=true,citecolor=blue,pdfcreator={},pdfproducer={}]{hyperref}
\modulolinenumbers[1]
\usepackage{amsmath}
\usepackage{amsfonts}
\usepackage{bm}
\usepackage{subfigure}
\usepackage{graphicx}
\allowdisplaybreaks
\usepackage{amsbsy}
\usepackage{amssymb}
\usepackage[ruled]{algorithm2e}
\usepackage{natbib}

\newcommand*\patchAmsMathEnvironmentForLineno[1]{%
\expandafter\let\csname old#1\expandafter\endcsname\csname #1\endcsname
\expandafter\let\csname oldend#1\expandafter\endcsname\csname end#1\endcsname
\renewenvironment{#1}%
{\linenomath\csname old#1\endcsname}%
{\csname oldend#1\endcsname\endlinenomath}}%
\newcommand*\patchBothAmsMathEnvironmentsForLineno[1]{%
\patchAmsMathEnvironmentForLineno{#1}%
\patchAmsMathEnvironmentForLineno{#1*}}%
\AtBeginDocument{%
\patchBothAmsMathEnvironmentsForLineno{equation}%
\patchBothAmsMathEnvironmentsForLineno{align}%
\patchBothAmsMathEnvironmentsForLineno{flalign}%
\patchBothAmsMathEnvironmentsForLineno{alignat}%
\patchBothAmsMathEnvironmentsForLineno{gather}%
\patchBothAmsMathEnvironmentsForLineno{multline}%
}

\newcommand{\bfzero}{\mathbf{0}}
\newcommand{\bfA}{\mathbf{A}}
\newcommand{\bfB}{\mathbf{B}}
\newcommand{\bfI}{\mathbf{I}}
\newcommand{\bfC}{\mathbf{C}}
\graphicspath{{./Figures/}}
\journal{Geophysical Journal International}

\begin{document}


\begin{frontmatter}

\title{Optimal damping ratios of multi-axial perfectly matched layers for elastic-wave modeling in general 
anisotropic media}

\author[a]{Kai Gao}
\author[a]{Lianjie Huang}

\address[a]{Geophysics Group, Los Alamos National Laboratory, Los Alamos, NM 87545}

\begin{abstract}
The conventional Perfectly Matched Layer (PML) is unstable for certain 
kinds of anisotropic media. This instability is intrinsic and independent 
of PML formulation or implementation.  The Multi-axial PML (MPML) 
removes such instability using a nonzero damping coefficient in the 
direction parallel with the interface between a PML and the investigated 
domain. The damping ratio of MPML is the ratio between the damping 
coefficients along the directions parallel with and perpendicular to the 
interface between a PML and the investigated domain. No quantitative 
approach is available for obtaining these damping ratios for general 
anisotropic media. We develop a quantitative approach to determining 
optimal damping ratios to not only stabilize PMLs, but also minimize the 
artificial reflections from MPMLs. Numerical tests based on finite-difference method show that our new 
method can effectively provide a set of optimal MPML damping ratios 
for elastic-wave propagation in 2D and 3D
general anisotropic media.  
\end{abstract}

\begin{keyword}
Anisotropic medium, elastic-wave propagation, Multi-axial Perfectly Matched Layers (MPML), damping ratio.
\end{keyword}

\end{frontmatter}

\section{Introduction}
Elastic-wave modeling usually needs to absorb 
outgoing wavefields at boundaries of an investigated domain. Two main categories of boundary absorbers have been 
developed: one is called the Absorbing Boundary Condition (ABC) 
\cite[e.g.,][]{Clayton_Engquist_1977,Reynolds_1978,Liao_etal_1984,Cerjan_1985,Higdon_1986,Higdon_1987,Long_Liow_1990,Peng_Toksoz_1994}, 
and the other is termed the Perfectly Matched Layer (PML) 
\cite[e.g.,][]{Berenger_1994,Hastings_etal_1996,Collino_Tsogka_2001}. In 
some literature, PML is considered as one of ABCs. However, there are 
fundamental differences in the construction of PML and its variants 
compared with traditional ABCs, we therefore differentiate them in names.  
For a brief summary, please refer to \cite{Hastings_etal_1996} and other 
relevant references. 

The PML approach was first introduced by \cite{Berenger_1994} for 
electromagnetic-wave modeling, and has been widely used in elastic-wave modeling because of its simplicity and 
superior absorbing capability 
\cite[e.g.,][]{Collino_Tsogka_2001,Koma_Tromp_2003,Drossaert_Gian_2007}.  
Various improved PML methods for elastic-wave modeling have been 
developed, such as non-splitting convolutional PML (CPML) to enahce 
absorbing capability for grazing incident waves 
\cite[]{Komatitsch_Martin_2007,Martin_Komatitsch_2009}, and CPML with 
auxiliary differential equation (ADE-PML) for modeling with a high-order 
time accuracy formulation \cite[]{Zhang_Shen_2010,Martin_etal_2010}.  
However, a well-known problem of PML and its variants/improvements is 
that numerical modeling with PML is unstable in certain kinds of anisotropic media
for long-time wave propagation.

To address the instability problem of PML, \cite{Becache_etal_2003} 
analyzed the PML for 2D anisotropic media and found that,  if there 
exists points where the $i^{\text{th}}$ component of group velocity 
$\mathbf{v}$ has an opposite direction relative to the $i^{\text{th}}$ component 
of wavenumber $\mathbf{k}$, i.e., $v_i k_i<0$ (no summation rules applied), then the $x_i$-direction 
PML is unstable. The original version of this aforementioned  condition was expressed 
with so-called ``slowness vector'' defined by \cite{Becache_etal_2003}, 
but it can be recast in such form according to the definition of the 
``slowness vector'' in eq.~(45) of \cite[]{Becache_etal_2003}. This PML 
instability is intrinsic and independent of PML/CPML formulations adopted 
for wavefield modelings. To make PML stable, elasticity parameters of 
an  anisotropic medium need to satisfy certain inequality relations 
\cite[]{Becache_etal_2003}.  These restrictions limit the applicability  
of PML for arbitrary anisotropic media. 

\cite{Fajardo-Papa_2008} presented an explanation for the instability of 
conventional PML. They recast the elastic wave equations in PML to an 
autonomous system and found that the PML instability is caused by the 
fact that the PML coefficient matrix having one or more 
eigenvalues with positive imaginary parts. They showed that PML 
becomes stable when adding appropriate nonzero damping coefficients to 
PML in the direction parallel with the PML/non-PML interface. The ratio 
between the PML damping coefficients along the directions parallel with 
and perpendicular to the PML/non-PML interface is called the damping 
ratio.
The resulting PML with nonzero damping ratios is termed the Multi-axial 
PML (MPML).

A key step in the stability analysis of PML is to derive eigenvalue 
derivatives of the damped system coefficient matrix.  
\cite{Fajardo-Papa_2008} derived expressions of the eigenvalue 
derivatives for anisotropic media. However, these expressions are valid 
only for two-dimensional isotropic media and anisotropic media with up to 
hexagonal/orthotropic anisotropy, that is, $C_{11}\neq0$, $C_{33}\neq0$, 
$C_{55}\neq0$, $C_{15}=C_{35}=0$, and $C_{13}$ can either be zero or 
nonzero depending on medium properties. Furthermore, although 
\cite{Fajardo-Papa_2008}  showed that the nonzero damping ratios can 
stabilize PML, they did not present a method to select the appropriate dampoing ratios.  
Adding these nonzero damping coefficients makes the 
PML no longer ``perfectly matched'', and the larger are the ratios, the 
stronger the artificial reflections become 
\cite[]{Dmitriev_Lisitsa_2011}. This increase is linear. Therefore, it 
is necessary to find a set of ``optimal'' damping ratios to not only  
ensure the stability of MPML, but to also eliminate artificial reflections 
as much as possible. 

We develop a new method to determine the optimal MPML damping ratios for 
general anisotropic media. We show that, even for 
a two-dimensional anisotropic medium with nonzero $C_{15}$ and $C_{35}$, 
the MPML stability analysis is complicated, and new equations must be 
derived to calculate both the eigenvalues of the undamped system and the 
eigenvalue derivatives of the damped system. The resulting expressions 
are functions of all nonzero $C_{ij}$ components as well as the  
wavenumber $\mathbf{k}$. For 3D general anisotropic media, we find that 
such an analytic procedure becomes practically impossible because it requires definite analytic expressions of eigenvalues 
and eigenvalue derivatives. In the 3D 
case,  the dimension of the asymmetric system coefficient matrix is up to 
$27 \times 27$, and therefore a purely numerical approach should be 
employed. We present two algorithms with slightly different forms but 
essentially the same logic, to determine the optimal damping ratios for 
2D and 3D MPMLs. With these algorithms, it is possible to stabilize PML 
for any kind of anisotropic media without using a trial-and-error method.  
Our new algorithms enable us to use MPML for finite-difference modeling 
of elastic-wave propagation in 2D and 3D general anisotropic media 
where all elastic parameters $C_{ij}$ may be nonzero. These algorithms are also 
applicable to other elastic-wave modeling methods such as spectral-element method \cite[e.g.,][]{Koma_etal_2000}
and discontinuous Galerkin finite-element method \cite[e.g.,][]{Puente_etal_2007}. 

Our paper is organized as follows. In the Methodology section, we derive 
the equations for the eigenvalue derivatives for 2D and 3D general 
anisotropic media. We also present two algorithms to obtain the optimal 
MPML damping ratios. To validate our algorithms, we give six numerical  
examples in the Results section, including three 2D anisotropic 
elastic-wave modeling examples and three 3D anisotropic elastic-wave 
modeling examples, and show that our algorithms can give appropriate 
damping ratios for both 2D and 3D modeling in general anisotropic 
media.

\section{Methodology}
\subsection{Optimal damping ratios of 2D MPML}
In this section, we concentrate our analysis on the $x_1x_3$-plane. This 
analysis is also valid for the $x_1x_2$- and $x_2x_3$-planes. We assume 
that $C_{15}$ and $C_{35}$ are generally nonzero for an anisotropic 
medium. The 2D elastic-wave equations in the stees-velocity 
form are given by \cite[e.g.,][]{Carcione_2007},
\begin{align} \rho \frac{\partial  \mathbf{v}}{\partial t} &=  
   \boldsymbol{\Lambda} \boldsymbol{\sigma}+\mathbf{f},  \label{eq:elastic_2d_v}\\
\frac{\partial\boldsymbol{\sigma}}{\partial t} & = \bfC \boldsymbol{\Lambda}^{\mathrm{T}} \mathbf{v}, \label{eq:elastic_2d_sigma}
\end{align}
where $\boldsymbol{\sigma}=(\sigma_{11},\sigma_{33},\sigma_{13})^{\mathrm{T}}$ is the  stress wavefield, $\mathbf{v}=(v_1,v_3)^{\mathrm{T}}$ is the particle velocity wavefield, $\mathbf{f}$ is the external force, $\rho$ is the mass density, $\bfC$ is the elasticity tensor in Voigt notation defined as
\begin{equation}
\bfC=\left(
\begin{array}{ccc}
	C_{11} & C_{13} & C_{15} \\
	C_{13} & C_{33} & C_{35} \\
	C_{15} & C_{35} & C_{55} \\
\end{array} \right),
\end{equation}
and $\boldsymbol{\Lambda}$ is the differential operator matrix defined as
\begin{equation}
\boldsymbol{\Lambda}=\left(
\begin{array}{ccc}
	\frac{\partial}{\partial x_1} & 0 & \frac{\partial}{\partial x_3} \\
	0 & \frac{\partial}{\partial x_3} &  \frac{\partial}{\partial x_1} \\
\end{array} \right).
\end{equation}
In the following analysis, we ignore the external force term $\mathbf{f}$ without loss of generality. 

Using the convention in \cite{Fajardo-Papa_2008} for isotropic and 
VTI/HTI/othotropic media, the undamped system of 
eqs.~\eqref{eq:elastic_2d_v}--\eqref{eq:elastic_2d_sigma} can be 
also written as
\begin{align}
\rho\frac{\partial}{\partial t}\left[\begin{array}{c}
v_{1}\\
v_{3}
\end{array}\right] & = \left[\begin{array}{c}
\frac{\partial}{\partial x_{1}}\sigma_{11}+\frac{\partial}{\partial x_{3}}\sigma_{13}\\
\frac{\partial}{\partial x_{1}}\sigma_{13}+\frac{\partial}{\partial x_{3}}\sigma_{33}
\end{array}\right], \\
\frac{\partial}{\partial t}\left[\begin{array}{c}
\sigma_{11}\\
\sigma_{33}\\
\sigma_{13}
\end{array}\right] & = \left[\begin{array}{cccc}
C_{11} & C_{13} & C_{15} & C_{15}\\
C_{13} & C_{33} & C_{35} & C_{35}\\
C_{15} & C_{35} & C_{55} & C_{55}
\end{array}\right]\left[\begin{array}{c}
\frac{\partial}{\partial x_{1}}v_{1}\\
\frac{\partial}{\partial x_{3}}v_{3}\\
\frac{\partial}{\partial x_{1}}v_{3}\\
\frac{\partial}{\partial x_{3}}v_{1}
\end{array}\right].
\end{align}
Equivalently, the system of the above two equations can be written as
\begin{align}
\frac{\partial}{\partial t}\mathbf{v} & = \mathbf{D}_{1}\frac{\partial}{\partial x_{1}}\boldsymbol{\sigma}+\mathbf{D}_{3}\frac{\partial}{\partial x_{3}}\boldsymbol{\sigma},\\
\frac{\partial}{\partial t}\mathbf{\boldsymbol{\sigma}} & = \bfC_{1}\frac{\partial}{\partial x_{1}}\mathbf{v}+\bfC_{3}\frac{\partial}{\partial x_{3}}\mathbf{v},
\end{align}
where 
\begin{align}
\mathbf{v} & = (v_{1},v_{3})^{\mathrm{T}},\\
\boldsymbol{\sigma} & = (\sigma_{11},\sigma_{33},\sigma_{13})^{\mathrm{T}},\\
\mathbf{D}_{1} & = \rho^{-1}\left[\begin{array}{ccc}
1 & 0 & 0\\
0 & 0 & 1
\end{array}\right],\\
\mathbf{D}_{3} & = \rho^{-1}\left[\begin{array}{ccc}
0 & 0 & 1\\
0 & 1 & 0
\end{array}\right],\\
\bfC_{1} & = \left[\begin{array}{cc}
C_{11} & C_{15}\\
C_{13} & C_{35}\\
C_{15} & C_{55}
\end{array}\right],\\
\bfC_{3} & = \left[\begin{array}{cc}
C_{15} & C_{13}\\
C_{35} & C_{33}\\
C_{55} & C_{55}
\end{array}\right].
\end{align}

In the conventional 2D PML, each field variable is split into two 
orthogonal components that are perpendicular to and parallel with the 
interface between the PML and the investigated domain, and the system of 
wave equations in PML can be written as \cite[]{Fajardo-Papa_2008}
\begin{equation}\label{eq:elastic_wave_damped}
\frac{\partial \boldsymbol{\Psi}}{\partial t} = \bfA \boldsymbol{\Psi},
\end{equation}
with $\bfA=\bfA_0 + \bfB$, and 
\begin{align}
\bfA_0 & = \left[\begin{array}{cccc}
	\bfzero_{33} & \bfzero_{33} & \frac{\partial}{\partial x_{1}}\bfC_{1} & \frac{\partial}{\partial x_{1}}\bfC_{1}\\
	\bfzero_{33} & \bfzero_{33} & \frac{\partial}{\partial x_{3}}\bfC_{3} & \frac{\partial}{\partial x_{3}}\bfC_{3}\\
	\frac{\partial}{\partial x_{1}}\mathbf{D}_{1} & \frac{\partial}{\partial x_{1}}\mathbf{D}_{1} & \bfzero_{22} & \bfzero_{22}\\
	\frac{\partial}{\partial x_{3}}\mathbf{D}_{3} & \frac{\partial}{\partial x_{3}}\mathbf{D}_{3} & \bfzero_{22} & \bfzero_{22}
\end{array}\right],\\
\bfB & = \left[\begin{array}{cccc}
	-d_{1}\bfI_{3} & \bfzero_{33} & \bfzero_{32} & \bfzero_{32}\\
	\bfzero_{33} & -d_{3}\bfI_{3} & \bfzero_{32} & \bfzero_{32}\\
	\bfzero_{23} & \bfzero_{23} & -d_{1}\bfI_{2} & \bfzero_{22}\\
	\bfzero_{23} & \bfzero_{23} & \bfzero_{22} & -d_{3}\bfI_{2}
\end{array}\right],
\end{align}
where $\bfzero_{mn}$ is the $m \times n$ zero matrix, $\bfI_m$ is the $m\times m$ identity matrix, and 
\begin{equation}
\boldsymbol{\Psi} = (\sigma_{11}^{(1)},\sigma_{33}^{(1)},\sigma_{13}^{(1)}, \sigma_{11}^{(3)},\sigma_{33}^{(3)},\sigma_{13}^{(3)} , v_1^{(1)}, v_3^{(1)}, v_1^{(3)}, v_3^{(3)})^{\mathrm{T}}
\end{equation} represents the split wavefield variables in PML. In the 
damping matrix $\bfB$, $d_1$ and $d_3$ represent the PML damping 
coefficients along the $x_1$- and $x_3$-axis, respectively. The PML 
damping coefficients depend on the thickness of PML, the desired 
reflection coefficient and the P-wave velocity at the PML/non-PML 
interface. Usually, they vary with the distance from a location inside the PML to the 
PML/non-PML interface according to the power of two or the power of 
three \cite[e.g.,][]{Collino_Tsogka_2001}.

Transforming system \eqref{eq:elastic_wave_damped} into the wavenumber 
domain leads to
\begin{equation}
\frac{\partial \mathbf{U}}{\partial t} = \tilde{\bfA} \mathbf{U},
\end{equation}
where $\mathbf{U}=\mathcal{F} [\boldsymbol{\Psi}]$ is the Fourier transform of the split filed variables $\boldsymbol{\Psi}$, $\tilde{\bfA}=\tilde{\bfA}_0 + \bfB$, and $\tilde{\bfA}_0$ now is
\begin{equation}
\tilde{\bfA}_0 = \left[\begin{array}{cccc}
\bfzero_{33} & \bfzero_{33} & ik_1 \bfC_{1} & ik_1 \bfC_{1}\\
\bfzero_{33} & \bfzero_{33} & ik_3\bfC_{3} & ik_3\bfC_{3}\\
ik_1 \mathbf{D}_{1} & ik_1 \mathbf{D}_{1} & \bfzero_{22} & \bfzero_{22}\\
ik_3\mathbf{D}_{3} & ik_3\mathbf{D}_{3} & \bfzero_{22} & \bfzero_{22}
\end{array}\right],
\end{equation}
where $k_1$ and $k_3$ are respectively the $x_1$- and $x_3$-components of 
wavenumber vector $\mathbf{k}$. 

As demonstrated by the stability theory of autonomous system in 
\cite{Fajardo-Papa_2008}, for a stable PML in an elastic medium, either 
isotropic or anisotropic, all the eigenvalues of the system matrix 
$\tilde{\bfA}$ should have non-positive imaginary parts. In conventional PML, the outgoing wavefield is damped only along the 
direction perpendicular to the PML/non-PML interface, and the damping 
matrix $\bfB$ for PML in the $x_1$ and $x_3$-directions can be 
respectively written as
\begin{align}
\bfB_1 & = \left[\begin{array}{cccc}
-d_{1}\bfI_{3} & \bfzero_{33} & \bfzero_{32} & \bfzero_{32}\\
\bfzero_{33} & \bfzero_{33} & \bfzero_{32} & \bfzero_{32}\\
\bfzero_{23} & \bfzero_{23} & -d_{1}\bfI_{2} & \bfzero_{22}\\
\bfzero_{23} & \bfzero_{23} & \bfzero_{22} & \bfzero_{22}
\end{array}\right], \\
	\bfB_3 & = \left[\begin{array}{cccc}
	\bfzero_{33} & \bfzero_{33} & \bfzero_{32} & \bfzero_{32}\\
	\bfzero_{33} & -d_{3}\bfI_{3} & \bfzero_{32} & \bfzero_{32}\\
	\bfzero_{23} & \bfzero_{23} & \bfzero_{22} & \bfzero_{22}\\
	\bfzero_{23} & \bfzero_{23} & \bfzero_{22} & -d_{3}\bfI_{2}
        \end{array}\right],
\end{align}
resulting in an unstable PML. \cite{Fajardo-Papa_2008} analyzed the 
derivatives of eigenvalues of $\tilde{\bfA}$ with respect to the damping 
parameters $d_1$ and $d_3$, and showed that if an appropriate damping 
ratio $\xi_1$ or $\xi_3$ is added along the direction parallel with the 
PML/non-PML interface, i.e.,
\begin{align}
\bfB_1 (\xi_1) & = \left[\begin{array}{cccc}
-d_{1}\bfI_{3} & \bfzero_{33} & \bfzero_{32} & \bfzero_{32}\\
\bfzero_{33} & -\xi_1 d_1 \bfI_{33} & \bfzero_{32} & \bfzero_{32}\\
\bfzero_{23} & \bfzero_{23} & -d_{1}\bfI_{2} & \bfzero_{22}\\
\bfzero_{23} & \bfzero_{23} & \bfzero_{22} & - \xi_1 d_1 \bfI_{22}
\end{array}\right], \\
\bfB_3 (\xi_3)& = \left[\begin{array}{cccc}
-d_3 \xi_3 \bfI_{33} & \bfzero_{33} & \bfzero_{32} & \bfzero_{32}\\
\bfzero_{33} & -d_{3}\bfI_{3} & \bfzero_{32} & \bfzero_{32}\\
\bfzero_{23} & \bfzero_{23} & -d_3 \xi_3 \bfI_{22} & \bfzero_{22}\\
\bfzero_{23} & \bfzero_{23} & \bfzero_{22} & -d_{3}\bfI_{2}
\end{array}\right],
\end{align}
then the PML becomes stable. The underlying principle of such stability 
comes from the fact that, after adding nonzero damping ratios $\xi_1$ and 
$\xi_3$, the eigenvalue derivatives of $\tilde{\bfA}$ have negative 
values along all $\mathbf{k}$ directions and consequently, all the 
relevant eigenvalues of $\tilde{\bfA}_0$ have negative imaginary parts  
(other eigenvalues are a pure zero), making autonomous 
system~\eqref{eq:elastic_wave_damped} stable. 

A key step for determing such damping ratios $\xi_1$ and $\xi_3$ is to compute the eigenvalue derivatives of 
$\tilde{\bfA}$. \cite{Fajardo-Papa_2008} adopted the following procedure: 
\begin{enumerate}
\item Calculate eigenvalues $e$ of undamped system coefficient matrix 
$\tilde{\bfA}_0$, of which six are a pure zero, and the rest four are 
pure imaginary numbers as functions of elasticity coefficient $\bfC$ and 
wavenumber $\mathbf{k}$;
\item Calculate the eigenvalue derivative of $\tilde{\bfA} 
= \tilde{\bfA}_0 + \tilde{\bfB}_1(\xi_1)$ or $\tilde{\bfA} 
= \tilde{\bfA}_0 + \tilde{\bfB}_3(\xi_3)$ with respect $d_1$ or $d_3$ at 
$d_1=0$ or $d_3=0$.  These eigenvalue derivatives are functions of 
elasticity coefficient $\bfC$, wavenumber $\mathbf{k}$ and damping ratio 
$\xi_1$ or $\xi_3$;
\item Choose an appropriate value to g
ensure that the values of the eigenvalue derivatives are negative in the 
range of $(0,\pi/2]$ (the direction of wavenumber $\mathbf{k}$).
\end{enumerate}

Both eigenvalues of $\tilde{\bfA}_0$ and eigenvalue derivatives of 
$\tilde{\bfA}$ are calculated analytically in the above procedure.  
Specially, Step 2 involves implicit differentiation operation and solving 
roots for high-order polynomials, and could not be accomplished 
numerically. 

We adopt the above procedure for obtaining optimal damping ratios for 2D 
general anisotropic media where $C_{15}$ or $C_{35}$ may be nonzero.  We first derive relevant expressions for the 
eigenvalues of $\tilde{\bfA}_0$ and the eigenvalue derivatives of 
$\tilde{\bfA}$.  Because our procedure is the same as that in 
\cite{Fajardo-Papa_2008}, we only show the resulting equations. The four 
nonzero eigenvalues of $\tilde{\bfA}_0(\bfC,\mathbf{k})$ are 
\begin{align}
e_1 (\bfC,\mathbf{k}) & = \pm \frac{i}{\sqrt{2} \rho}
\sqrt{P - \sqrt{Q}}, \label{eq:eigen_1}\\
e_2 (\bfC,\mathbf{k}) & =\pm \frac{i}{\sqrt{2} \rho}
\sqrt{P + \sqrt{Q}}, \label{eq:eigen_2}\\
P & = \rho [(C_{11} + C_{55}) k_1^2 + 
2 (C_{15} + C_{35}) k_1 k_3 + (C_{33} + 
C_{55}) k_3^2], \\
Q & = \rho^2 \{[(C_{11} + C_{55}) k_1^2 + 
2 (C_{15} + C_{35}) k_1 k_3 + (C_{33} + C_{55}) k_3^2]^2  \nonumber \\
&+ 
4 [(C_{15}^2 - C_{11} C_{55}) k_1^4 + 
2 (C_{13} C_{15} - C_{11} C_{35}) k_1^3 k_3 \nonumber \\
&+ (C_{13}^2 - C_{11} C_{33} - 
2 C_{15} C_{35} + 2 C_{13} C_{55}) k_1^2 k_3^2 \nonumber \\
& + 
2 (-C_{15} C_{33} + C_{13} C_{35}) k_1 k_3^3 + (C_{35}^2 - 
C_{33} C_{55})] k_3^4\},
\end{align}
where $C_{IJ}$ are components of the elasticity matrix and $\rho$ is the 
mass density. The two eigenvalues in $e_1$ or $e_2$ have the 
same length with different signs, and we take only the negative ones, 
i.e., \begin{align}
e_1 (\bfC,\mathbf{k}) & = -\frac{i}{\sqrt{2} \rho}
\sqrt{P - \sqrt{Q}}, \\
e_2 (\bfC,\mathbf{k}) & =-\frac{i}{\sqrt{2} \rho}
\sqrt{P + \sqrt{Q}}.
\end{align}
The choice of the signs of $e_1$ and $e_2$ does not affect the following  
stability analysis and optimal damping ratios.

The eigenvalue derivatives of $\tilde{\bfA}$ with respect to the damping 
coefficient $d_1$ at $d_1=0$ can be written as
\begin{align}\label{eq:eigenderiv_1}
\chi_1^{(l)} (\bfC,\mathbf{k}, \xi_1, e_l) &=(2 C_{15} C_{35} k_1^2 k_3^2 + 3 C_{15} C_{33} k_1 k_3^3 - 2 C_{35}^2 k_3^4 \nonumber \\
&+ 
2 C_{33} C_{55} k_3^4 - 2 C_{15}^2 k_1^4 \xi + 2 C_{15} C_{35} k_1^2 k_3^2 \xi_1 \nonumber \\
& + 
C_{15} C_{33} k_1 k_3^3 \xi_1 - C_{13}^2 k_1^2 k_3^2 (1 + \xi_1) \nonumber \\
& - 
C_{13} k_1 k_3 (C_{15} k_1^2 (1 + 3 \xi_1) + 
k_3 (2 C_{55} k_1 (1 + \xi_1) \nonumber \\
& + C_{35} k_3 (3 + \xi_1))) + 
e_l^2 (k_3 (3 C_{15} k_1 (1 + \xi_1)  \nonumber \\
&+ 3 C_{35} k_1 (1 + \xi_1) + 
C_{33} k_3 (2 + \xi_1)) \nonumber \\
&+ 
C_{55} (k_3^2 (2 + \xi_1) + k_1^2 (1 + 2 \xi_1))) \rho + 
2 e_l^4 (1 + \xi_1) \rho^2 \nonumber \\
&+ 
C_{11} k_1^2 (2 C_{55} k_1^2 \xi_1 + C_{33} k_3^2 (1 + \xi_1) + 
C_{35} k_1 k_3 (1 + 3 \xi_1) + 
e_l^2 (1 + 2 \xi_1) \rho)) \nonumber \\
&/(2 ((C_{15}^2 - C_{11} C_{55}) k_1^4 + 
2 (C_{13} C_{15} - C_{11} C_{35}) k_1^3 k_3 + (C_{13}^2 - C_{11} C_{33} - 2 C_{15} C_{35} \nonumber \\
&+ 
2 C_{13} C_{55}) k_1^2 k_3^2 + 
2 (-C_{15} C_{33} + C_{13} C_{35}) k_1 k_3^3 + (C_{35}^2 - C_{33} C_{55}) k_3^4) \nonumber \\
& - 
3 e_l^2 ((C_{11} + C_{55}) k_1^2 + 
2 (C_{15} + C_{35}) k_1 k_3 + (C_{33} + C_{55}) k_3^2) \rho - 4 e_l^4 \rho^2),
\end{align}
where subscript ``$l$'' is for the $l^{\text{th}}$ eigvenvalue, and $e_l$ 
stands for the $l^{\text{th}}$ eigenvalue of $\tilde{\bfA}_0$.  The 
eigenvalue derivatives of $\tilde{\bfA}$ with respect to the damping 
coefficient $d_3$ at $d_3=0$ is given by
\begin{align}\label{eq:eigenderiv_2}
\chi_3^{(l)}(\bfC,\mathbf{k},\xi_3,e_l)&=(-2 C_{15}^2 k_1^4 + 
k_3^2 (-2 (C_{35}^2 - C_{33} C_{55}) k_3^2 \xi_3 \nonumber \\
&- C_{13}^2 k_1^2 (1 + \xi_3) - 
C_{13} k_1 (2 C_{55} k_1 (1 + \xi_3) \nonumber \\
& + C_{35} k_3 (1 + 3 \xi_3))) + 
e_l^2 (k_3 (3 C_{35} k_1 (1 + \xi_3) \nonumber \\
& + C_{33} k_3 (1 + 2 \xi_3)) + 
C_{55} (k_1^2 (2 + \xi_3) + k_3^2 (1 + 2 \xi_3))) \rho + 
2 e_l^4 (1 + \xi_3) \rho^2 \nonumber \\
& + 
C_{15} k_1 k_3 (-C_{13} k_1^2 (3 + \xi_3) + 
k_3 (2 C_{35} k_1 (1 + \xi_3)  \nonumber \\
&+ C_{33} k_3 (1 + 3 \xi_3)) + 
3 e_l^2 (1 + \xi_3) \rho) + 
C_{11} k_1^2 (2 C_{55} k_1^2 \nonumber \\
&+ 
k_3 (C_{33} k_3 (1 + \xi_3) + C_{35} k_1 (3 + \xi_3)) + 
e_l^2 (2 + \xi_3) \rho)) \nonumber \\
&/(2 ((C_{15}^2 - C_{11} C_{55}) k_1^4 + 
2 (C_{13} C_{15} - C_{11} C_{35}) k_1^3 k_3 \nonumber \\
&+ (C_{13}^2 - C_{11} C_{33} - 2 C_{15} C_{35} + 
2 C_{13} C_{55}) k_1^2 k_3^2 + 
2 (-C_{15} C_{33} + C_{13} C_{35}) k_1 k_3^3 \nonumber \\
&+ (C_{35}^2 - C_{33} C_{55}) k_3^4) - 
3 e_l^2 ((C_{11} + C_{55}) k_1^2 \nonumber \\
&+ 
2 (C_{15} + C_{35}) k_1 k_3 + (C_{33} + C_{55}) k_3^2) \rho - 4 e_l^4 \rho^2).
\end{align}

There exists a subtle trade-off between the PML stability and artificail 
boundary reflections for anisotropic media. On one hand, it is necessary 
to introduce nonzero damping ratios $\xi_1$ and $\xi_3$ to stabilize PML. 
On the other hand, adding these nonzero damping ratios to PML makes PML 
no longer perfectly matched, and the larger are the damping ratios, the 
stronger the artificial boundary reflections become 
\cite[]{Dmitriev_Lisitsa_2011}. The original analysis of 
\cite{Fajardo-Papa_2008} only showed that a certain value of $\xi_1$ or 
$\xi_3$ can ensure that $\chi_1^{(l)}$ and $\chi_3^{(l)}$ are negative in 
all $\mathbf{k}$ directions. However, it did not provide a method to 
determine how large the damping ratios $\xi_1$ and $\xi_3$ are adequate 
for an arbitrary anisotropic medium. We therefore develop a procedure to   
determine the optimal damping ratios $\xi_1$ and $\xi_3$ to not only  
stabilize PMLs, but to also minimize resulting artificial boundary 
reflections.

We employ the following procedure described in 
Algorithm~\ref{alg:damping_ratio} to obtain the optimal damping ratios 
$\xi_1$ and $\xi_3$ of MPML for 2D general anisotropic media.

\vskip \baselineskip
\begin{algorithm}[H]
\DontPrintSemicolon
\SetKwInOut{Input}{input}
\SetKwInOut{Output}{output}

\Input{$\xi_i=0$, $\epsilon=-0.005$, $\Delta \xi=0.001$. }

\BlankLine
\For{$\theta \in(0,\pi]$}{
1) Calculate wavenumber $\mathbf{k}=(\sin\theta ,\cos \theta)$\;
2) Calculate eigenvalues $e_l$ ($l=1,2$) of $\tilde{\bfA}_0 
(\bfC,\mathbf{k})$ using eqs.~\eqref{eq:eigen_1} and 
\eqref{eq:eigen_2}\;
3) Calculate eigenvalue derivatives 
$\chi_i^{(l)}(\bfC,\mathbf{k},\xi_i,e_l)$\;
4) $\chi_{i,\max} = \max(\chi_i^{(1)},\chi_i^{(2)})$ using eq.~\eqref{eq:eigenderiv_1} or \eqref{eq:eigenderiv_2}\;
\If{$\chi_{i,\max} > \epsilon$}{
$\xi_i=\xi_i+\Delta \xi$\;
\textbf{go to} step 3\;
}
}

\BlankLine
\Output{$\xi_i$}

\label{alg:damping_ratio}
\caption{Determine the optimal damping ratio $\xi_i$ ($i=1, 3$) 
of MPML for 2D general anisotropic media}
\end{algorithm}
\vskip \baselineskip

We apply the above procedure to both the $x_1$- and $x_3$-directions to 
obtain the optimal values of $\xi_1$ and $\xi_3$.

Note that the eigenvalue derivatives along these two directions 
have different expressions, although the expressions for eigenvalues 
$e_l$ are the same for both the $x_1$- and $x_3$-directions. Therefore, the 
optimal damping ratios in the $x_1$- and $x_3$-directions might be different 
from one another. In addition, the searching range for the eigenvalue 
derivative should be $(0,\pi]$ instead of $(0,\pi/2]$. We verify these 
findings in numerical examples in the next section.

We call the aforementioned procedure based on analytic expressions of 
eigenvalues and eigenvalue derivatives the analytic approach. 

\subsection{Optimal damping ratios of 3D MPML}
Elastic-wave 
equations~\eqref{eq:elastic_2d_v}--\eqref{eq:elastic_2d_sigma} are also 
valid for 3D general anisotropic media, but with
\begin{align}
\mathbf{v} & = (v_1,v_2,v_3)^{\mathrm{T}}, \\
\boldsymbol{\sigma} & = (\sigma_{11},\sigma_{22},\sigma_{33},\sigma_{23},\sigma_{13},\sigma_{12})^{\mathrm{T}}, \\
\bfC &=\left[\begin{array}{cccccc}
C_{11} & C_{12} & C_{13} & C_{14} & C_{15} & C_{16}\\
C_{12} & C_{22} & C_{23} & C_{24} & C_{25} & C_{26}\\
C_{13} & C_{23} & C_{33} & C_{34} & C_{35} & C_{36}\\
C_{14} & C_{24} & C_{34} & C_{44} & C_{45} & C_{46}\\
C_{15} & C_{25} & C_{35} & C_{54} & C_{55} & C_{56}\\
C_{16} & C_{26} & C_{36} & C_{64} & C_{56} & C_{66}
\end{array}\right], \\
\boldsymbol{\Lambda}&=\left[\begin{array}{cccccc}
\frac{\partial}{\partial x_{1}} & 0 & 0 & 0 & \frac{\partial}{\partial x_{3}} & \frac{\partial}{\partial x_{2}}\\
0 & \frac{\partial}{\partial x_{2}} & 0 & \frac{\partial}{\partial x_{3}} & 0 & \frac{\partial}{\partial x_{1}}\\
0 & 0 & \frac{\partial}{\partial x_{3}} & \frac{\partial}{\partial x_{2}} & \frac{\partial}{\partial x_{1}} & 0
\end{array}\right].
\end{align}

Analogous to the 2D case, the 3D elastic-wave equations can be written 
using decomposed coefficient matrices $\bfC_i$ and $\mathbf{D}_i$ 
($i=1,2,3$) as
\begin{align}
\frac{\partial}{\partial t}\mathbf{v} & = \mathbf{D}_{1}\frac{\partial}{\partial x_{1}}\boldsymbol{\sigma}+\mathbf{D}_{2}\frac{\partial}{\partial x_{2}}\boldsymbol{\sigma}+\mathbf{D}_{3}\frac{\partial}{\partial x_{3}}\boldsymbol{\sigma},\\
\frac{\partial}{\partial t}\boldsymbol{\sigma} & = \mathbf{C}_{1}\frac{\partial}{\partial x_{1}}\mathbf{v}+\mathbf{C}_{2}\frac{\partial}{\partial x_{2}}\mathbf{v}+\mathbf{C}_{3}\frac{\partial}{\partial x_{3}}\mathbf{v}, 
\end{align} 
where
\begin{align}
\mathbf{C}_{1} &=\left[\begin{array}{ccc}
C_{11} & C_{16} & C_{15}\\
C_{12} & C_{26} & C_{25}\\
C_{13} & C_{36} & C_{35}\\
C_{14} & C_{46} & C_{45}\\
C_{15} & C_{56} & C_{55}\\
C_{16} & C_{66} & C_{56}
\end{array}\right], \\
\mathbf{C}_{2} &=\left[\begin{array}{ccc}
C_{16} & C_{12} & C_{14}\\
C_{26} & C_{22} & C_{24}\\
C_{36} & C_{23} & C_{34}\\
C_{46} & C_{24} & C_{44}\\
C_{56} & C_{25} & C_{54}\\
C_{66} & C_{26} & C_{64}
\end{array}\right], \\
\mathbf{C}_{3} &=\left[\begin{array}{ccc}
C_{15} & C_{14} & C_{13}\\
C_{25} & C_{24} & C_{23}\\
C_{35} & C_{34} & C_{33}\\
C_{45} & C_{44} & C_{34}\\
C_{55} & C_{45} & C_{35}\\
C_{65} & C_{46} & C_{36}
\end{array}\right], \\
\mathbf{D}_{1} &=\left[\begin{array}{cccccc}
1 & 0 & 0 & 0 & 0 & 0\\
0 & 0 & 0 & 0 & 0 & 1\\
0 & 0 & 0 & 0 & 1 & 0
\end{array}\right], \\
\mathbf{D}_{2} &=\left[\begin{array}{cccccc}
0 & 0 & 0 & 0 & 0 & 1\\
0 & 1 & 0 & 0 & 0 & 0\\
0 & 0 & 0 & 1 & 0 & 0
\end{array}\right], \\
\mathbf{D}_{3}&=\left[\begin{array}{cccccc}
0 & 0 & 0 & 0 & 1 & 0\\
0 & 0 & 0 & 1 & 0 & 1\\
0 & 0 & 1 & 0 & 0 & 0
\end{array}\right].
\end{align}

The system of wave equations in PMLs for the 3D case can be expressed in 
the form of an autonomous system in the wavenumber domain as
\begin{equation}
\frac{\partial \mathbf{U}}{\partial t} = \tilde{\bfA} \mathbf{U},
\end{equation}
where
\begin{align}
\mathbf{U}&=\mathcal{F}[\boldsymbol{\Psi}], \\
\boldsymbol{\Psi} & = (
\sigma_{11}^{(1)},\sigma_{22}^{(1)},\sigma_{33}^{(1)},
\sigma_{23}^{(1)},\sigma_{13}^{(1)},\sigma_{12}^{(1)}, 
\sigma_{11}^{(2)},\sigma_{22}^{(2)},\sigma_{33}^{(2)},
\sigma_{23}^{(2)},\sigma_{13}^{(2)},\sigma_{12}^{(2)}, \nonumber \\
&
\sigma_{11}^{(3)},\sigma_{22}^{(3)},\sigma_{33}^{(3)},
\sigma_{23}^{(3)},\sigma_{13}^{(3)},\sigma_{12}^{(3)}, v_1^{(1)}, v_2^{(1)}, v_3^{(1)}, 
v_1^{(2)}, v_2^{(2)}, v_3^{(2)}, 
v_1^{(3)}, v_2^{(3)}, v_3^{(3)})^{\mathrm{T}}, \\
\tilde{\bfA} & = \tilde{\bfA}_0 + \bfB, \\
\tilde{\mathbf{A}}_{0} &=\left[\begin{array}{cccccc}
\bfzero_{66} & \bfzero_{66} & \bfzero_{66} & ik_{1}\bfC_{1} & ik_{1}\bfC_{1} & ik_{1}\bfC_{1}\\
\bfzero_{66} & \bfzero_{66} & \bfzero_{66} & ik_{2}\bfC_{2} & ik_{2}\bfC_{2} & ik_{2}\bfC_{2}\\
\bfzero_{66} & \bfzero_{66} & \bfzero_{66} & ik_{3}\bfC_{3} & ik_{3}\bfC_{3} & ik_{3}\bfC_{3}\\
ik_{1}\rho^{-1}\mathbf{D}_{1} & ik_{1}\rho^{-1}\mathbf{D}_{1} & ik_{1}\rho^{-1}\mathbf{D}_{1} & \bfzero_{33} & \bfzero_{33} & \bfzero_{33}\\
ik_{2}\rho^{-1}\mathbf{D}_{2} & ik_{2}\rho^{-1}\mathbf{D}_{2} & ik_{2}\rho^{-1}\mathbf{D}_{2} & \bfzero_{33} & \bfzero_{33} & \bfzero_{33}\\
ik_{3}\rho^{-1}\mathbf{D}_{3} & ik_{3}\rho^{-1}\mathbf{D}_{3} & ik_{3}\rho^{-1}\mathbf{D}_{3} & \bfzero_{33} & \bfzero_{33} & \bfzero_{33}
\end{array}\right], \\
\mathbf{B} &=\left[\begin{array}{cccccc}
-d_{1}\mathbf{I}_{6} & \bfzero_{66} & \bfzero_{66} & \bfzero_{63} & \bfzero_{63} & \bfzero_{63}\\
\bfzero_{66} & -d_{2}\mathbf{I}_{6} & \bfzero_{66} & \bfzero_{63} & \bfzero_{63} & \bfzero_{63}\\
\bfzero_{66} & \bfzero_{66} & -d_{3}\mathbf{I}_{6} & \bfzero_{63} & \bfzero_{63} & \bfzero_{63}\\
\bfzero_{36} & \bfzero_{36} & \bfzero_{36} & -d_{1}\mathbf{I}_{3} & \bfzero_{33} & \bfzero_{33}\\
\bfzero_{36} & \bfzero_{36} & \bfzero_{36} & \bfzero_{33} & -d_{2}\mathbf{I}_{3} & \bfzero_{33}\\
\bfzero_{36} & \bfzero_{36} & \bfzero_{36} & \bfzero_{33} & \bfzero_{33} & -d_{3}\mathbf{I}_{3}
\end{array}\right],
\end{align}
and $d_i$ is the damping coefficient along the $x_i$-direction ($i=1,2,3$). 

To stabilize the PML in the $x_i$-direction, we need to employ nonzero 
damping ratios along the other two directions perpendicular to $x_i$.  
Therefore, for the $x_1$-, $x_2$- and $x_3$-directions, we respectively set 
the damping matrix to be
\begin{align}
\mathbf{B}_{1}(\xi_{1}) &=\left[\begin{array}{cccccc}
-d_{1}\mathbf{I}_{6} & \bfzero_{66} & \bfzero_{66} & \bfzero_{63} & \bfzero_{63} & \bfzero_{63}\\
\bfzero_{66} & -d_{1}\xi_{1}\mathbf{I}_{6} & \bfzero_{66} & \bfzero_{63} & \bfzero_{63} & \bfzero_{63}\\
\bfzero_{66} & \bfzero_{66} & -d_{1}\xi_{1}\mathbf{I}_{6} & \bfzero_{63} & \bfzero_{63} & \bfzero_{63}\\
\bfzero_{36} & \bfzero_{36} & \bfzero_{36} & -d_{1}\mathbf{I}_{3} & \bfzero_{33} & \bfzero_{33}\\
\bfzero_{36} & \bfzero_{36} & \bfzero_{36} & \bfzero_{33} & -d_{1}\xi_{1}\mathbf{I}_{3} & \bfzero_{33}\\
\bfzero_{36} & \bfzero_{36} & \bfzero_{36} & \bfzero_{33} & \bfzero_{33} & -d_{1}\xi_{1}\mathbf{I}_{3}
\end{array}\right],\label{eq:B1} \\
\mathbf{B}_{2}(\xi_{2}) &=\left[\begin{array}{cccccc}
-d_{2}\xi_{2}\mathbf{I}_{6} & \bfzero_{66} & \bfzero_{66} & \bfzero_{63} & \bfzero_{63} & \bfzero_{63}\\
\bfzero_{66} & -d_{2}\mathbf{I}_{6} & \bfzero_{66} & \bfzero_{63} & \bfzero_{63} & \bfzero_{63}\\
\bfzero_{66} & \bfzero_{66} & -d_{2}\xi_{2}\mathbf{I}_{6} & \bfzero_{63} & \bfzero_{63} & \bfzero_{63}\\
\bfzero_{36} & \bfzero_{36} & \bfzero_{36} & -d_{2}\xi_{2}\mathbf{I}_{3} & \bfzero_{33} & \bfzero_{33}\\
\bfzero_{36} & \bfzero_{36} & \bfzero_{36} & \bfzero_{33} & -d_{2}\mathbf{I}_{3} & \bfzero_{33}\\
\bfzero_{36} & \bfzero_{36} & \bfzero_{36} & \bfzero_{33} & \bfzero_{33} & -d_{2}\xi_{2}\mathbf{I}_{3}
\end{array}\right],\label{eq:B2} \\
\mathbf{B}_{3}(\xi_{3}) &=\left[\begin{array}{cccccc}
-d_{3}\xi_{3}\mathbf{I}_{6} & \bfzero_{66} & \bfzero_{66} & \bfzero_{63} & \bfzero_{63} & \bfzero_{63}\\
\bfzero_{66} & -d_{3}\xi_{3}\mathbf{I}_{6} & \bfzero_{66} & \bfzero_{63} & \bfzero_{63} & \bfzero_{63}\\
\bfzero_{66} & \bfzero_{66} & -d_{3}\mathbf{I}_{6} & \bfzero_{63} & \bfzero_{63} & \bfzero_{63}\\
\bfzero_{36} & \bfzero_{36} & \bfzero_{36} & -d_{3}\xi_{3}\mathbf{I}_{3} & \bfzero_{33} & \bfzero_{33}\\
\bfzero_{36} & \bfzero_{36} & \bfzero_{36} & \bfzero_{33} & -d_{3}\xi_{3}\mathbf{I}_{3} & \bfzero_{33}\\
\bfzero_{36} & \bfzero_{36} & \bfzero_{36} & \bfzero_{33} & \bfzero_{33} & -d_{3}\mathbf{I}_{3}
\end{array}\right].\label{eq:B3}
\end{align}
In the above 3D MPML damped matrices, we employ the same damping ratio 
along the two directions parallel with the PML/non-PML interface. For 
instance, for damping along the $x_1$-direction, we use the same nonzero 
damping ratio $\xi_1$ for the $x_2$- and $x_3$-directions, so both the 
damping coefficients along the $x_2$- and $x_3$-directions in 3D MPML are 
$\xi_1 d_1$. Using different damping ratios along different directions may also 
stabilize PML, but searching 
optimal values of damping ratios becomes even more complicated.

We develop a new approach to computing the eigenvalue derivatives of 
damped matrix $\tilde{\mathbf{A}}$ using eqs~\eqref{eq:B1}-\eqref{eq:B3} 
for 3D general anisotropic media. Matrix $\tilde{\mathbf{A}}$ (as 
well as $\tilde{\bfA}_0$) has a dimension of $27\times27$, resulting in 
an order of 27 of the characteristic polynomial of $\tilde{\bfA}$ or 
$\tilde{\bfA}_0$. Therefore, it is very  difficult, if not impossible, to 
derive analytic expressions for the eigenvalues and eigenvalue
derivatives, particularly for media with all $C_{ij}\neq0$.

We therefore adopt a numerical approach to solving for the eigenvalues
and eigenvalue derivatives. Using the 
definitions of eigenvalues and eigenvectors of matrix $\tilde{\bfA}$, we 
have
\begin{equation}\label{eq:right_eigen}
(\tilde{\mathbf{A}}-\lambda\mathbf{I}_{27})\mathbf{P}=\bfzero,
\end{equation}
where $\lambda$ is the eigenvalue of $\tilde{\mathbf{A}}$,
and the columns of $\mathbf{P}$ are the eigenvectors of 
$\tilde{\mathbf{A}}$. In addition, we have
\begin{equation}\label{eq:left_eigen}
\mathbf{Q}^{\mathrm{T}}(\tilde{\mathbf{A}}-\lambda\mathbf{I}_{27})=\bfzero,
\end{equation}
where the columns of $\mathbf{Q}^{\mathrm{T}}$ are the left eigenvectors of $\tilde{\mathbf{A}}$. 

Differentiating equation \eqref{eq:right_eigen} with respect to damping 
parameter $d_i$ gives
\begin{equation}\label{eq:eigen_df}
\left(\frac{\partial \tilde{\bfA}}{\partial d_i}-\frac{\partial \lambda}{\partial d_i} \bfI_{27}\right) \mathbf{P} + (\tilde{\bfA}-\lambda \bfI_{27}) \frac{\partial \mathbf{P}}{\partial d_i} =0. 
\end{equation}
Multiplying both sides of equation \eqref{eq:eigen_df} with 
$\mathbf{Q}^{\mathrm{T}}$ leads to
\begin{equation}
\mathbf{Q}^{\mathrm{T}}\left(\frac{\partial \tilde{\bfA}}{\partial d_i}-\frac{\partial \lambda}{\partial d_i} \bfI_{27}\right) \mathbf{P} + \mathbf{Q}^{\mathrm{T}} (\tilde{\bfA}-\lambda \bfI_{27}) \frac{\partial \mathbf{P}}{\partial d_i} =0,
\end{equation} 
which implies
\begin{equation}
\mathbf{Q}^{\mathrm{T}} \frac{\partial \tilde{\bfA}}{\partial d_i} \mathbf{P} =\mathbf{Q}^{\mathrm{T}}  \frac{\partial \lambda}{\partial d_i} \bfI_{27} \mathbf{P}.
\end{equation}
Therefore,
\begin{equation}
\frac{\partial \lambda}{\partial d_i} = \dfrac{\mathbf{Q}^{\mathrm{T}} \dfrac{\partial \tilde{\bfA}}{\partial d_i} \mathbf{P}}{\mathbf{Q}^{\mathrm{T}}  \bfI_{27} \mathbf{P}}.
\end{equation}

Because $\tilde{\bfA}=\tilde{\bfA}_0+\bfB_{i}(\xi_i)$ and 
$\tilde{\bfA}_0$ is irrelevant to $d_i$, the eigenvalue derivative along 
$x_i$-axis can be written as
\begin{equation}\label{eq:egd_3}
\chi_i (\xi_i)= \dfrac{\mathbf{Q}^{\mathrm{T}} \mathbf{R}_i(\xi_i) \mathbf{P}}{\mathbf{Q}^{\mathrm{T}}  \mathbf{P}},
\end{equation}
where 
\begin{align}
\mathbf{R}_{1}(\xi_{1}) &=\left[\begin{array}{cccccc}
-\mathbf{I}_{6} & \bfzero_{66} & \bfzero_{66} & \bfzero_{63} & \bfzero_{63} & \bfzero_{63}\\
\bfzero_{66} & -\xi_{1}\mathbf{I}_{6} & \bfzero_{66} & \bfzero_{63} & \bfzero_{63} & \bfzero_{63}\\
\bfzero_{66} & \bfzero_{66} & -\xi_{1}\mathbf{I}_{6} & \bfzero_{63} & \bfzero_{63} & \bfzero_{63}\\
\bfzero_{36} & \bfzero_{36} & \bfzero_{36} & -\mathbf{I}_{3} & \bfzero_{33} & \bfzero_{33}\\
\bfzero_{36} & \bfzero_{36} & \bfzero_{36} & \bfzero_{33} & -\xi_{1}\mathbf{I}_{3} & \bfzero_{33}\\
\bfzero_{36} & \bfzero_{36} & \bfzero_{36} & \bfzero_{33} & \bfzero_{33} & -\xi_{1}\mathbf{I}_{3}
\end{array}\right], \\
\mathbf{R}_{2}(\xi_{2}) &=\left[\begin{array}{cccccc}
-\xi_{2}\mathbf{I}_{6} & \bfzero_{66} & \bfzero_{66} & \bfzero_{63} & \bfzero_{63} & \bfzero_{63}\\
\bfzero_{66} & -\mathbf{I}_{6} & \bfzero_{66} & \bfzero_{63} & \bfzero_{63} & \bfzero_{63}\\
\bfzero_{66} & \bfzero_{66} & -\xi_{2}\mathbf{I}_{6} & \bfzero_{63} & \bfzero_{63} & \bfzero_{63}\\
\bfzero_{36} & \bfzero_{36} & \bfzero_{36} & -\xi_{2}\mathbf{I}_{3} & \bfzero_{33} & \bfzero_{33}\\
\bfzero_{36} & \bfzero_{36} & \bfzero_{36} & \bfzero_{33} & -\mathbf{I}_{3} & \bfzero_{33}\\
\bfzero_{36} & \bfzero_{36} & \bfzero_{36} & \bfzero_{33} & \bfzero_{33} & -\xi_{2}\mathbf{I}_{3}
\end{array}\right], \\
\mathbf{R}_{3}(\xi_{3}) &=\left[\begin{array}{cccccc}
-\xi_{3}\mathbf{I}_{6} & \bfzero_{66} & \bfzero_{66} & \bfzero_{63} & \bfzero_{63} & \bfzero_{63}\\
\bfzero_{66} & -\xi_{3}\mathbf{I}_{6} & \bfzero_{66} & \bfzero_{63} & \bfzero_{63} & \bfzero_{63}\\
\bfzero_{66} & \bfzero_{66} & -\mathbf{I}_{6} & \bfzero_{63} & \bfzero_{63} & \bfzero_{63}\\
\bfzero_{36} & \bfzero_{36} & \bfzero_{36} & -\xi_{3}\mathbf{I}_{3} & \bfzero_{33} & \bfzero_{33}\\
\bfzero_{36} & \bfzero_{36} & \bfzero_{36} & \bfzero_{33} & -\xi_{3}\mathbf{I}_{3} & \bfzero_{33}\\
\bfzero_{36} & \bfzero_{36} & \bfzero_{36} & \bfzero_{33} & \bfzero_{33} & -\mathbf{I}_{3}
\end{array}\right].
\end{align}

These equations indicate that we only need to obtain the eigenvalues and 
the left and right eigenvectors of matrix 
$\tilde{\bfA}(\bfC,\mathbf{k},\xi_i)$ for obtaining the optimal damping 
ratios of MPML in 3D general anisotropic media. This can be achieved 
using a linear algebra library such as LAPACK, and the procedure is 
summarized in Algorithm~\ref{alg:damping_ratio_3}. 

\vskip\baselineskip
\begin{algorithm}[H]
\DontPrintSemicolon
\SetKwInOut{Input}{input}
\SetKwInOut{Output}{output}

\Input{$\xi_i=0$, $\epsilon=-0.005$, $\Delta \xi=0.001$. }

\BlankLine
\For{$\theta \in(0,\pi]$}{
\For{$\phi \in(0,\pi]$}{
1) Calculate wavenumber $\mathbf{k}=(\cos\phi \sin\theta ,\sin\phi \sin\theta,\cos \theta)$\;
2) Calculate the left and right eigenvectors of 
$\tilde{\bfA} (\bfC,\mathbf{k},\xi_i)$ using a numerical 
eigensolver\;
3) Calculate the eigenvalue derivatives $\chi_i^{(l)}$ ($l=1,2,3$) according to eq.~\eqref{eq:egd_3}\;
4) $\chi_{i,\max} = \max(\chi_i^{(1)},\chi_i^{(2)},\chi_i^{(3)})$ \;
\If{$\chi_{i,\max} > \epsilon$}{
$\xi_i=\xi_i+\Delta \xi$\;
\textbf{go to} step 2\;
}
}
}

\BlankLine
\Output{$\xi_i$}

\label{alg:damping_ratio_3}
\caption{Determine the optimal damping ratio $\xi_i$ ($i=1,2,3$) 
for MPML in 3D general anisotropic media}
\end{algorithm}
\vskip\baselineskip

In the above algorithm, it is not necessary to seek analytic forms of the 
left/right eigenvectors, which is generally impossible for matrix 
$\tilde{\bfA}$. In our following numerical tests, we calculate the 
left/right eigenvectors with the Intel Math Kernel Library wrapper for 
LAPACK. The above numerical approach is obviously applicable to the 2D 
case with trivial modifications. Therefore, for 2D MPML, one can use 
either the analytic approach or the numerical approach, yet for 3D MPML, 
one can use only the numerical approach.

\section{Results}

We use three examples of 2D anisotropic media and three examples of 3D 
anisotropic media to validate the effectiveness of our new algorithms for 
calculating optimal damping ratios in 2D and 3D MPMLs. In the following, 
when presenting an elasticity matrix, we write only the upper triangle 
part of this matrix, but it should be clear that the elasticity matrix is 
essentially symmetric. We also assume that all the elasticity matrices 
have units of GPa, and all the media have mass density values of 
1000~kg/m$^3$ for convenience. 

\subsection{MPML for 2D anisotropic media}

To validate our new algorithm for determining the optimal damping ratios 
in MPML for 2D general anisotropic media, we consider a transversely 
isotropic medium with a horizontal symmetry axis (HTI medium), 
a transversely isotropic medium with a tilted symmetry axis (TTI medium), and a transversely isotropic medium with 
a vertical symmetry axis (VTI medium) with serious qS triplication 
in both $x_1$- and $x_3$-directions. 

For the HTI medium example, we use a well-known example with elasticity matrix \cite[]{Becache_etal_2003,Fajardo-Papa_2008}:
\begin{equation}
\bfC=\left[\begin{array}{ccc}
4 & 7.5 & 0\\
& 20 & 0\\
& & 2
\end{array}\right].
\label{eq:hti_2d}
\end{equation}
Note that this medium is considered as an orthotropic medium in 
\cite{Becache_etal_2003} and \cite{Fajardo-Papa_2008}. However, it could 
also be considered as an HTI medium on the $x_1x_3$-plane.  The only 
differences are that $C_{11}=C_{22}$ and $C_{44}=C_{55}$ for a 3D HTI 
medium, while there exists no such equality restrictions for an 
orthotropic medium.

\begin{figure}
\centering
\includegraphics[width=0.45\textwidth]{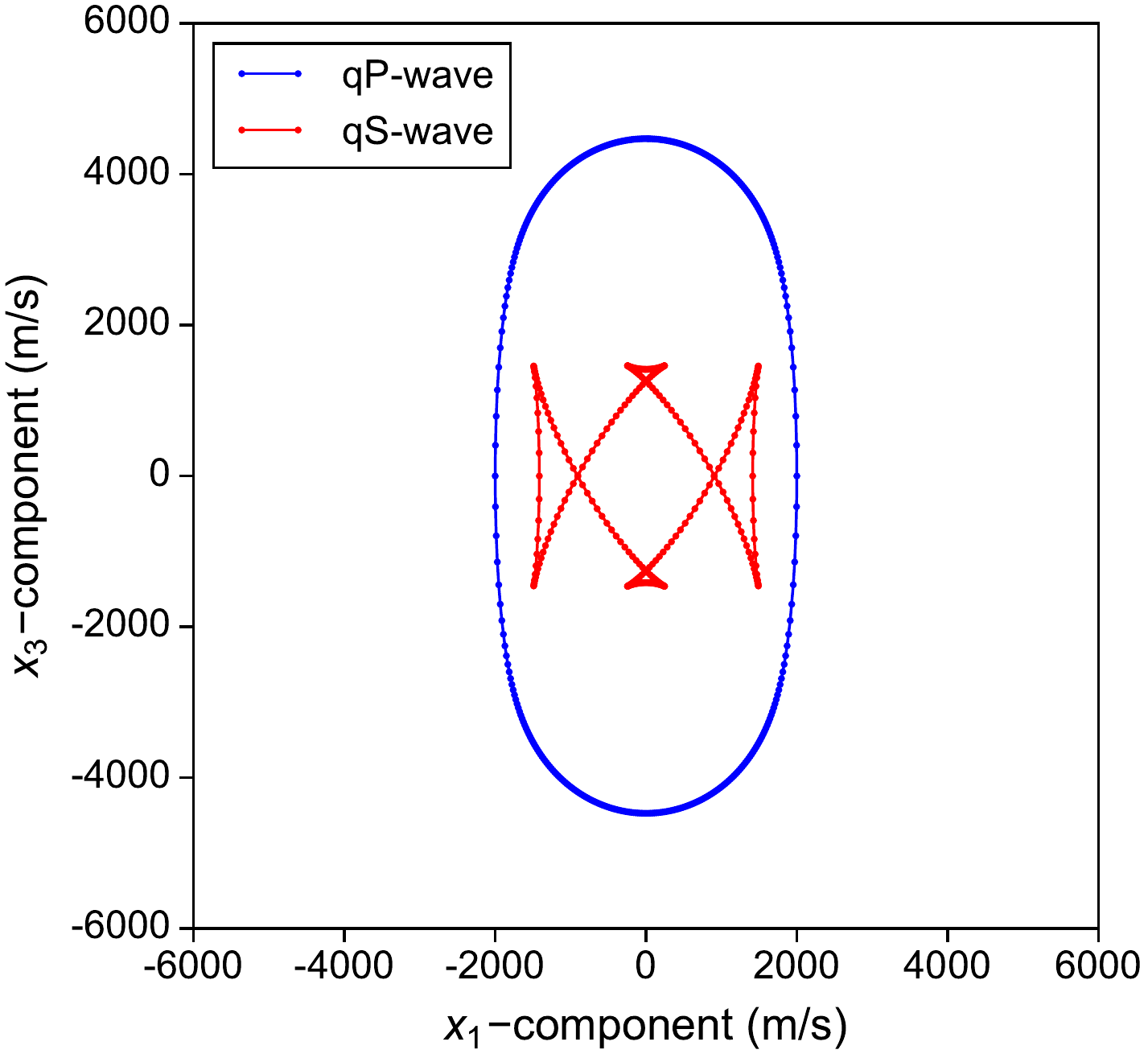}
\caption{Wavefront curves in the 2D HTI medium with elasticity matrix \eqref{eq:hti_2d}.} 
\label{fig:hti_2d_curve}
\end{figure}

The wavefront curves of qP- and qS-waves in Fig.~\ref{fig:hti_2d_curve}  
show the anisotropy characteristics of this HTI medium.
We employ Algorithm~\ref{alg:damping_ratio} to determine the optimal 
damping ratios in MPML along the $x_1$- and $x_3$-directions, leading to
\begin{equation}
\xi_1 =0.108, \qquad \xi_3=0.259.
\label{eq:hti_2d_ratio}
\end{equation}

In \cite{Fajardo-Papa_2008}, the suggested values of damping ratios are 
$\xi_1=0.30$ and $\xi_3=0.25$ for this HTI medium. Their 
suggested value for damping ratio $\xi_1$ is much larger than the optimal 
damping ratio given in eq.~\eqref{eq:hti_2d_ratio}, while their 
suggested value for $\xi_3$ is similar to the optimal 
damping ratio. 

\begin{figure}
\centering
\subfigure[]{\includegraphics[width=0.45\textwidth]{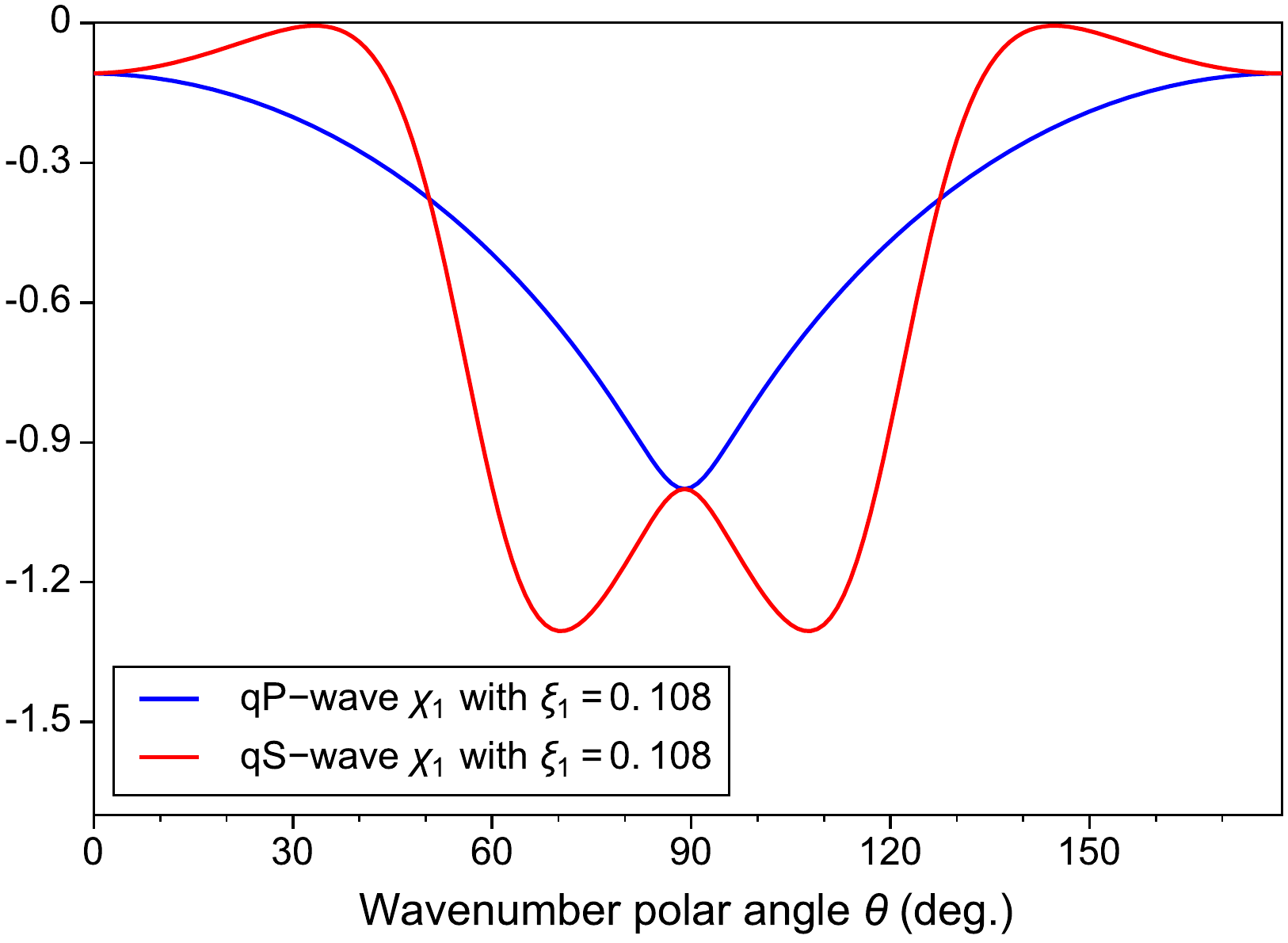}}
\subfigure[]{\includegraphics[width=0.45\textwidth]{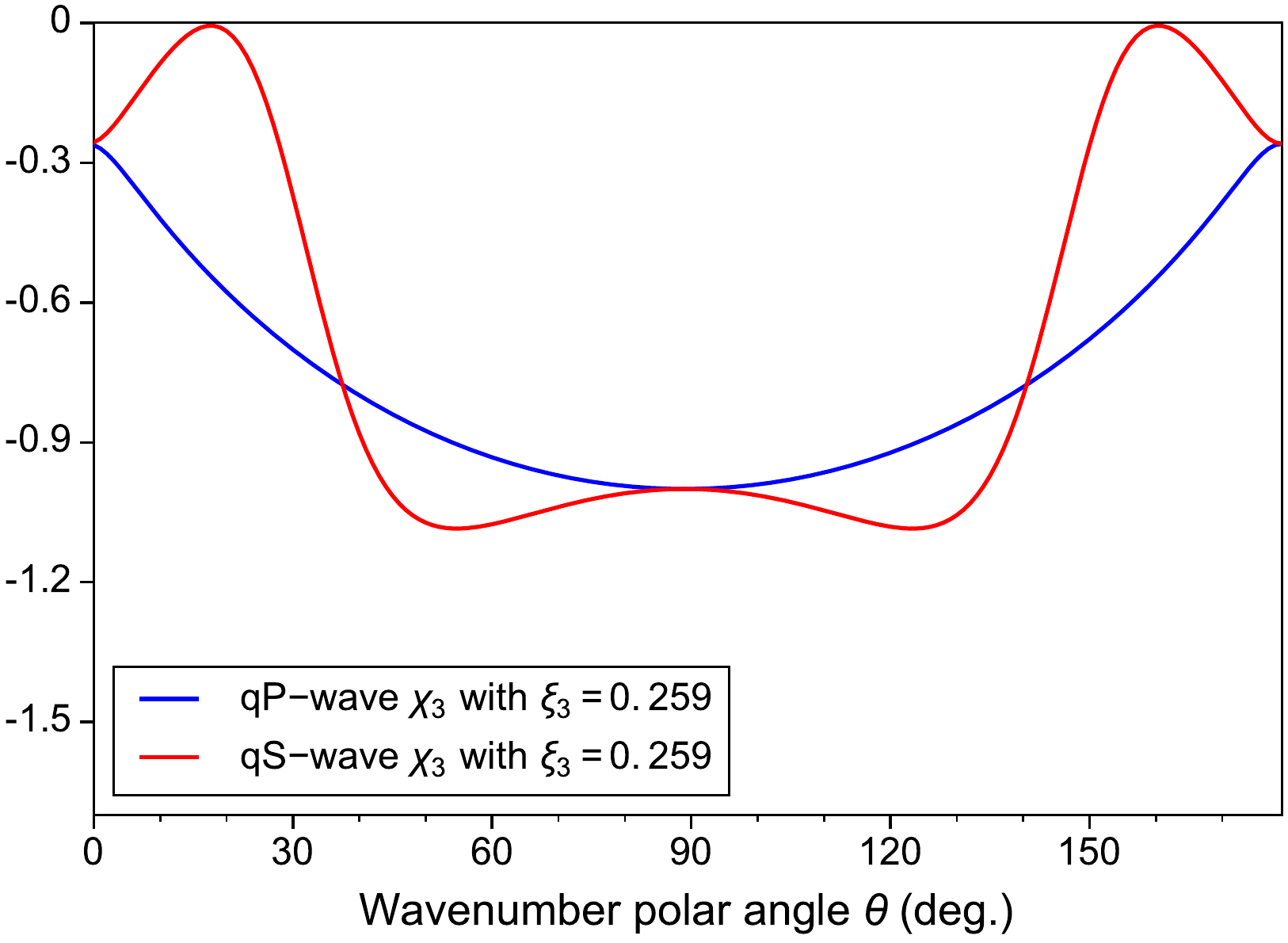}}
\caption{Eigenvalue derivatives of $\tilde{\bfA}$ of MPML in the (a) $x_1$- 
and (b) $x_3$-directions under calculated optimal damping ratios in eq.~\eqref{eq:hti_2d_ratio} for the 2D 
HTI medium with elasticity matrix \eqref{eq:hti_2d}. }
\label{fig:hti_2d_deriv}
\end{figure}

Figure~\ref{fig:hti_2d_deriv} plots the values of eigenvalue derivatives 
of $\tilde{\bfA}$ under the optimal damping ratios in the $x_1$- and $x_3$-directions. In both panels, the blue curves 
represent the qP-wave eigenvalue derivatives, and the red curves are for 
the qS-wave eigenvalue derivatives. Clearly, the qS-wave gives rise to the large damping ratios 
along both axes. Note that we set the threshold 
$\epsilon=-0.005$, therefore, in both panels of 
Fig.~\ref{fig:hti_2d_deriv}, the maximum values of the eigenvalue 
derivatives are $-0.005$. 

We validate the effectiveness of our new MPML in numerical modeling of 
anisotropic elastic-wave progation. We use the rotated-staggered grid (RSG) 
finite-difference method \cite[]{Saenger-etal_2000} to solve the 
stress-velocity form elastic-wave equations 
\eqref{eq:elastic_2d_v}--\eqref{eq:elastic_2d_sigma}. The RSG 
finite-difference method has 16th-order accuracy in space with optimal  
finite-difference coefficients \cite[]{Liu_2014}. We compute the 
wavefield energy decay curves of our wavefield modelings to validate 
the effectiveness of MPML.

\begin{figure}
\centering
\includegraphics[width=0.65\textwidth]{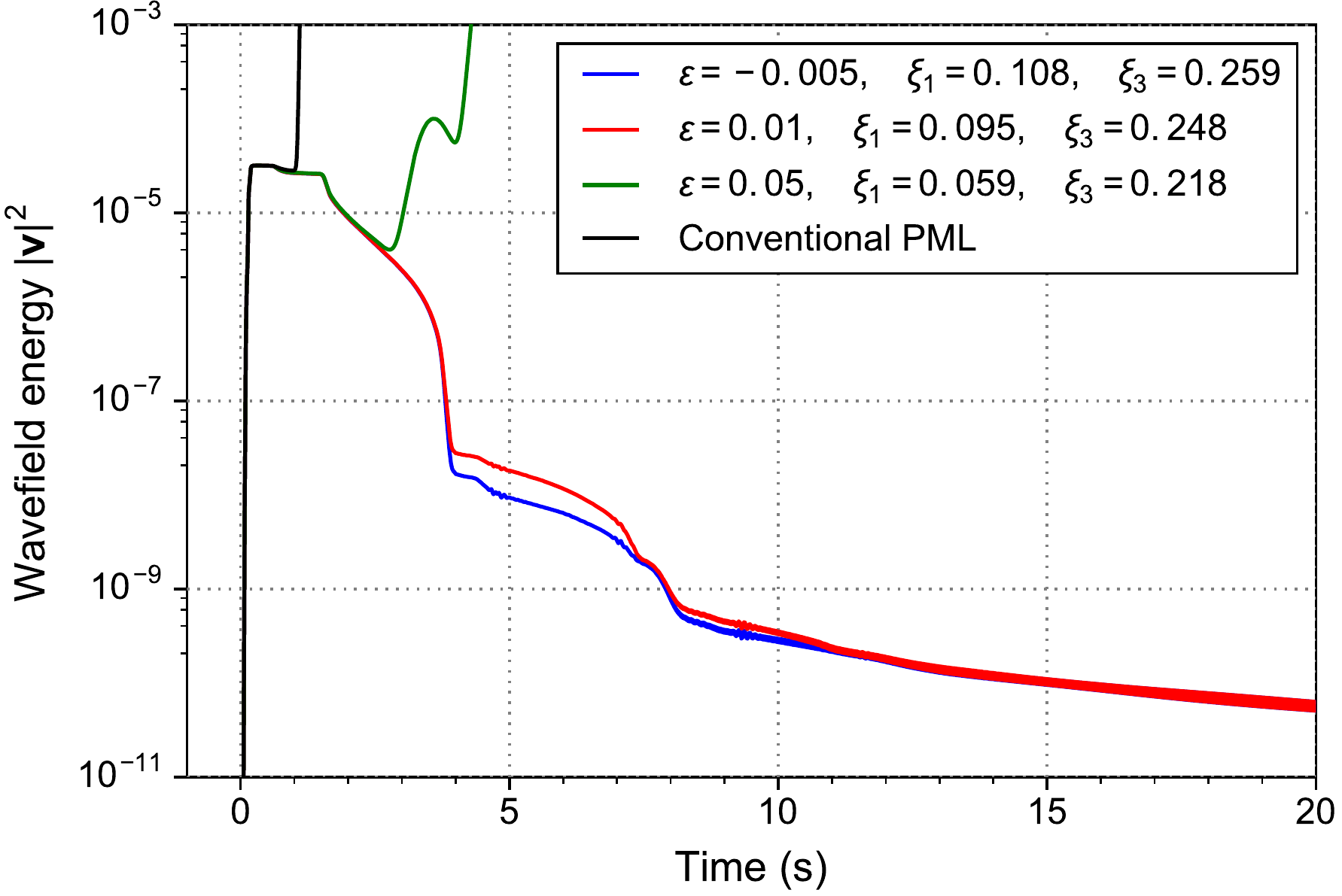}
\caption{Wavefield energy decay curves under different eigenvalue derivative thresholds for the 2D HTI medium with elasticity matrix \eqref{eq:hti_2d} within 20~s. Conventional PML can be considered as a special case of MPML with $\epsilon \gg 0$ or equivalently $\xi_1=\xi_3=0$.}
\label{fig:hti_2d_energy}
\end{figure}

In our numerical modeling, the model is defined in a $400\times 
400$ grid, and a PML of 30-node 
thickness are padded around the model domain. The grid size is 10~m in 
both the $x_1$- and $x_3$-directions. A vertical force vector source is 
located at the center of the computational domain, and a Ricker wavelet 
with a 10~Hz central frequency is used as the source time function. We 
simulate wave propagation for 20~s with a  time interval of 1~ms, which 
is smaller than what is required to satisfy the stability condition 
(about 1.54~ms). Figure~\ref{fig:hti_2d_energy} shows the resulting 
wavefield energy curve under the optimal damping ratios in eq.~\eqref{eq:hti_2d_ratio} 
, together with three others under different eigenvalue 
derivative threshold $\epsilon$ values, or equivalently, different 
damping ratios.  Figure~\ref{fig:hti_2d_energy} shows that within the 
20~s of wave propagation, the MPML with our calculated damping ratios 
$\xi_1=0.108$ and $\xi_3=0.259$ is stable.
These damping ratios are obtained under 
threshold value $\epsilon=-0.005$, meaning that the damping ratios have 
to ensure the 
eigenvalues derivatives $\chi_1$ and $\chi_3$ are not larger than 
$-0.005$ in the entire range of wavenumber direction.

We test the behavior of MPML under threshold $\epsilon=0.01$, or 
equivalently, $\xi_1=0.095$ and $\xi_3=0.248$, and show in 
Fig.~\ref{fig:hti_2d_energy} that the numerical modeling is stable. We 
further increase the threshold $\epsilon$ to be 0.05, and the MPML 
becomes unstable quickly after about 2~s.  Finally, the conventional PML, 
which is equivalent to MPML under $\xi_1=\xi_3=0$, becomes unstable even 
earlier (before 1~s).  These tests indicate that a small positive 
threshold $\epsilon$ may still result in a stable MPML.  However, there 
is no simple method to determine how large this positive $\epsilon$ to 
ensure the numerical stability. In this HTI medium case, 
$\epsilon=0.01$ results in stability while $\epsilon=0.05$ results in 
instability. Because a negative threshold resulting in stable MPML is 
consistent with the stability theory presented by 
\cite{Fajardo-Papa_2008}, we therefore should choose a negative threshold 
for the calculation of the optimal damping ratios to ensure  that the 
resulting MPML is stable. This is also verified in the hereinafter numerical  
examples.

\begin{figure}
\centering
\includegraphics[width=0.45\textwidth]{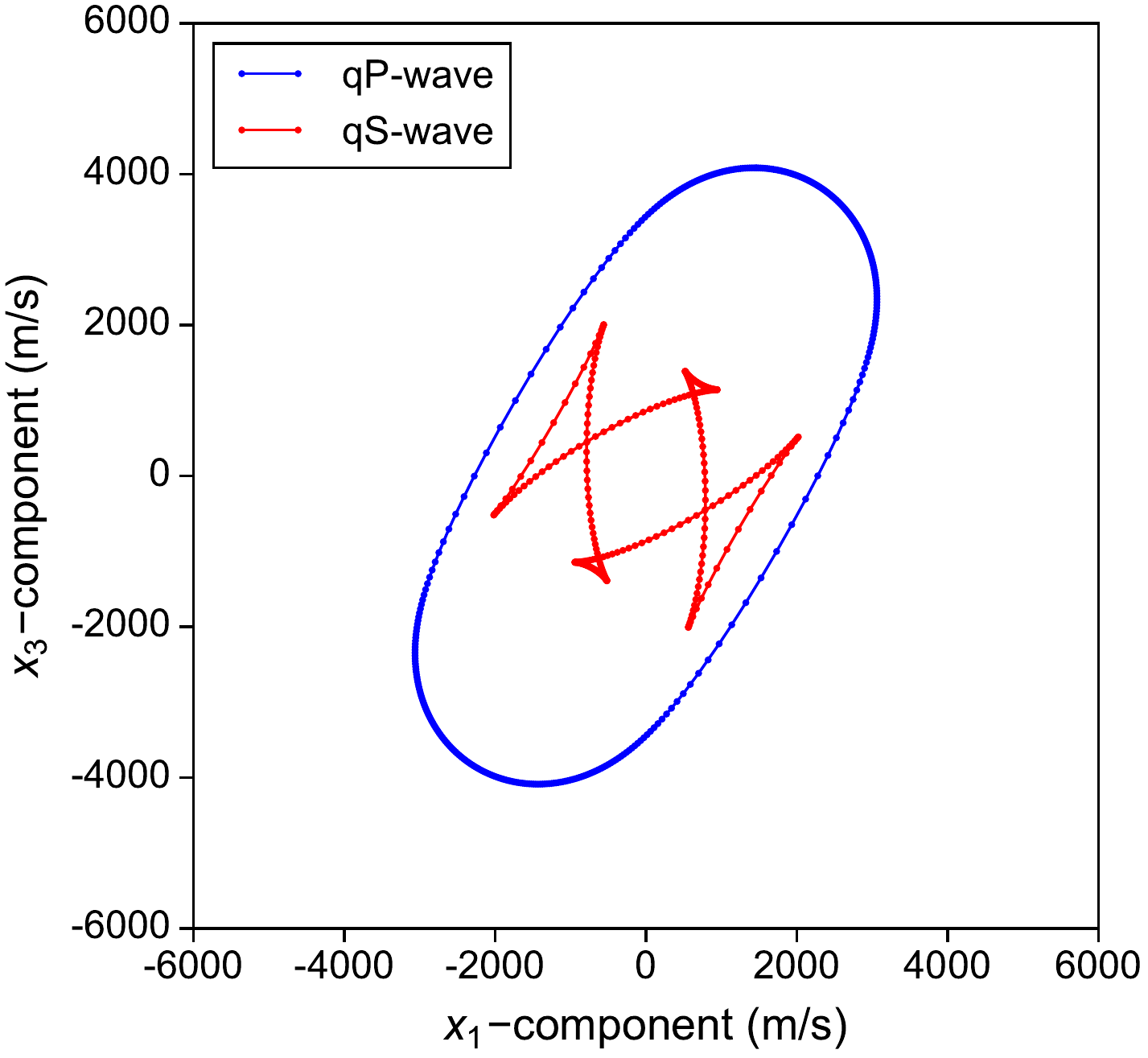}
\caption{Wavefront curves in the 2D HTI medium with elasticity matrix \eqref{eq:tti_2d}. }
\label{fig:tti_2d_curve}
\end{figure}

Next, we rotate the aforementioned HTI medium with respect to the $x_2$-axis 
clockwise by $\pi/6$ to obtain a TTI medium represented by the following 
elasticity matrix:
\begin{equation}
\bfC=\left[\begin{array}{ccc}
7.8125 & 7.6875 & 3.35585 \\
&15.8125 & 3.57235  \\
& & 2.1875
\end{array}\right],
\label{eq:tti_2d}
\end{equation}
with unit GPa. The rotation can be accomplished by rotation matrix 
\cite[e.g.,][]{Slawinski_2010}. The wavefront curves in this TTI medium is 
shown in Fig.~\ref{fig:tti_2d_curve}. Although this TTI medium is the 
rotation result of the HTI medium in the previous numerical example, it 
is not obvious how to change the damping ratios accordingly.  
We obtain the following optimal damping ratios of MPML under 
$\epsilon=-0.005$ using Algorithm~\ref{alg:damping_ratio}:
\begin{equation}
\xi_1=0.157, \qquad \xi_3=0.226.
\label{eq:tti_2d_ratio}
\end{equation}

\begin{figure}
\centering
\subfigure{\includegraphics[width=0.45\textwidth]{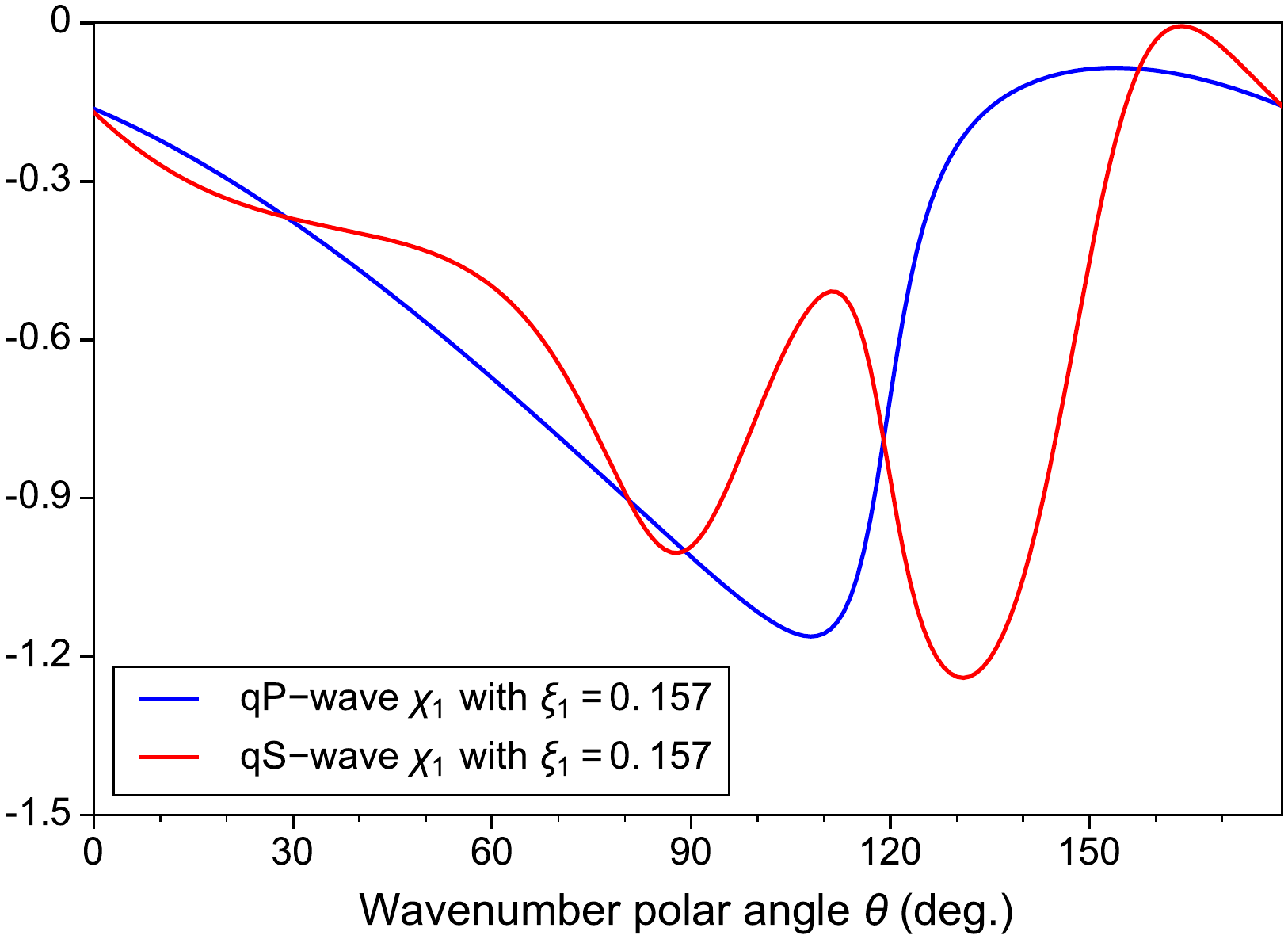}}
\subfigure{\includegraphics[width=0.45\textwidth]{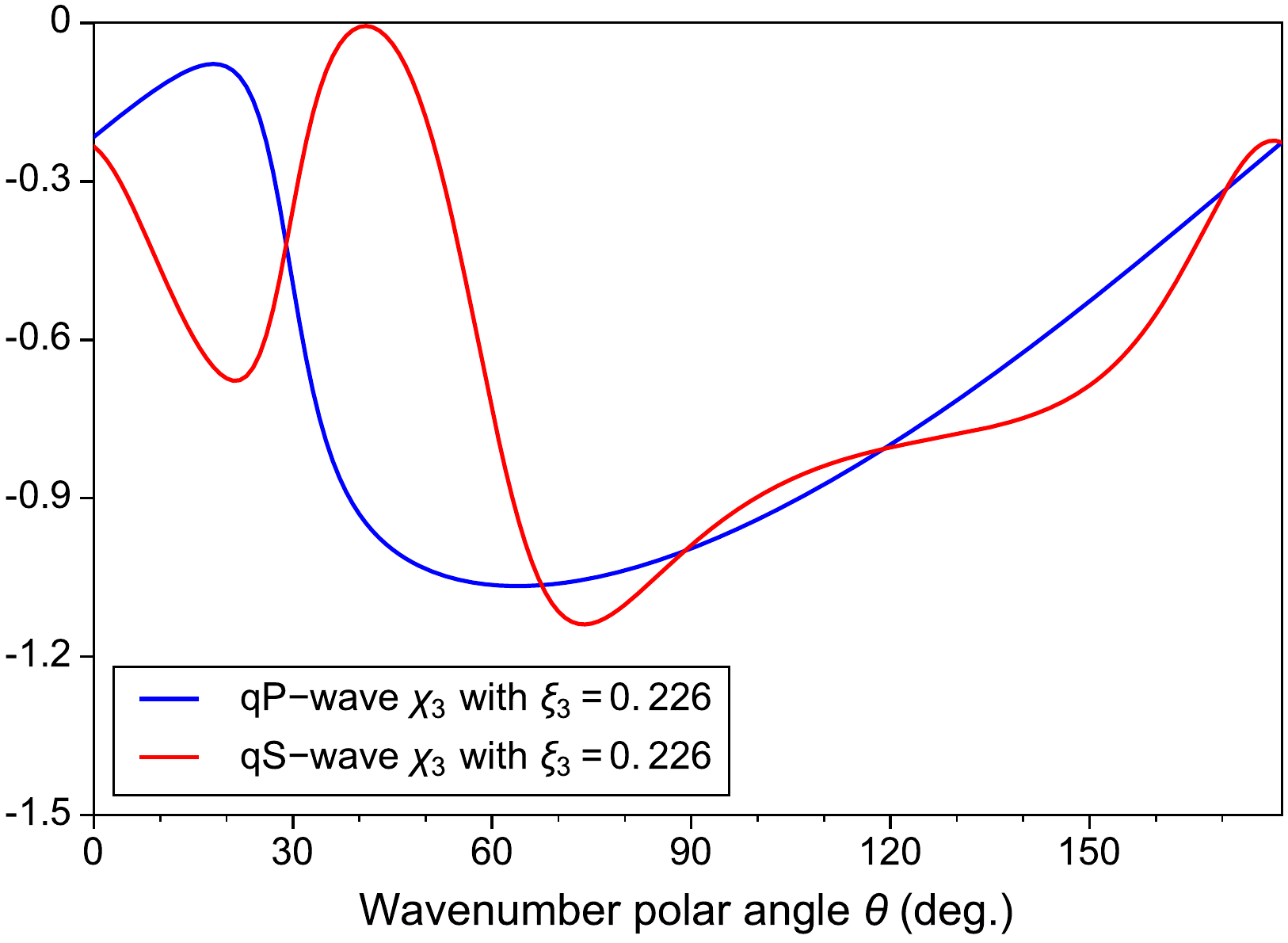}}
\caption{Eigenvalue derivatives of $\tilde{\bfA}$ of MPML in the (a) $x_1$- and (b) $x_3$-directions under calculated optimal damping ratios in eq.~\eqref{eq:tti_2d_ratio} for the 2D TTI medium with elasticity matrix \eqref{eq:tti_2d}. }
\label{fig:tti_2d_deriv}
\end{figure}

The eigenvalue derivatives under this set of damping ratios are shown in 
Fig.~\ref{fig:tti_2d_deriv}. The 
eigenvalue derivative curves are no longer symmetric with respect to 
$\theta=\pi/2$ (or has a period of $\pi/2$) as those for the 2D HTI 
medium (Fig.~\ref{fig:hti_2d_deriv}). Instead, they are periodic every 
$\pi$ angle, corresponding to the fact that there is always at least one 
symmetric axis for whatever kind of 2D anisotropic medium in the axis plane. These curves also indicate that, for 2D general 
anisotropic medium (TTI medium in this example), it is necessary to 
determine the values of eigenvalue derivatives within the range of 
$(0,\pi]$ instead of $(0,\pi/2]$.  Using only the range $(0,\pi/2]$ can 
lead to a totally incorrect optimal value of $\xi_1$, since the maximum value of 
$\chi_1$ in the range $(0,\pi/2]$ is smaller than that in the range 
$(\pi/2,\pi]$ for this TTI medium. In other words, even though $\chi_1$ in $(0,\pi/2]$ 
indicates a stable MPML, the MPML may still be unstable since $\chi_1$ may be larger than zero in $(\pi/2,\pi]$. Therefore, for anisotropic media with symmetric axis not aligned 
with a coordinate axis, it is necessary to consider the values 
of $\chi_i$ in wavenumber direction $\theta \in (0,\pi]$. This statement is also true for 3D 
anisotropic media as shown in the hereinafter 3D numerical examples.  

\begin{figure}
	\centering
	\includegraphics[width=0.65\textwidth]{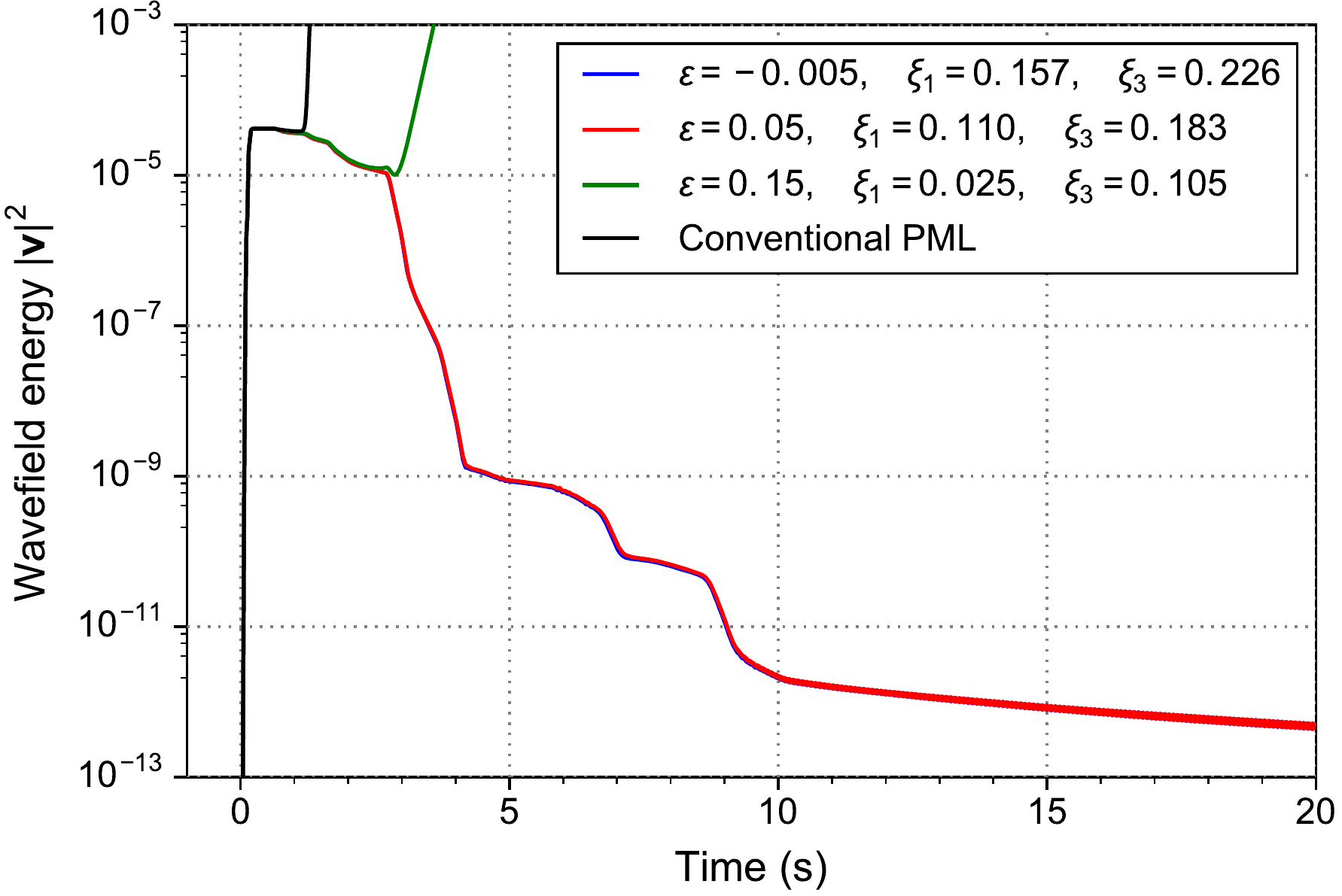}
	\caption{Wavefield energy decay curves under different eigenvalue derivative thresholds in the 2D TTI medium with elasticity matrix \eqref{eq:tti_2d} within 20~s.}
\label{fig:tti_2d_energy}
\end{figure}

Figure~\ref{fig:tti_2d_energy} displays the wavefield energy decay curves 
for this TTI medium under the optimal damping ratios, as well as under 
damping ratios calculated with positive eigenvalue derivative thresholds.  
In this example, the wavefield energy decays gradually within 
20~s for both cases with $\epsilon=-0.005$ and $\epsilon=0.05$. The 
numerical modeling with $\epsilon=0.15$ becomes unstable. For 
comparison, in the previous HTI case, $\epsilon=0.05$ results in an 
unstable MPML. These results further demonstrate that a positive eigenvalue derivative threshold should not be chosen to  
calculate the damping ratios, although a small positive $\epsilon$ 
might result in stable MPML. In contrast, a negative $\epsilon$ can always 
ensure the stability of MPML.

\begin{figure}
\centering
\includegraphics[width=0.45\textwidth]{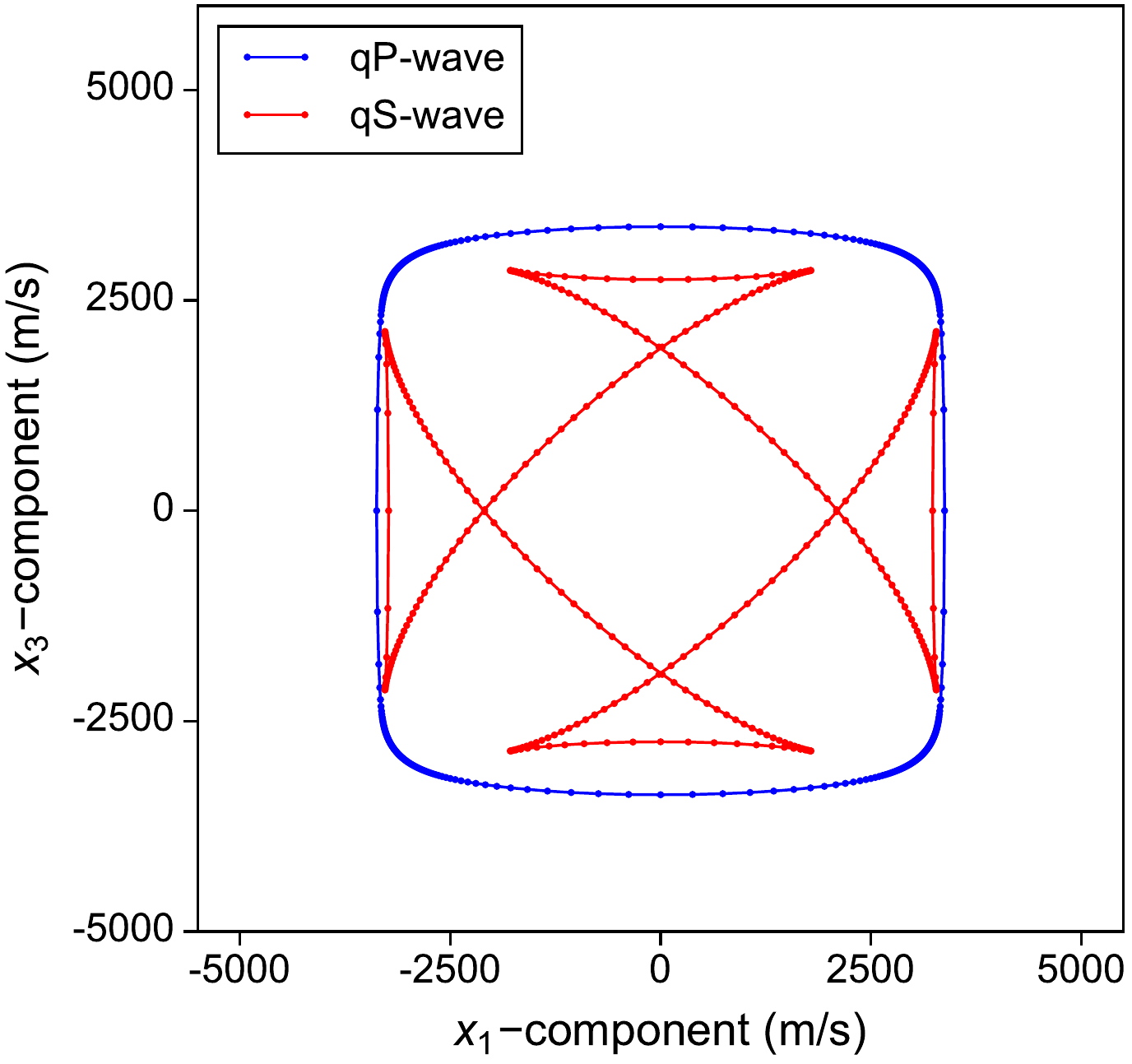}
\caption{Wavefront curves in the 2D VTI medium with elasticity matrix \eqref{eq:vti_2d}. }
\label{fig:vti_2d_curve}
\end{figure}

\begin{figure}
\centering
\subfigure{\includegraphics[width=0.45\textwidth]{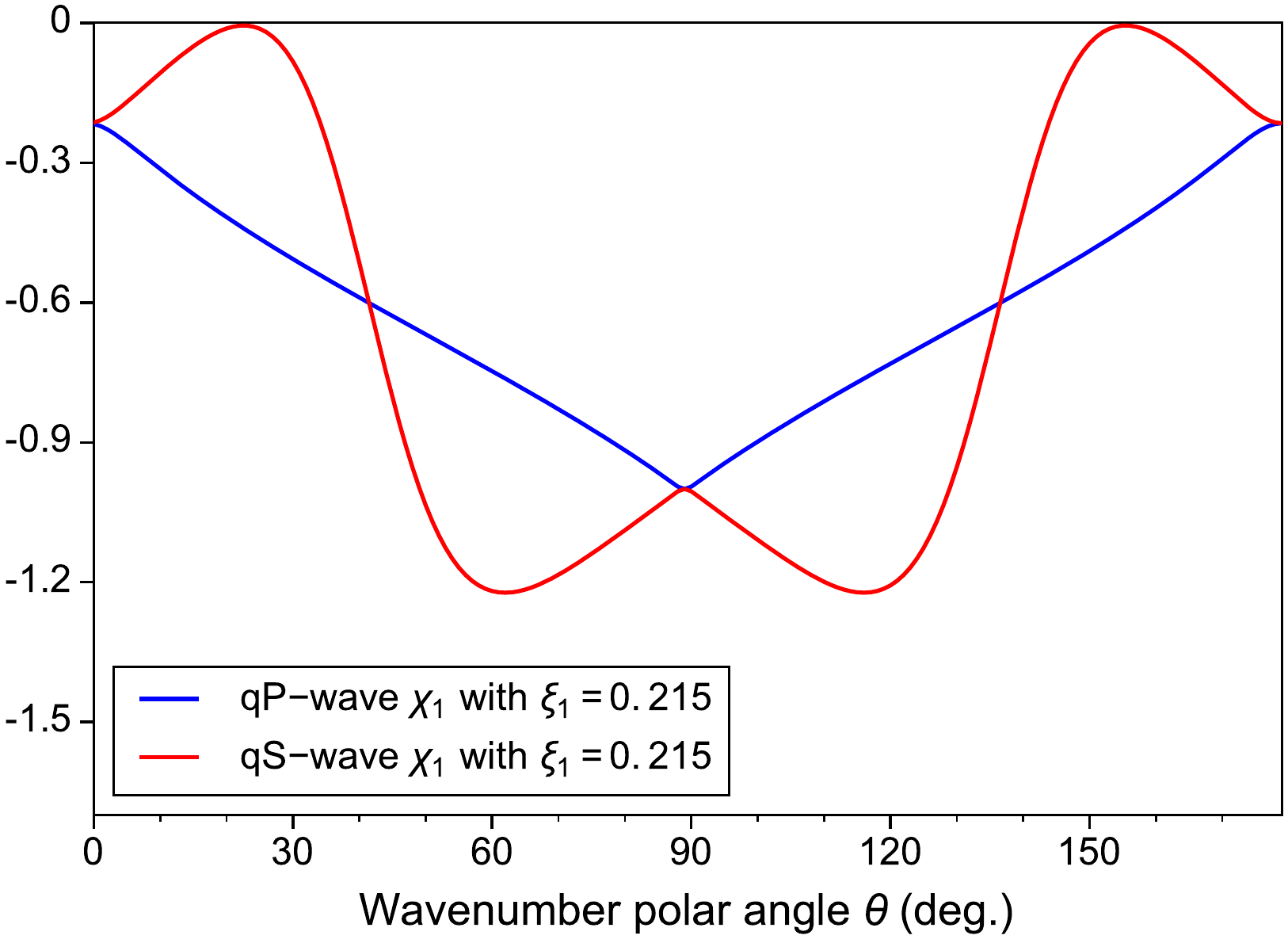}}
\subfigure{\includegraphics[width=0.45\textwidth]{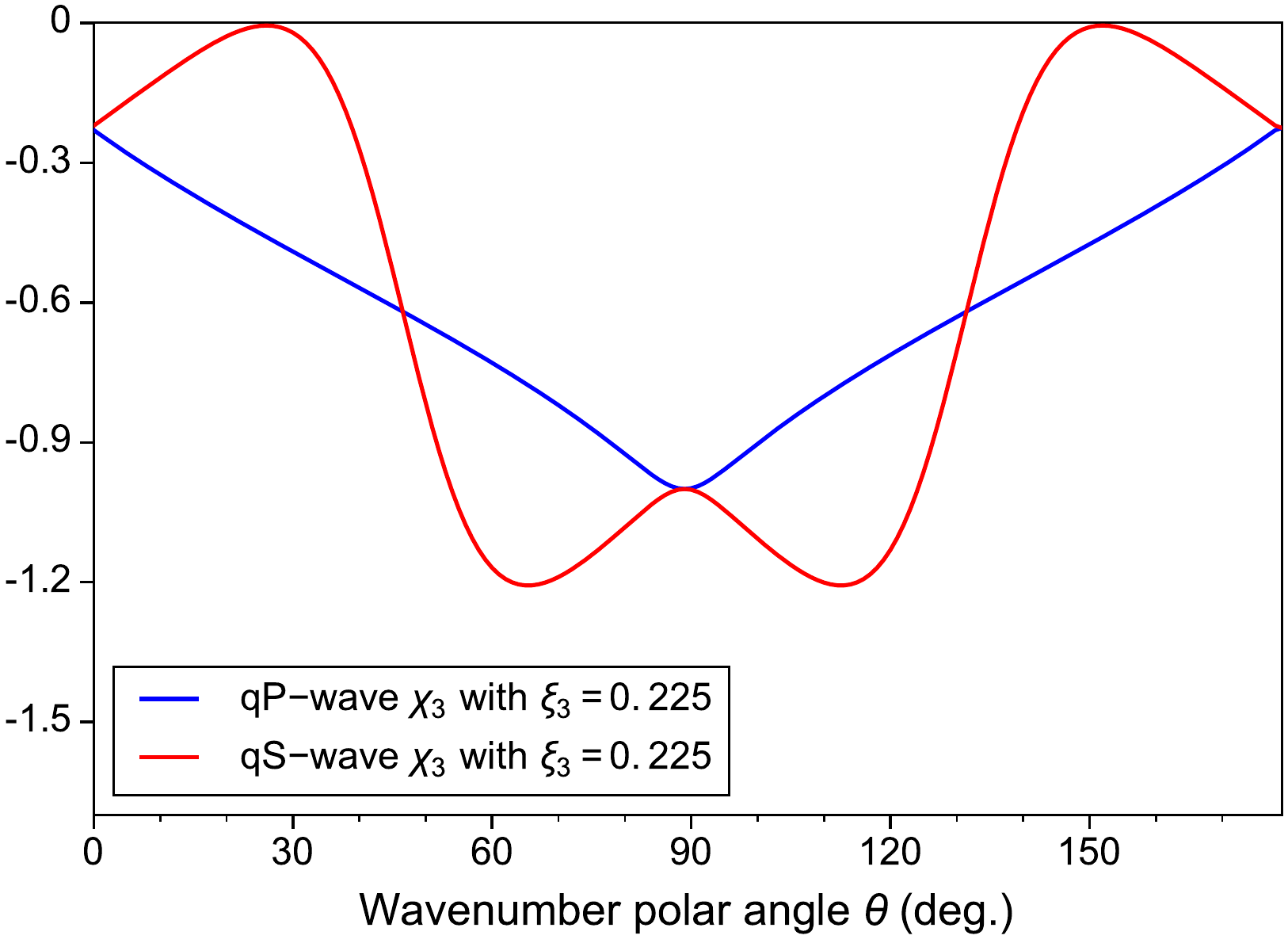}}
\caption{Eigenvalue derivatives of $\tilde{\bfA}$ of MPML in the (a) $x_1$- and (b) $x_3$-directions under calculated optimal damping ratios in eq.~\eqref{eq:vti_2d_ratio} for the 2D VTI medium with elasticity matrix \eqref{eq:vti_2d}. }
\label{fig:vti_2d_deriv}
\end{figure}

\begin{figure}
\centering
\includegraphics[width=0.65\textwidth]{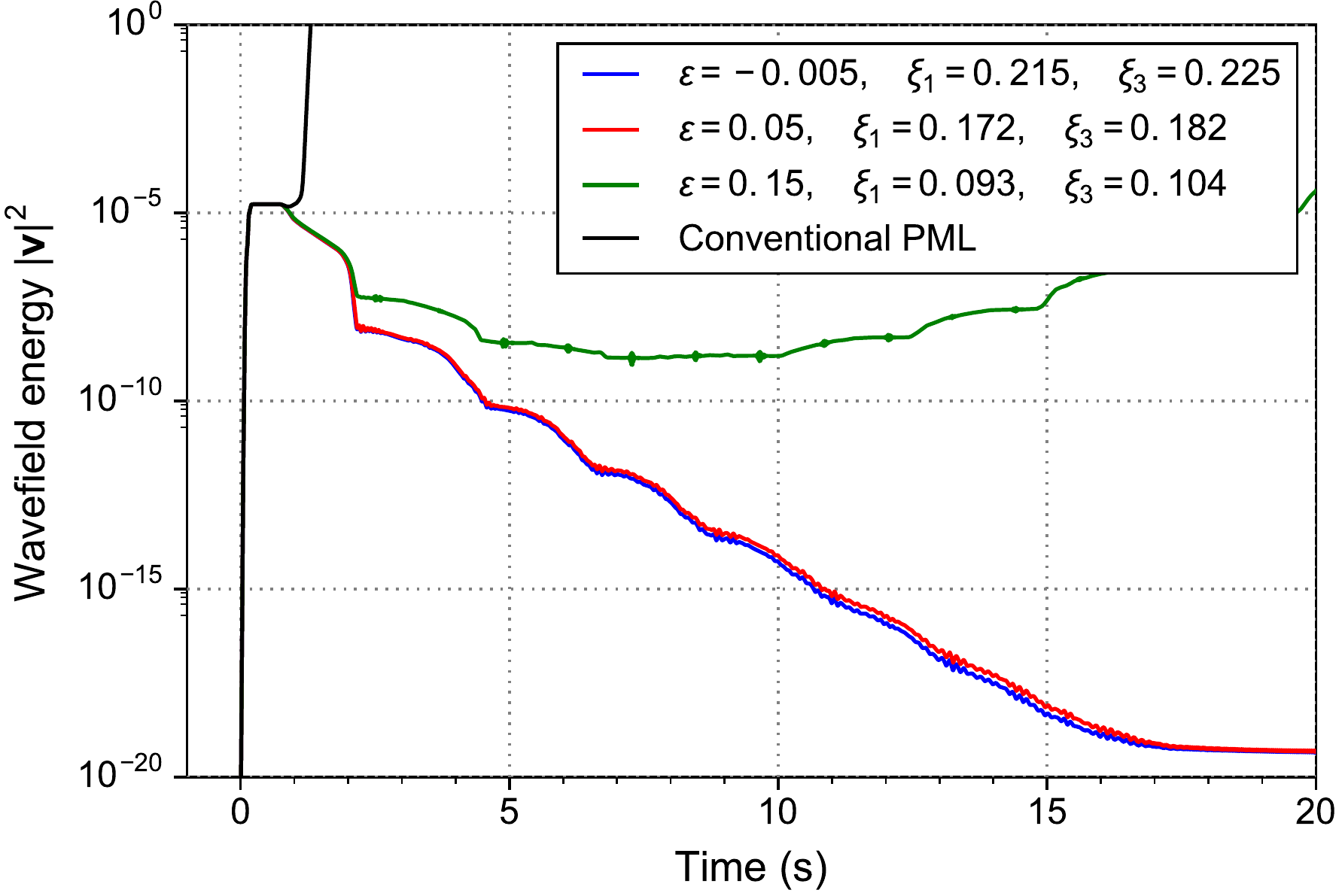}
\caption{Wavefield energy decay curves under different eigenvalue derivative thresholds in the 2D VTI medium with elasticity matrix \eqref{eq:vti_2d} within 20~s.}
\label{fig:vti_2d_energy}
\end{figure}

Our next numerical example uses a VTI medium defined 
by
\begin{equation}
\bfC=\left[\begin{array}{ccc}
10.4508 & 4.2623 & 0\\
& 7.5410 & 0\\
& & 11.3934
\end{array}\right].
\label{eq:vti_2d}
\end{equation}
The wavefront curves for this VTI medium are depicted in 
Fig.~\ref{fig:vti_2d_curve}.  The special feature of this VTI medium is 
that, in both the $x_1$- and $x_3$-directions, there exists serious qS-wave 
triplication phenomena.  We obtain the following optimal damping ratios 
using Algorithm~\ref{alg:damping_ratio}:
\begin{equation}
\xi_1= 0.215, \qquad \xi_3=0.225.
\label{eq:vti_2d_ratio}
\end{equation}
The corresponding eigenvalue derivatives in the $x_1$- and $x_3$-directions 
are displayed in Fig.~\ref{fig:vti_2d_deriv}. Again, it is the qS-wave that causes the damping ratios to be large to 
stabilize PML. Figure~\ref{fig:vti_2d_energy} depicts the wavefield energy 
decay curves under different eigenvalue derivative thresholds.  Similar 
with that of 2D TTI medium example, a positive threshold 0.05 can still 
stabilize PML, yet a value of 0.15 makes the MPML unstable. We choose 
a negative value of $\epsilon$ to ensure a stable MPML. 

\subsection{MPML for 3D anisotropic media}

For 3D anisotropic media, we need to determine the optimal MPML damping 
ratios along all three coordinate directions. We use three different 
anisotropic media (a quasi-VTI medium, a quasi-TTI medium and a triclinic 
medium) to demonstrate the determination of optimal MPML damping ratios.  

\begin{figure}
\centering
\subfigure[]{\includegraphics[width=0.45\textwidth]{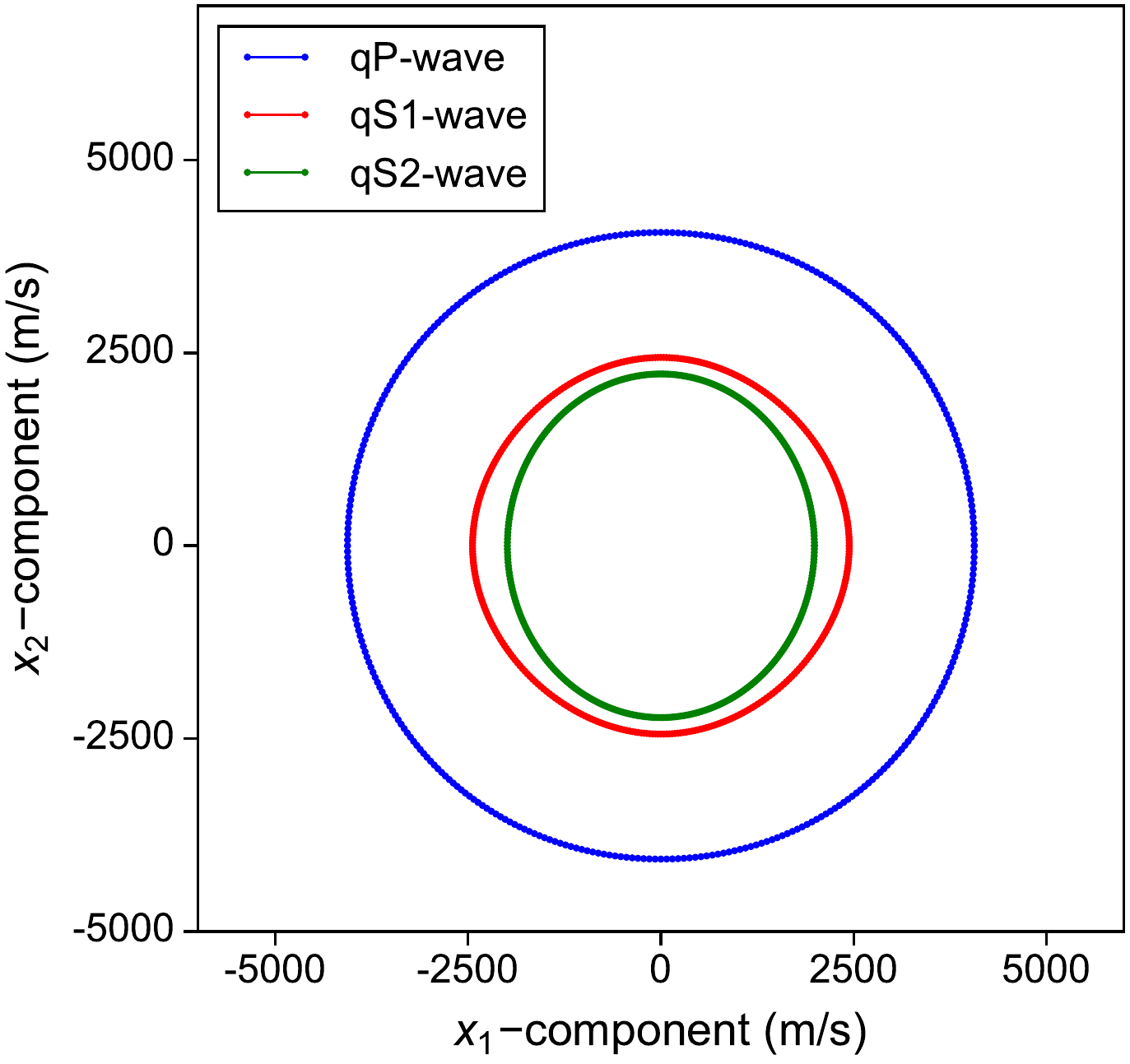}}
\subfigure[]{\includegraphics[width=0.45\textwidth]{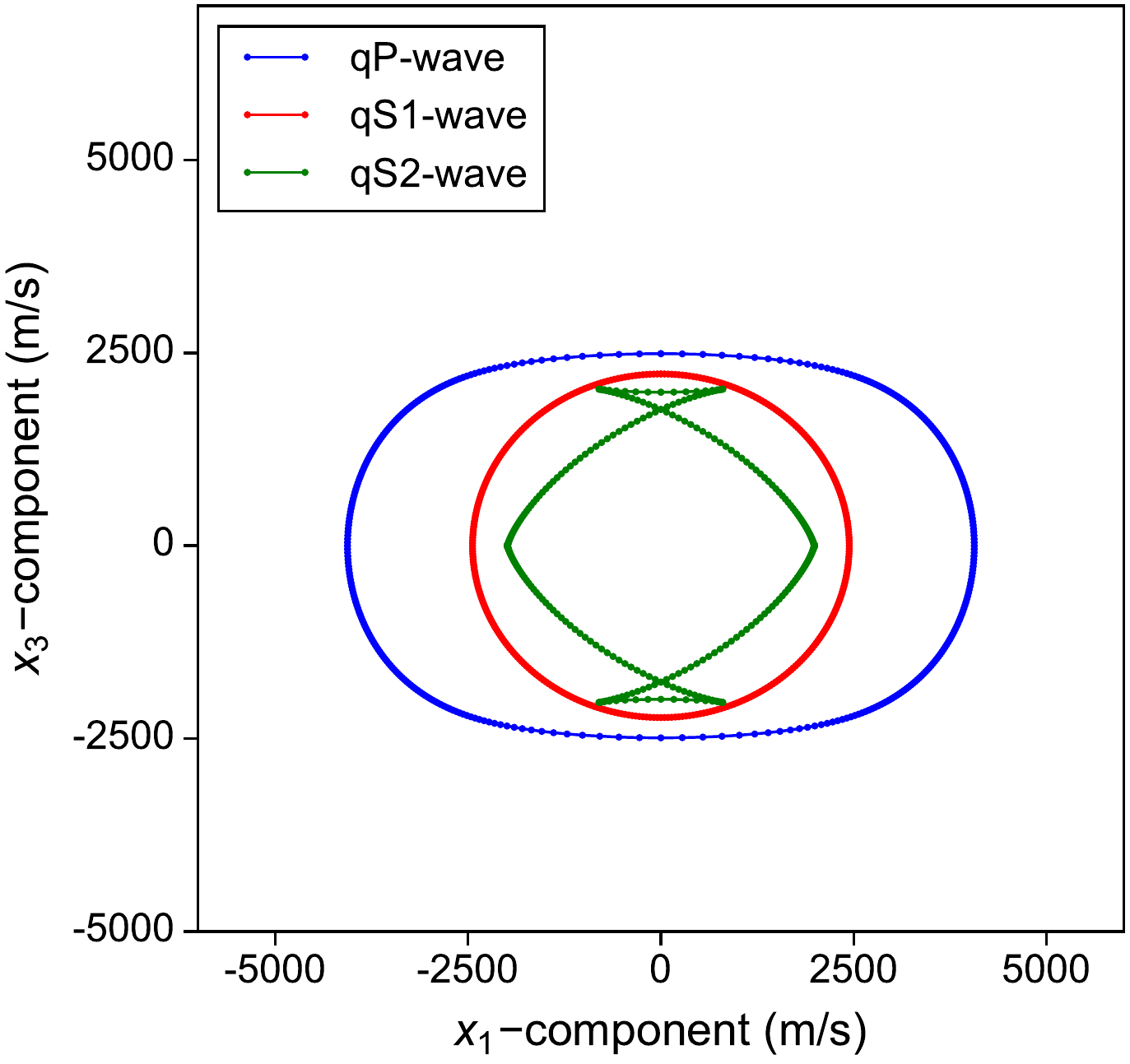}}
\subfigure[]{\includegraphics[width=0.45\textwidth]{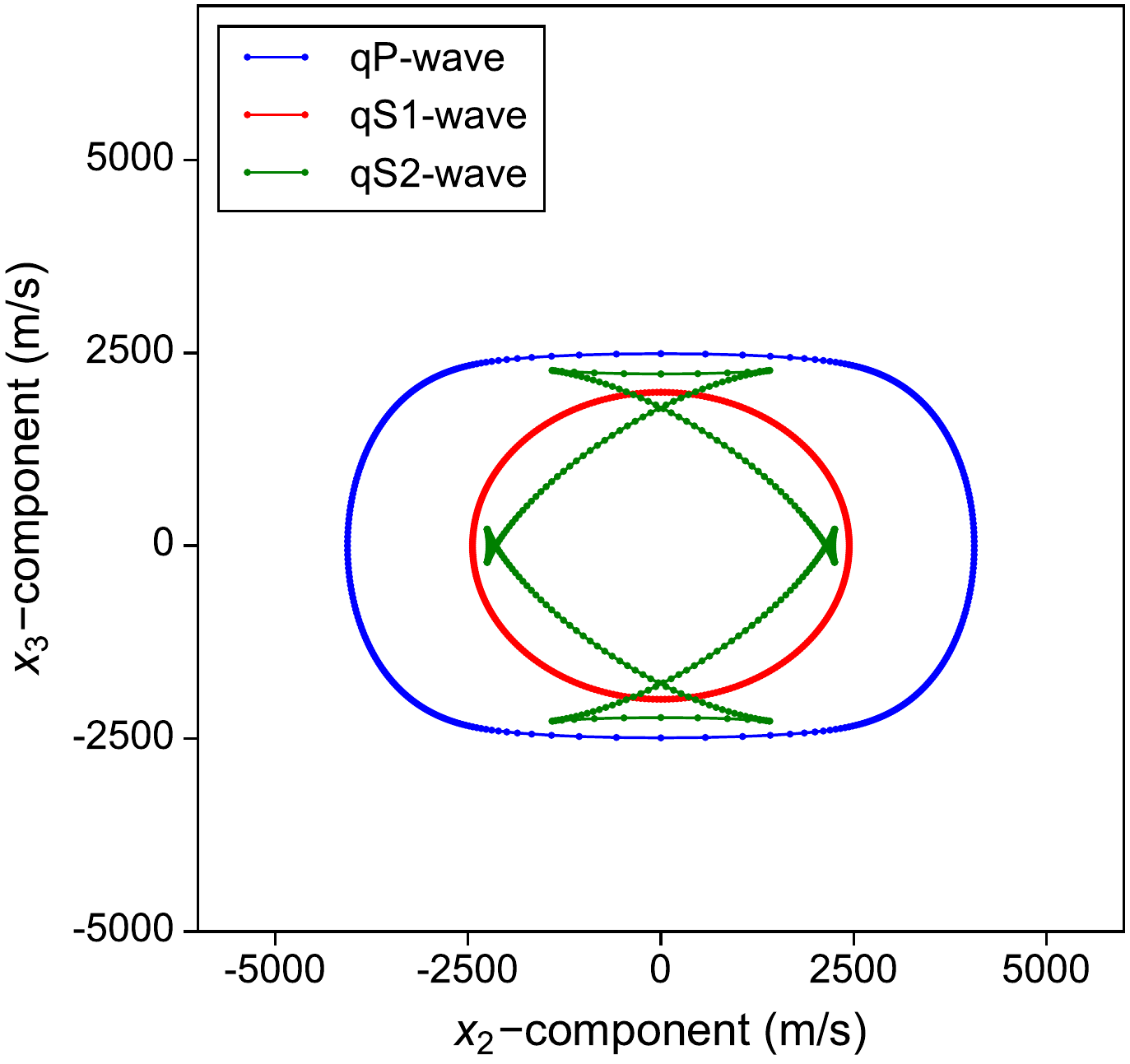}}
\caption{Wavefront curves in the 3D quasi-VTI medium with elasticity matrix \eqref{eq:vti_3d} on the (a) $x_1x_2$ (b) $x_1x_3$ and (c) $x_2x_3$ axis plane. qS1 and qS2 represents the two qS-waves. }
\label{fig:vti_3d_curve}
\end{figure}

We first use a 3D anisotropic medium represented by the elasticity matrix
\begin{equation}
\bfC =\left[\begin{array}{cccccc}
16.5 & 5 & 5 & 0  & 0&0 \\
& 16.5 & 5 & 0 & 0& 0\\
& & 6.2 & 0& 0& 0\\
& & & 4.96 & 0 & 0\\
& & & & 3.96 &0 \\
& & & & & 5.96\end{array}\right]. 
\label{eq:vti_3d}
\end{equation}
This elasticity matrix is modified from the elasticity matrix of zinc (a 
VTI medium, or hexagonal anisotropic medium) to increase the complexity of 
the resulting wavefronts and the characteristics of the eigenvalue 
derivatives along all three directions.  This modified elastic matrix 
still represents a physically feasible medium since it is easy to verify that 
it satisfies the following stability condition for anisotropic media \cite[]{Slawinski_2010}:
\begin{equation}
\det \left[\begin{array}{ccc}
C_{11} & \cdots & C_{1n}\\
\vdots & \ddots & \vdots\\
C_{1n} & \cdots & C_{nn}
\end{array}\right] >0,
\end{equation}
where $n=1,2,\cdots,6$. We call this anisotropic medium the quasi-VTI 
medium. Figure~\ref{fig:vti_3d_curve} shows the wavefront curves of this 
quasi-VTI medium on three axis planes.  

For comparison, a standard 3D VTI medium can be expressed by its five 
independent elasticity constants as \cite[e.g.,][]{Slawinski_2010}
\begin{equation}
\mathbf{C}=\left[\begin{array}{cccccc}
C_{11} & C_{12} & C_{13} & 0 & 0 & 0\\
& C_{11} & C_{13} & 0 & 0 & 0\\
&  & C_{33} & 0 & 0 & 0\\
&  &  & C_{44} & 0 & 0\\
&  &  &  & C_{44} & 0\\
&  &  &  &  & \frac{C_{11}-C_{12}}{2}
\end{array}\right].
\end{equation}

\begin{figure}
\centering
\subfigure[]{\includegraphics[width=0.3\textwidth]{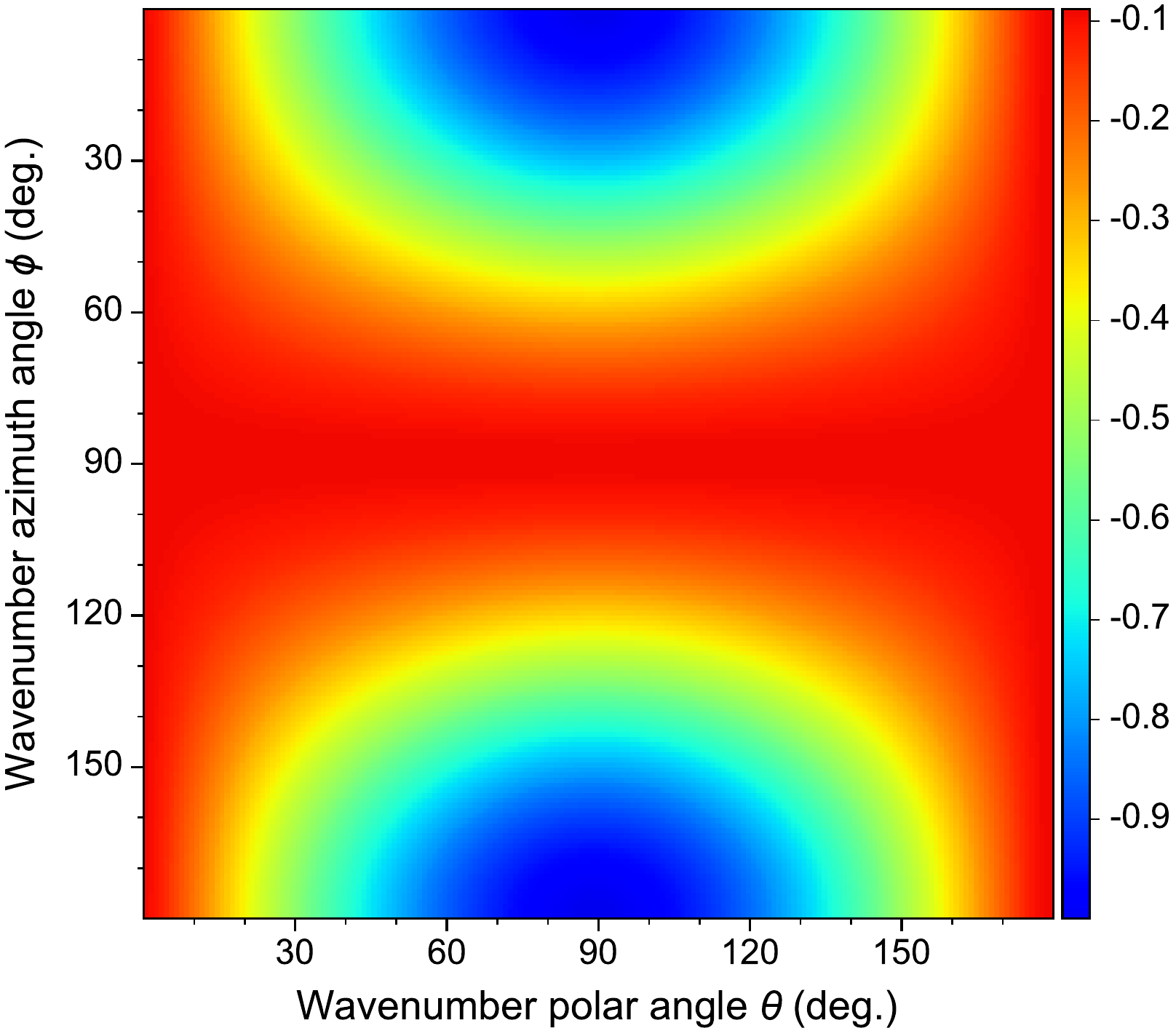}}
\subfigure[]{\includegraphics[width=0.3\textwidth]{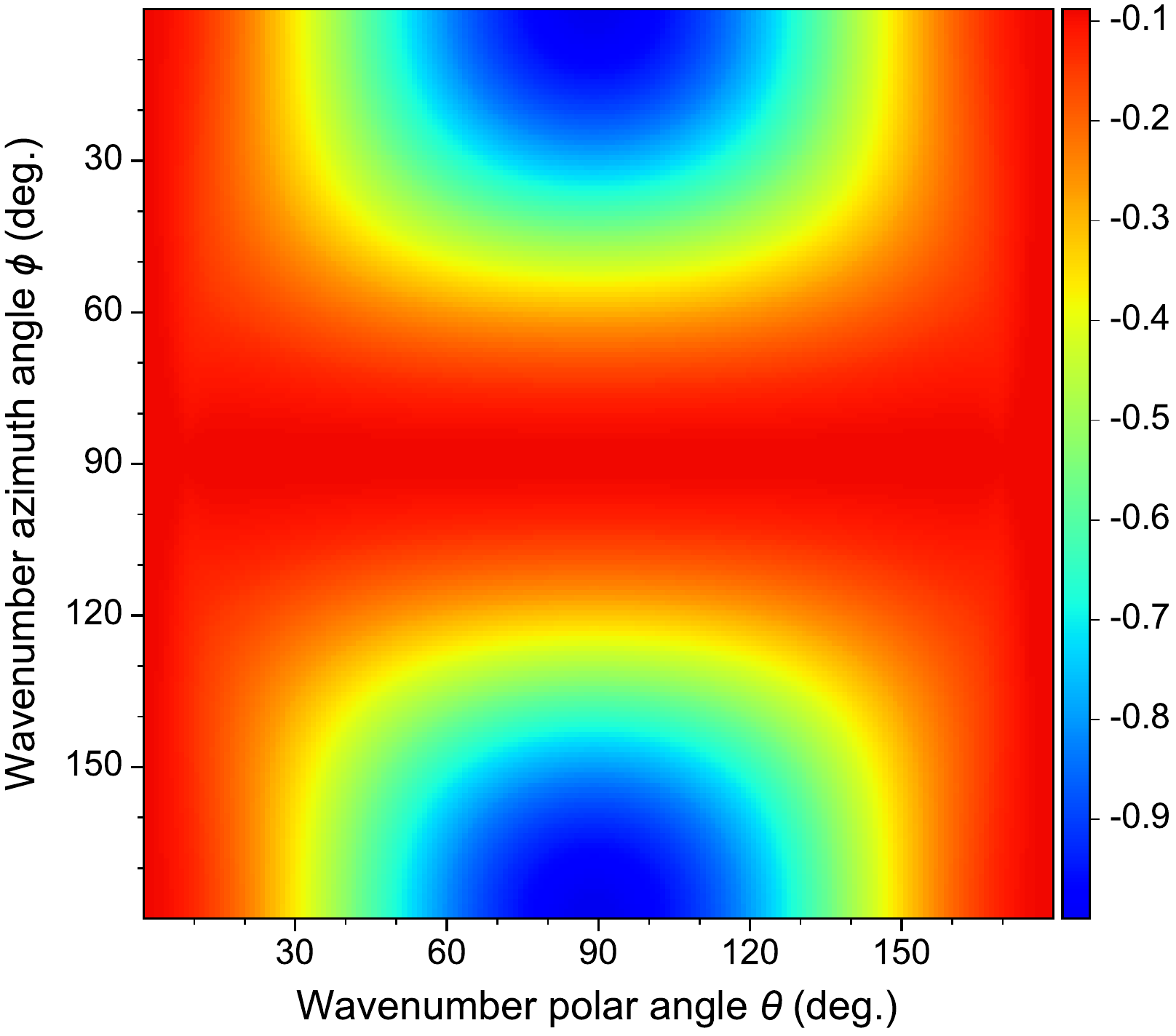}}
\subfigure[]{\includegraphics[width=0.3\textwidth]{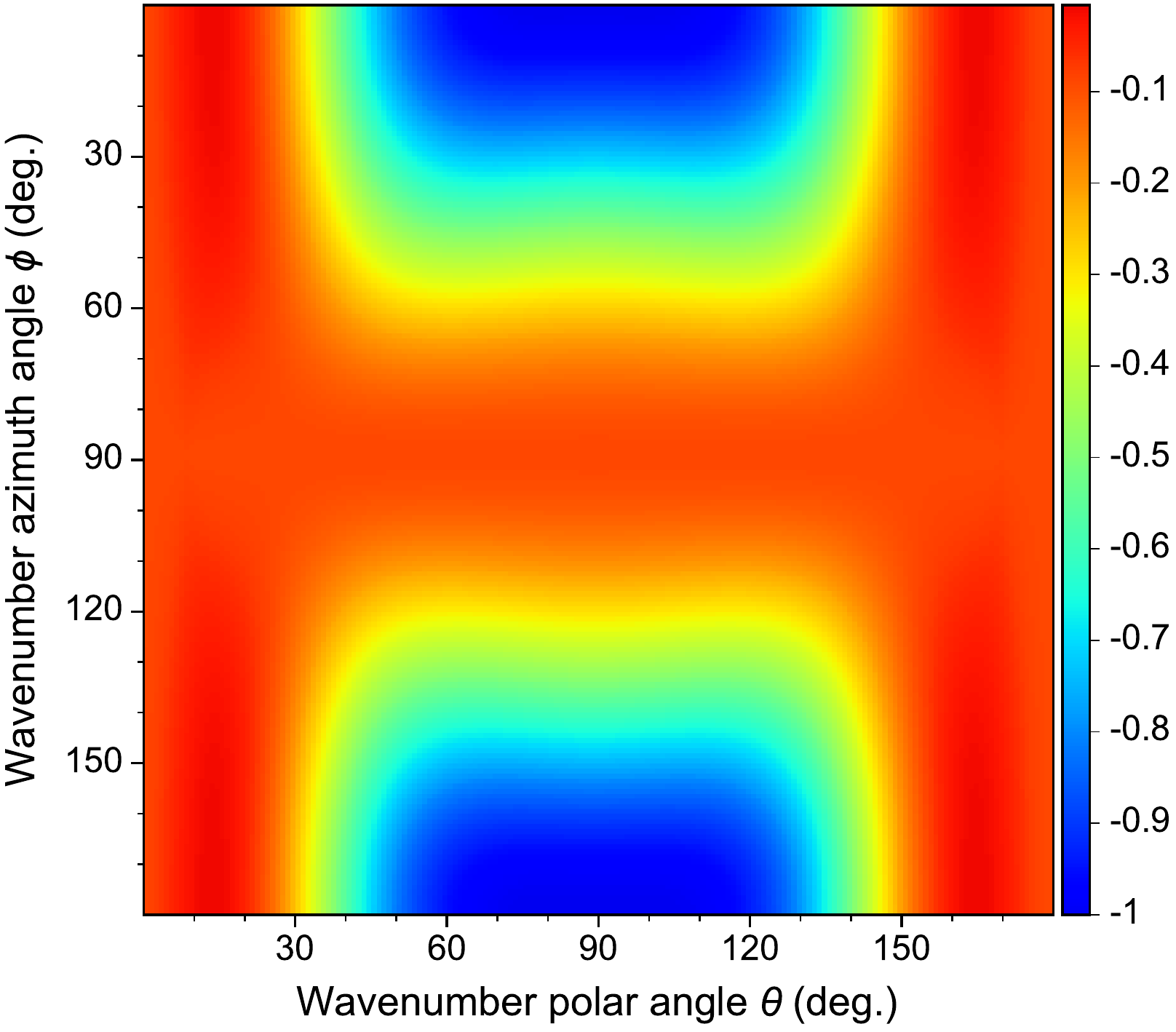}}
\subfigure[]{\includegraphics[width=0.3\textwidth]{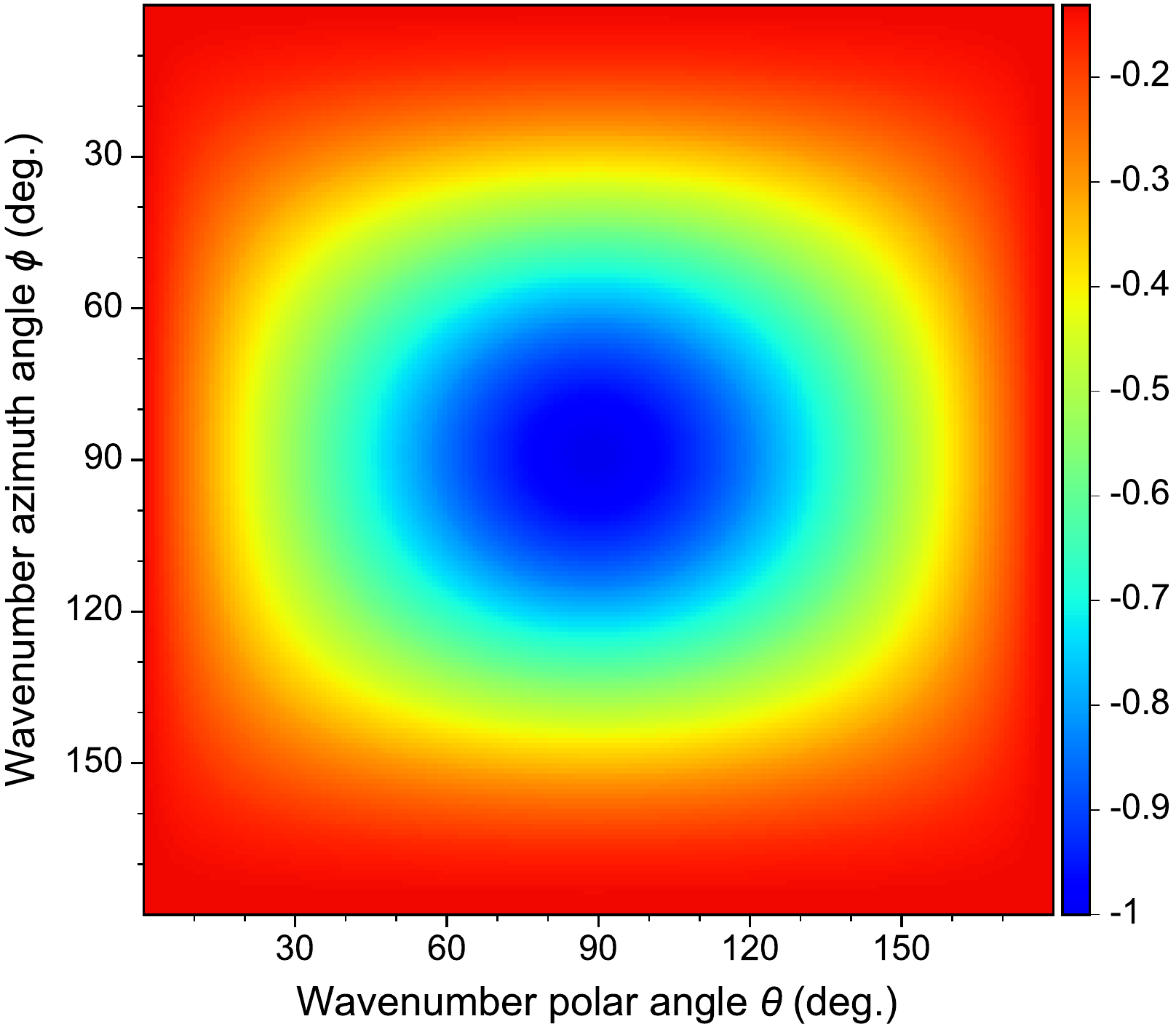}}
\subfigure[]{\includegraphics[width=0.3\textwidth]{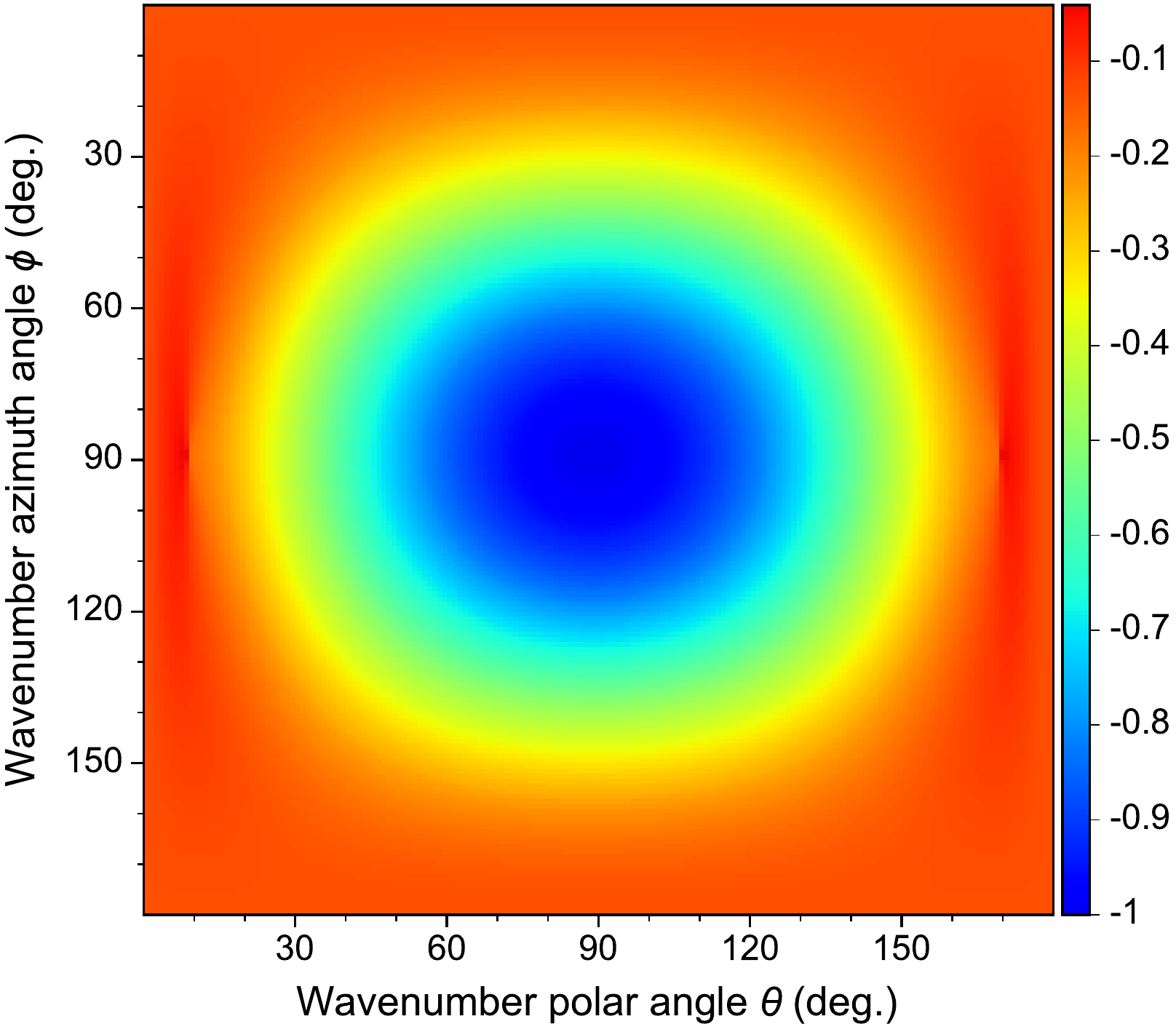}}
\subfigure[]{\includegraphics[width=0.3\textwidth]{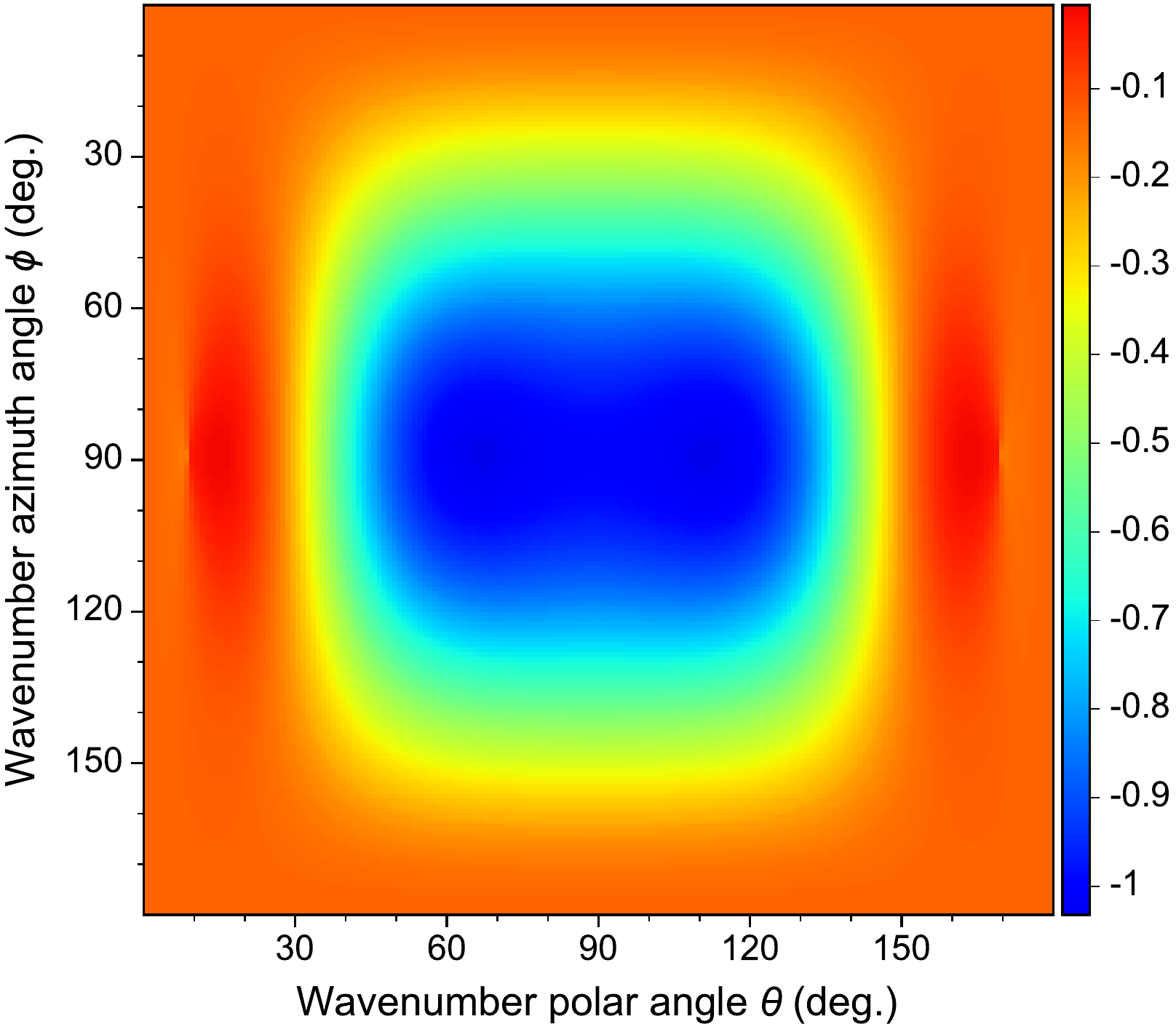}}
\subfigure[]{\includegraphics[width=0.3\textwidth]{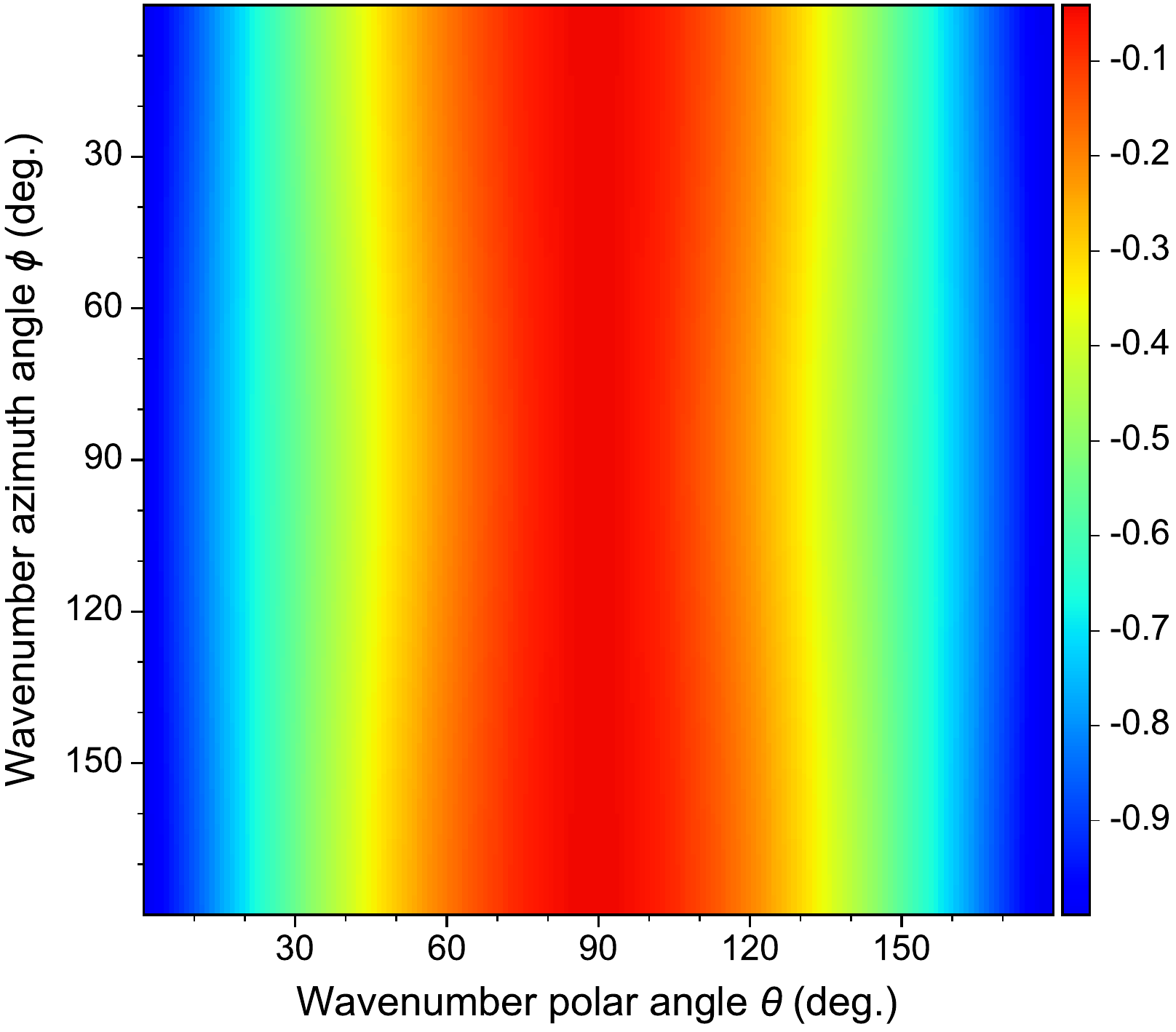}}
\subfigure[]{\includegraphics[width=0.3\textwidth]{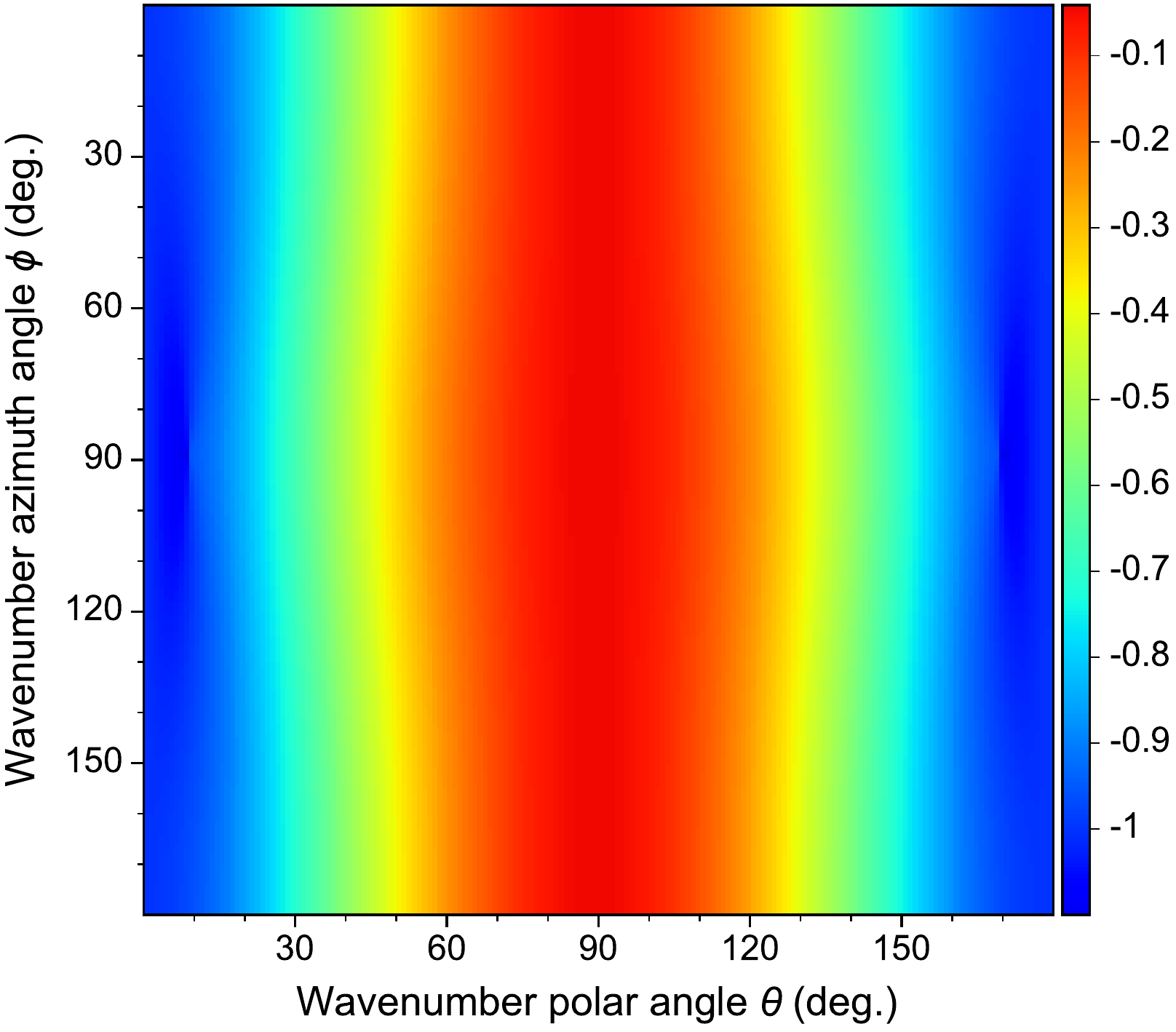}}
\subfigure[]{\includegraphics[width=0.3\textwidth]{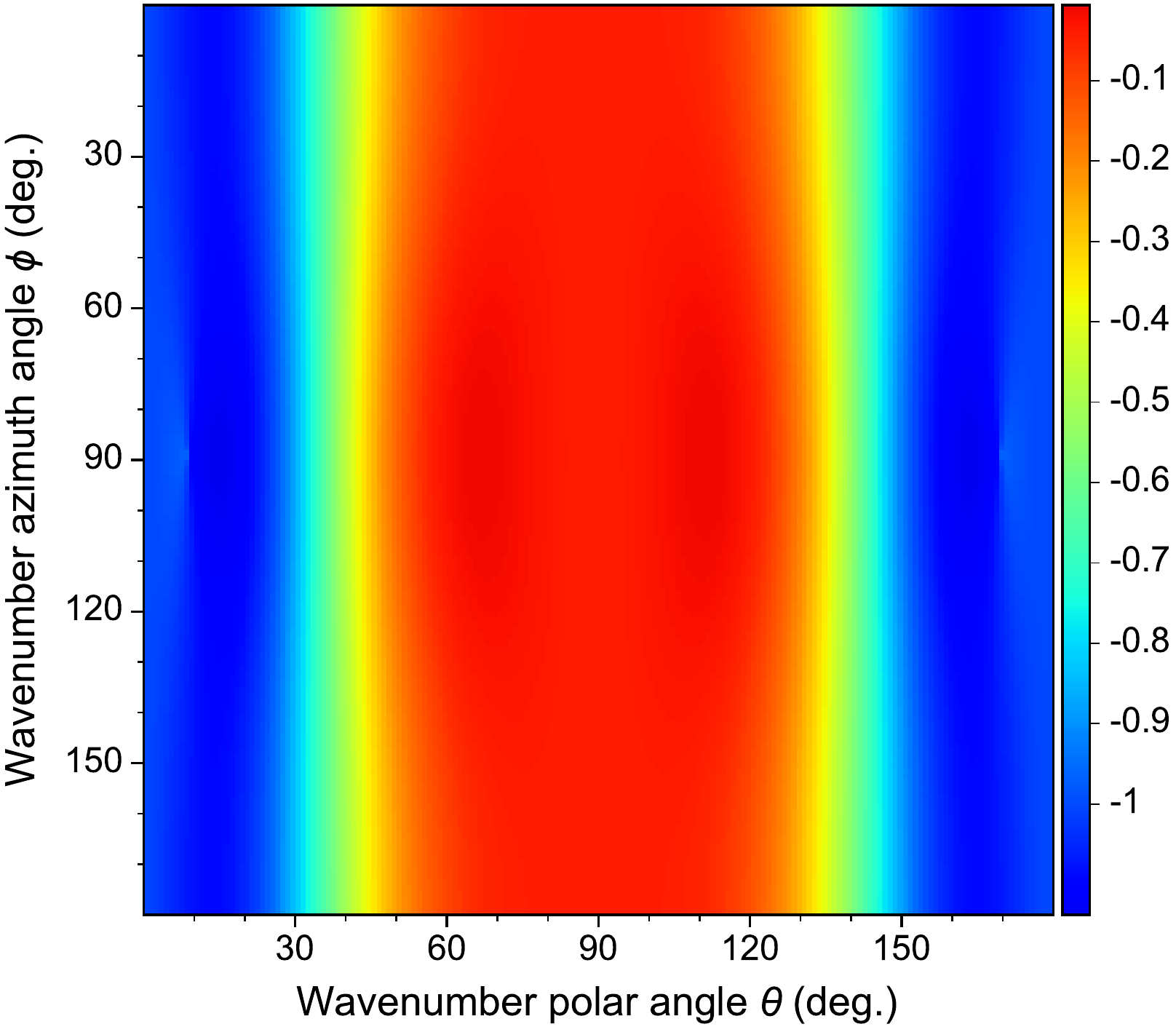}}
\caption{Eigenvalue derivatives of $\tilde{\bfA}$ of MPML in the (a)-(c) $x_1$-, (d)-(f) $x_2$-, and (g)-(i) $x_3$-directions under calculated optimal damping ratios in eq.~\eqref{eq:vti_3d_ratio} for the 3D quasi-VTI medium with elasticity matrix \eqref{eq:vti_3d}. (a), (d) and (g) represent qP-wave, (b), (e) and (h) represent qS1-wave, (c), (f) and (i) represent qS2-wave.}
\label{fig:vti_3d_deriv}
\end{figure}

We calculate the optimal damping ratios for MPML using 
Algorithm~\ref{alg:damping_ratio_3}, and obtain
\begin{equation}
\xi_1=0.088,\qquad \xi_2= 0.131, \qquad \xi_3= 0.041.
\label{eq:vti_3d_ratio}
\end{equation}
We plot the eigenvalue derivatives in the polar angle range $(0,\pi]$ and 
azimuth angle range $(0,\pi]$ for three axis directions in 
Fig.~\ref{fig:vti_3d_deriv}. Since the three symmetric axes of this VTI medium 
are aligned with three coordinate axes, the 
three eigenvalue derivatives are symmetric with respect to both $\theta=\pi/2$ 
and $\phi=\pi/2$ lines. 

\begin{figure}
\centering
\includegraphics[width=0.65\textwidth]{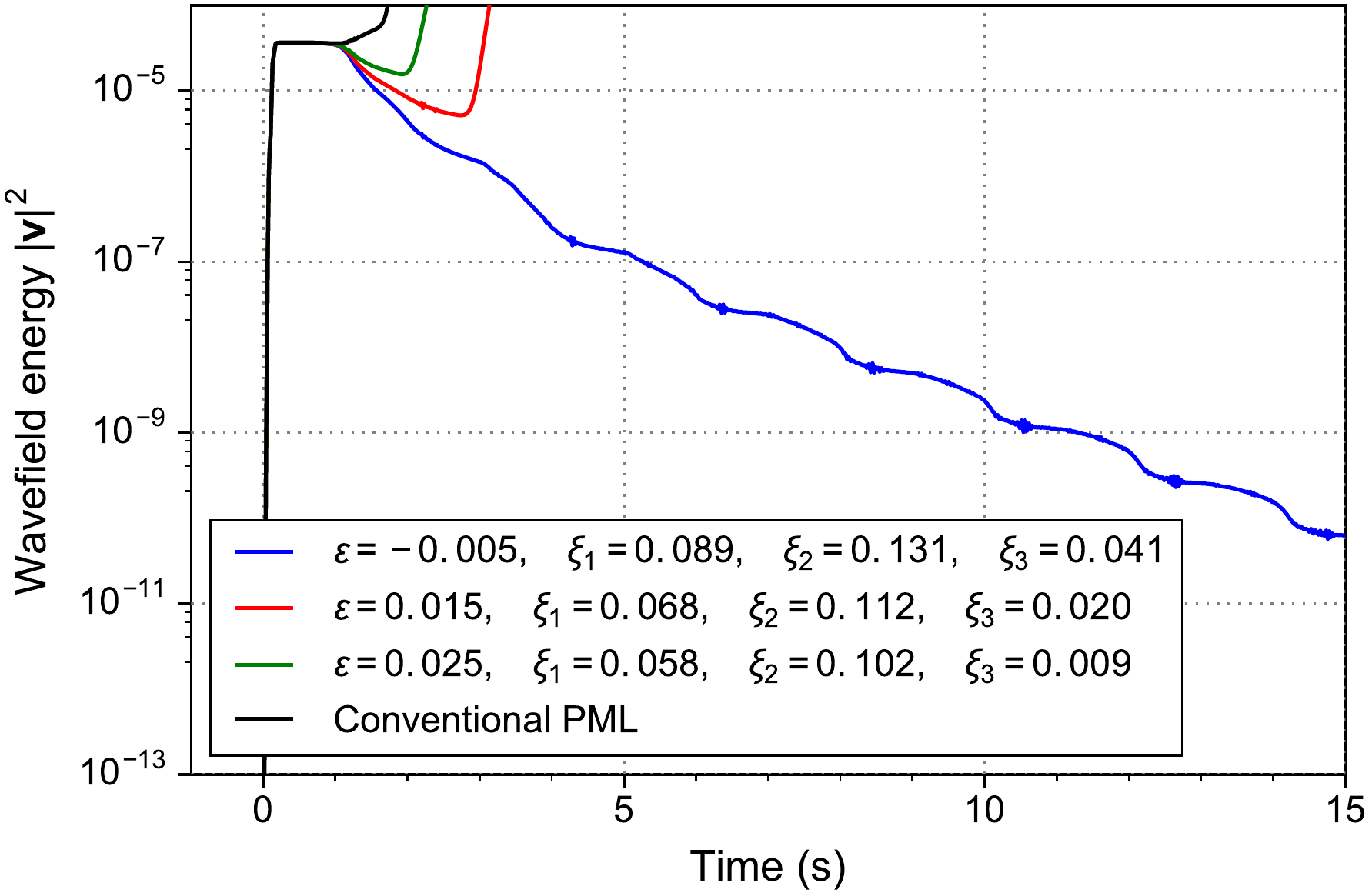}
\caption{Wavefield energy decay curve in the 3D VTI medium with elasticity matrix \eqref{eq:vti_3d}.}
\label{fig:vti_3d_energy}
\end{figure}

We conduct numerical wavefield modeling to verify the stability of MPML 
under these optimal damping ratios. The model is defined on a $400\times 
400 \times 400$ grid with a grid size of 10~m in all three 
directions. The thickness of PML layer is 25 grids. A vertical force 
vector is located at the center of the computational domain, and the 
source time function is a Ricker wavelet with a central frequency of 
10~Hz.  The time step size is 1~ms, which is smaller than the 
stability-required time step of 1.69~ms.  A total of 15,000 time steps, 
i.e., 15~s, are simulated, and the wavefield energy curve is shown in 
Fig.~\ref{fig:vti_3d_energy}. The blue curve in 
Fig.~\ref{fig:vti_3d_energy} is for the case with the optimal damping 
ratios in eq.~\eqref{eq:vti_3d_ratio}. We also carry out a wavefield modeling with the eigenvalue 
derivative threshold $\epsilon=0.015$ and $\epsilon=0.025$. The MPMLs under these thresholds are 
unstable according to the corresponding wavefield energy variation curves 
in Fig.~\ref{fig:vti_3d_energy}.  As in the 2D MPML case, we should 
always choose a negative $\epsilon$ to stabilize MPML for 3D 
anisotropic media.

\begin{figure}
\centering
\subfigure[]{\includegraphics[width=0.45\textwidth]{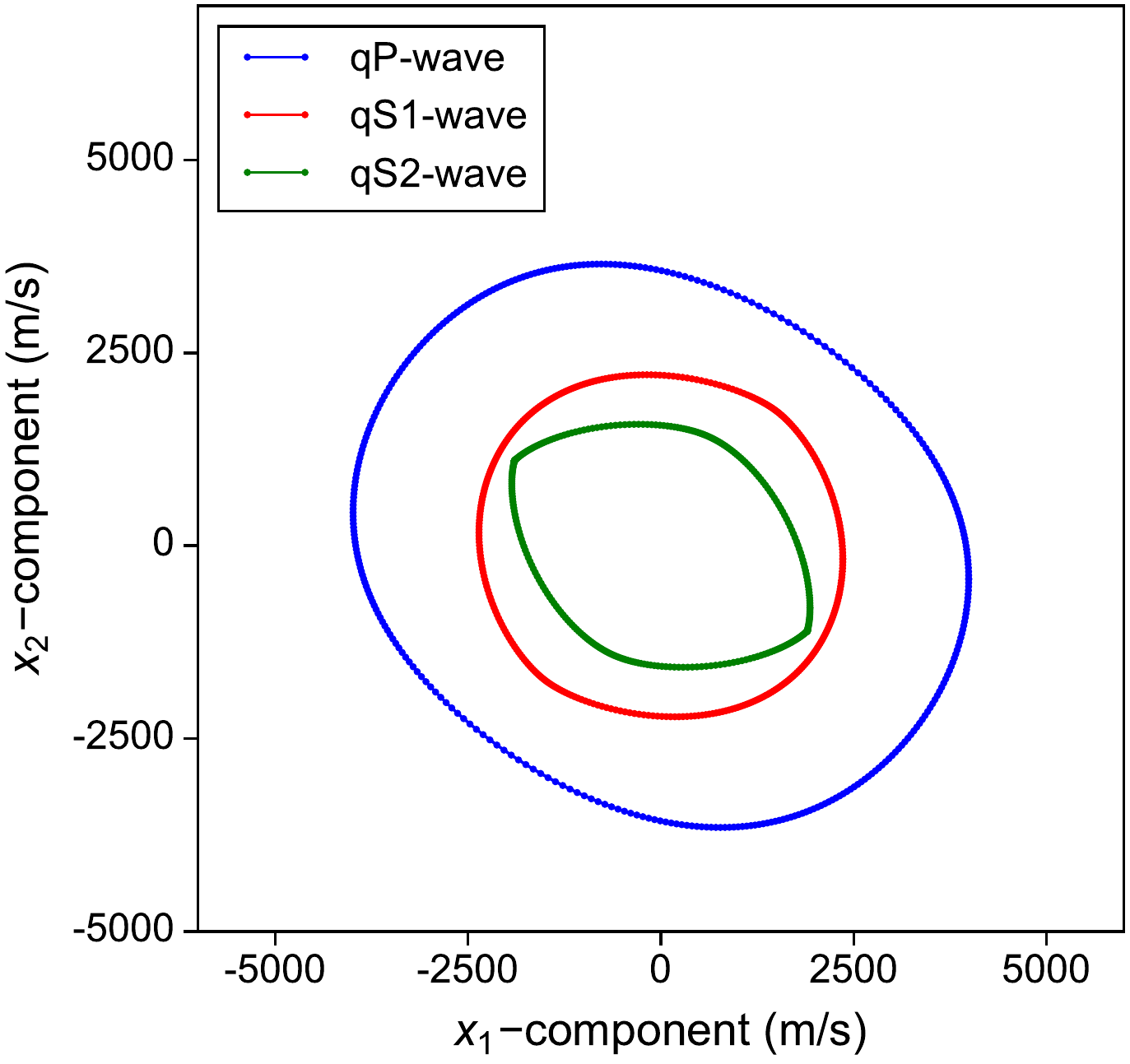}}
\subfigure[]{\includegraphics[width=0.45\textwidth]{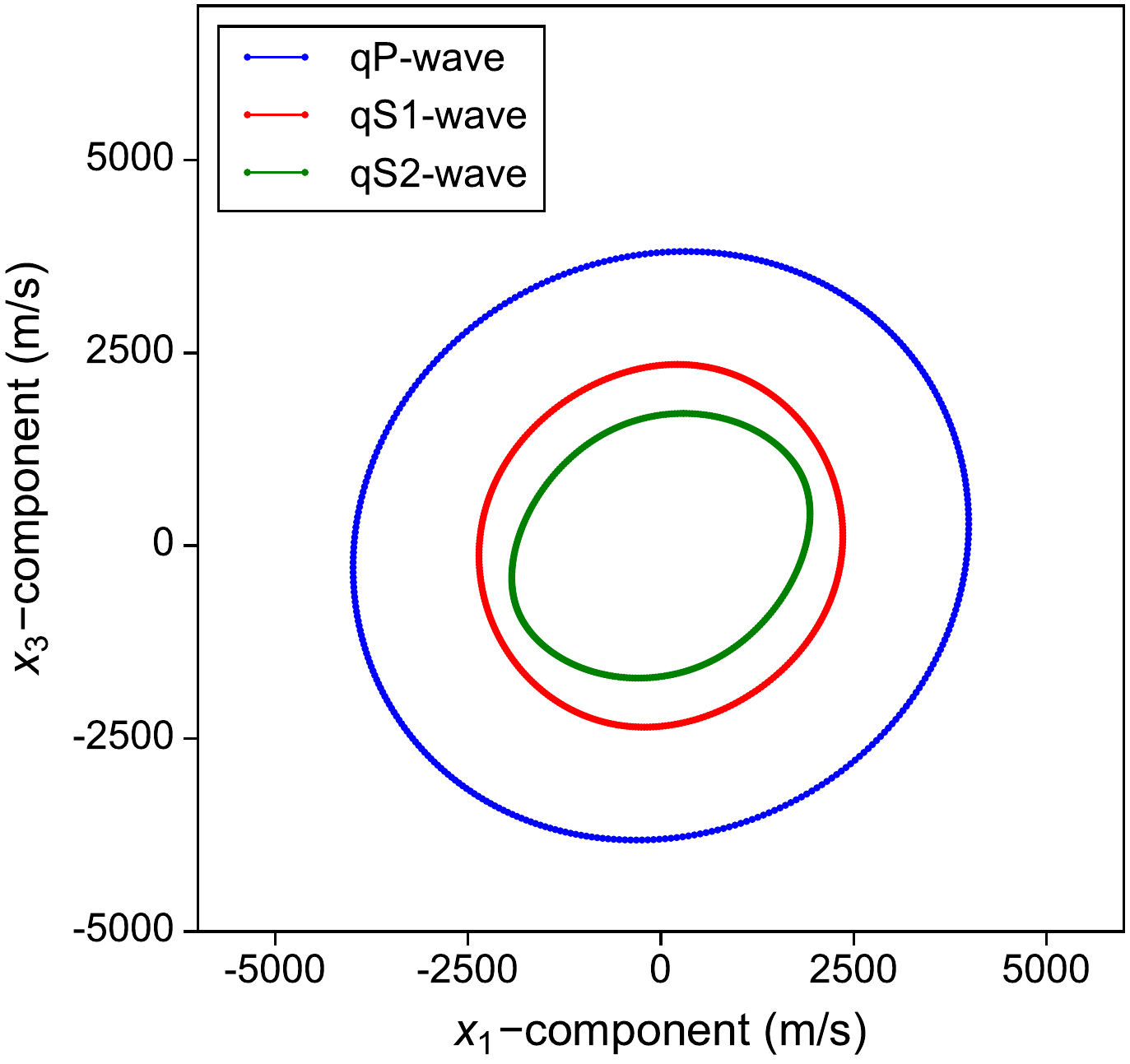}}
\subfigure[]{\includegraphics[width=0.45\textwidth]{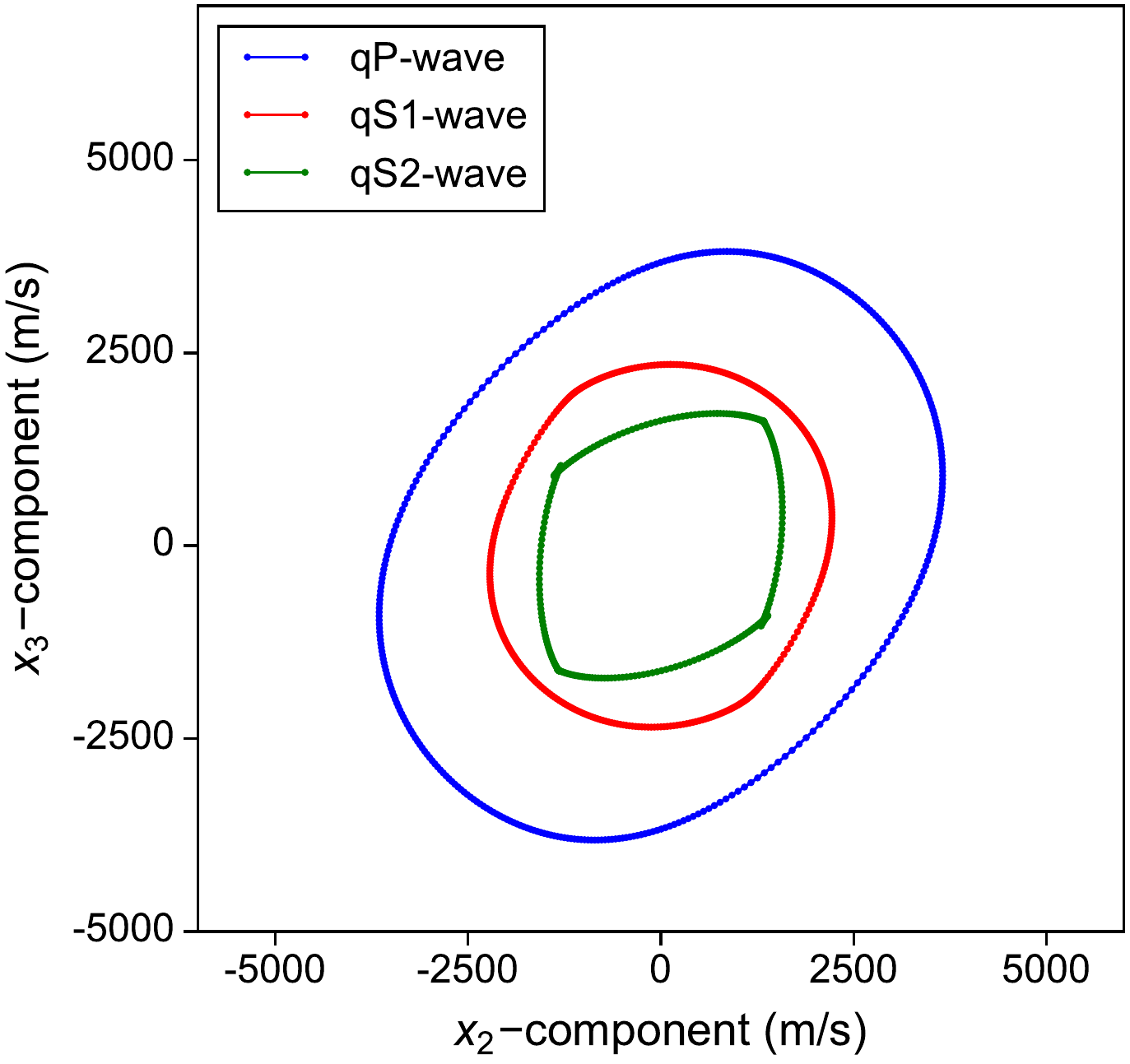}}
\caption{Wavefront curves in the 3D quasi-TTI medium with elasticity matrix \eqref{eq:tti_3d} on the (a) $x_1x_2$ (b) $x_1x_3$ and (c) $x_2x_3$ axis plane. qS1 and qS2 represents the two qS-waves. qS-wave wavefronts seem to be less complicated compared with those of the 3D quasi-VTI medium \eqref{eq:vti_3d} only because the qS-wave triplications are now out of axis planes after rotation.}
\label{fig:tti_3d_curve}
\end{figure}

Our next numerical example uses a rotation version of the aforementioned  
quasi-VTI medium. We rotate the quasi-VTI medium \eqref{eq:vti_3d} with 
respect to the $x_1$-axis by 30~degrees, the $x_2$-axis by 50~degrees, and 
the $x_3$-axis by 25~degrees, and the resulting elasticity matrix for this 
quasi-TTI medium is given by
\begin{equation}
\bfC =\left[\begin{array}{cccccc}
15.7930    &4.1757    &4.9651   & 0.1582   & 0.6529   &-1.0343 \\
&12.5979   & 4.1844    &2.0903   & 0.8186  & -1.7513\\
& &14.1587   & 1.9573    &0.7643  & -0.1606\\
&&&3.7879  & -0.8909    &0.8065\\
&&&&5.0750    &0.7668\\
&&&&&4.3423
\end{array}\right].
\label{eq:tti_3d} 
\end{equation}
The corresponding wavefront curves are shown in 
Fig.~\ref{fig:tti_3d_curve}.

\begin{figure}
\centering
\subfigure[]{\includegraphics[width=0.3\textwidth]{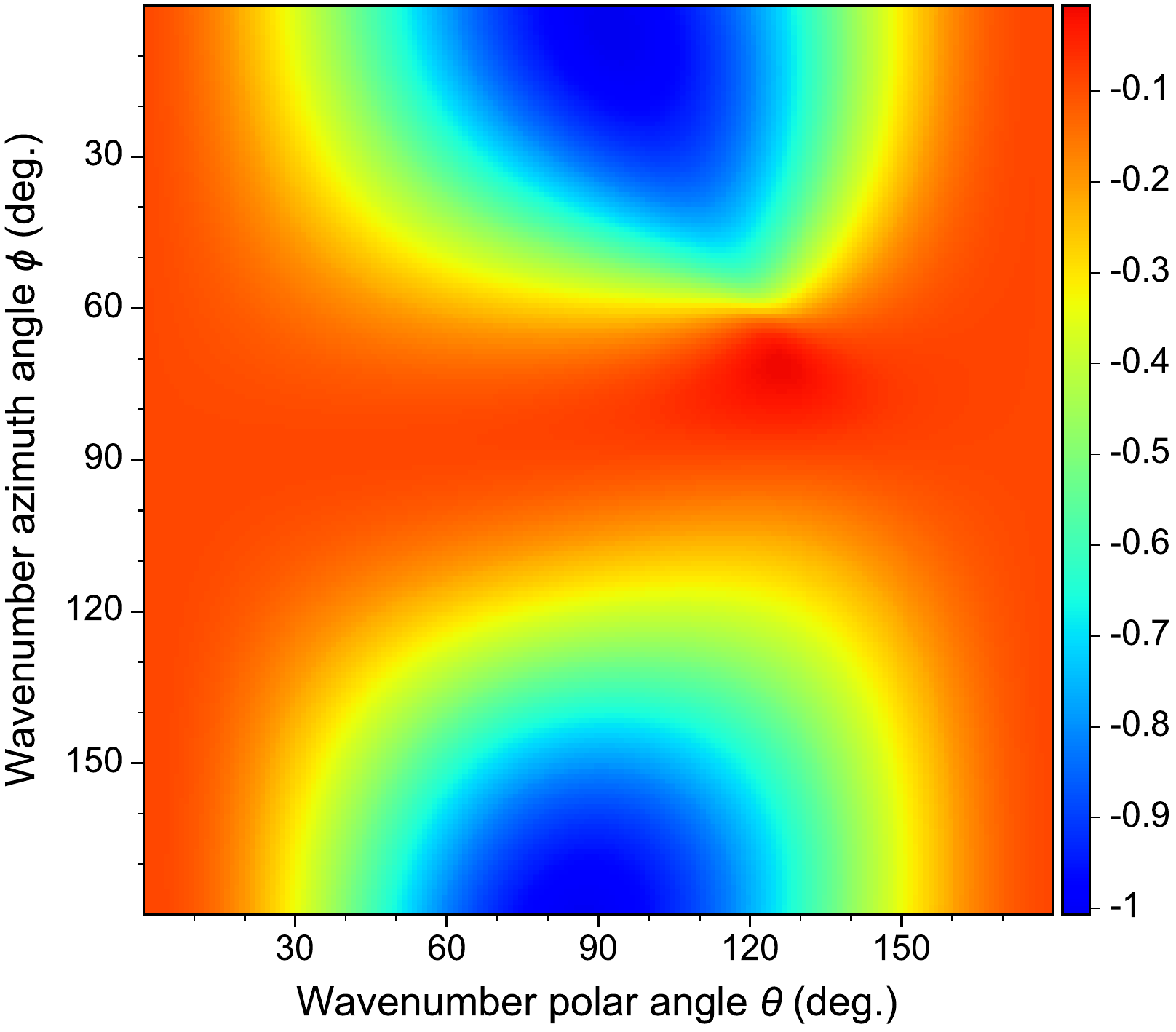}}
\subfigure[]{\includegraphics[width=0.3\textwidth]{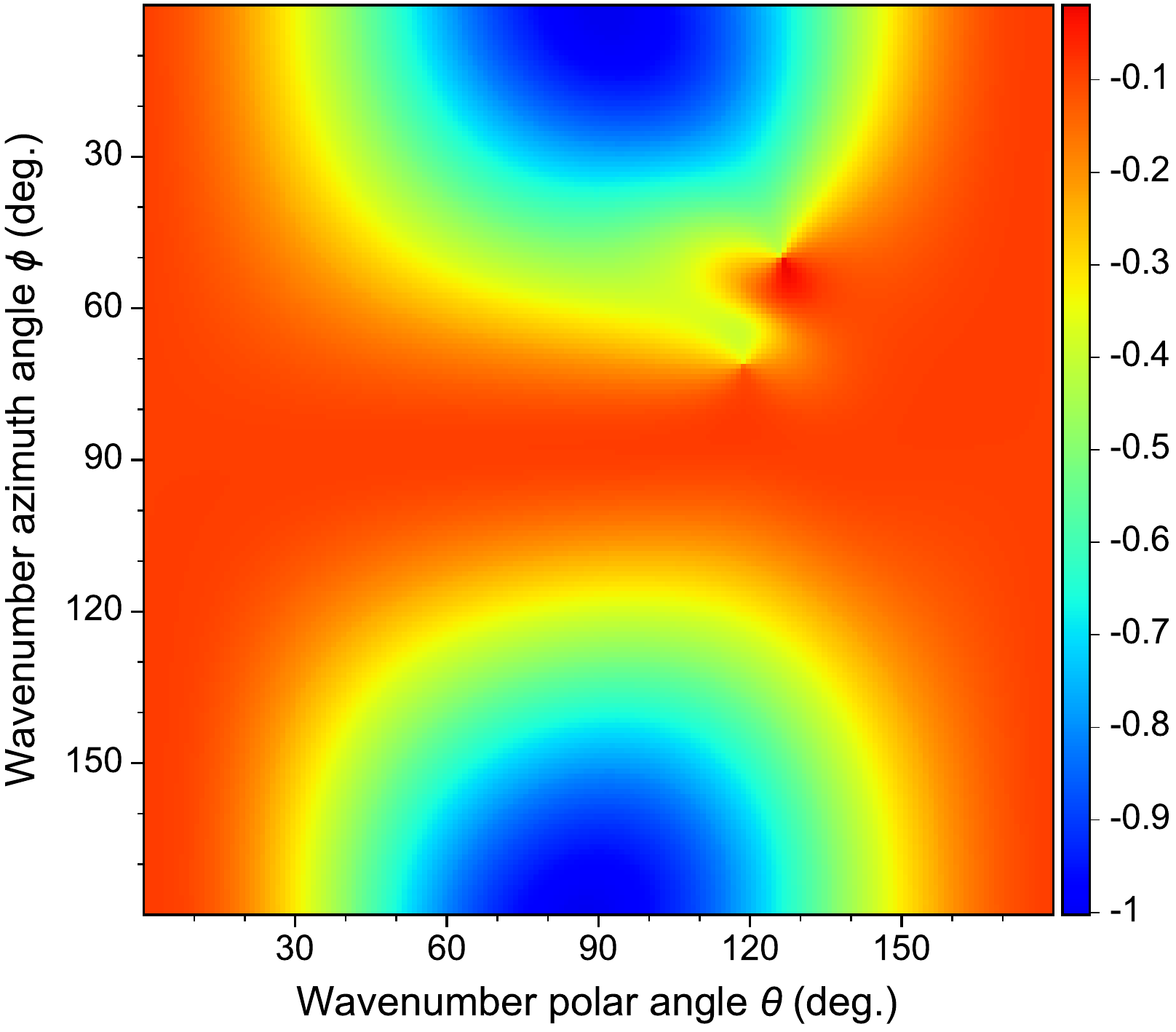}}
\subfigure[]{\includegraphics[width=0.3\textwidth]{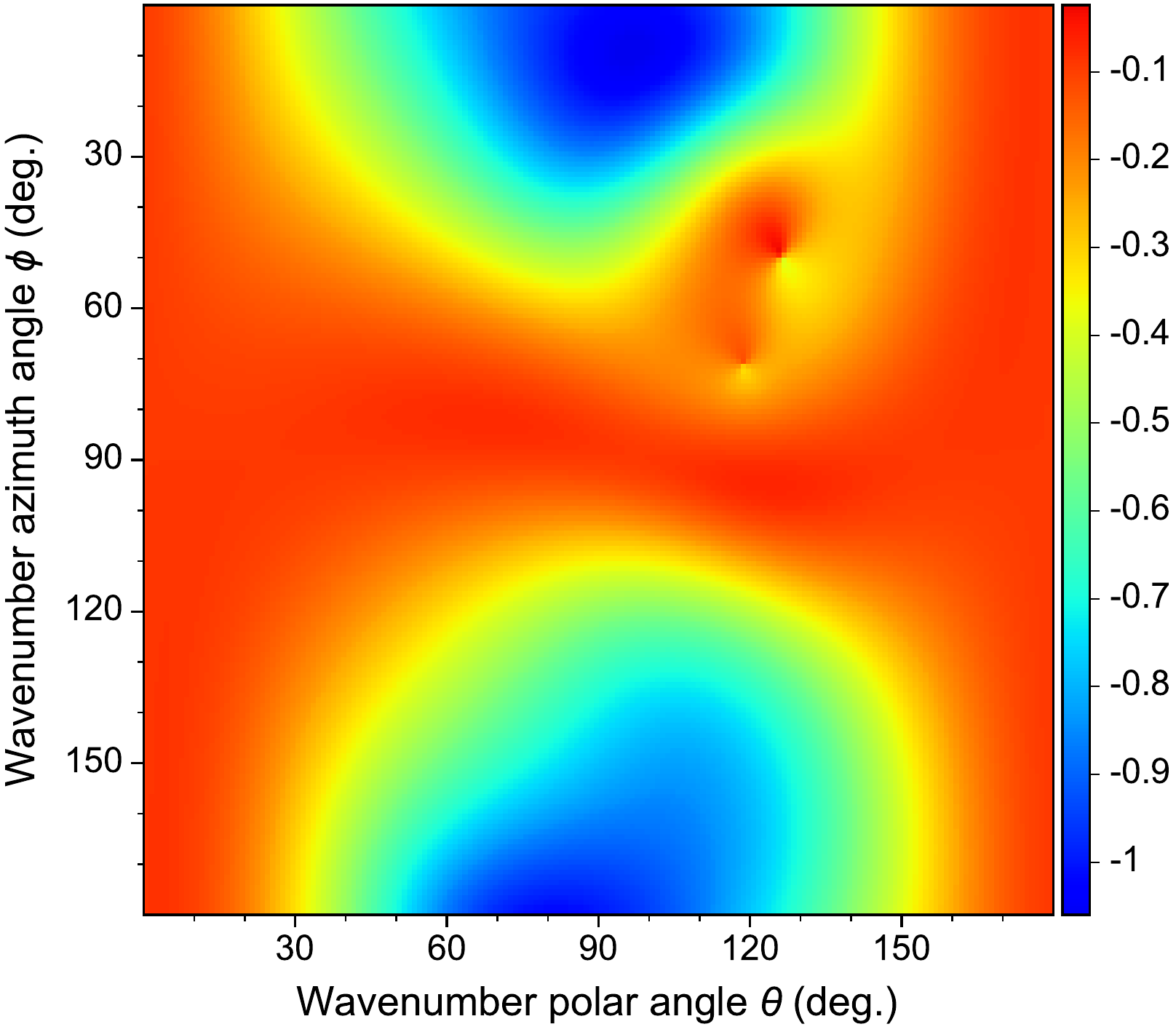}}
\subfigure[]{\includegraphics[width=0.3\textwidth]{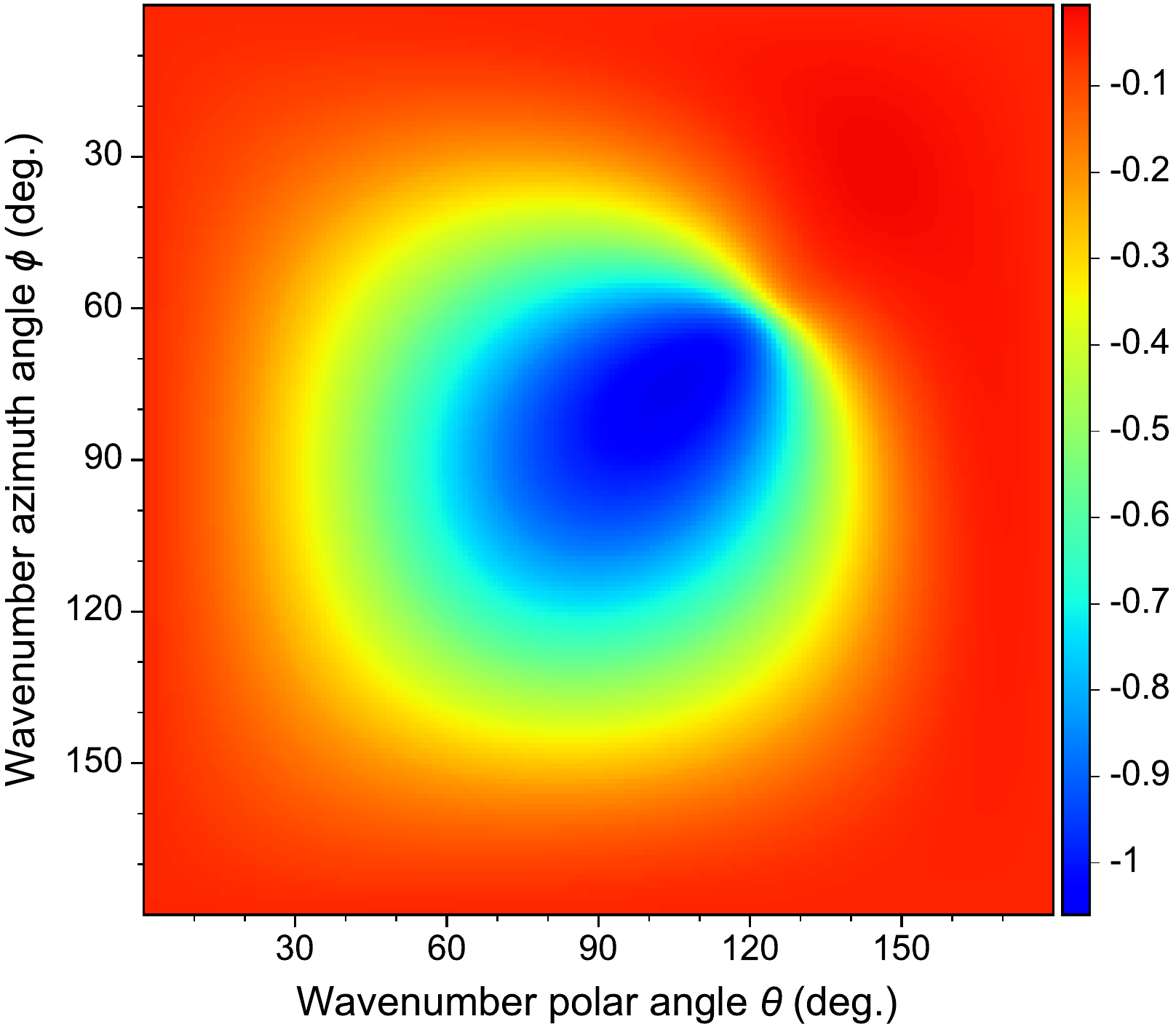}}
\subfigure[]{\includegraphics[width=0.3\textwidth]{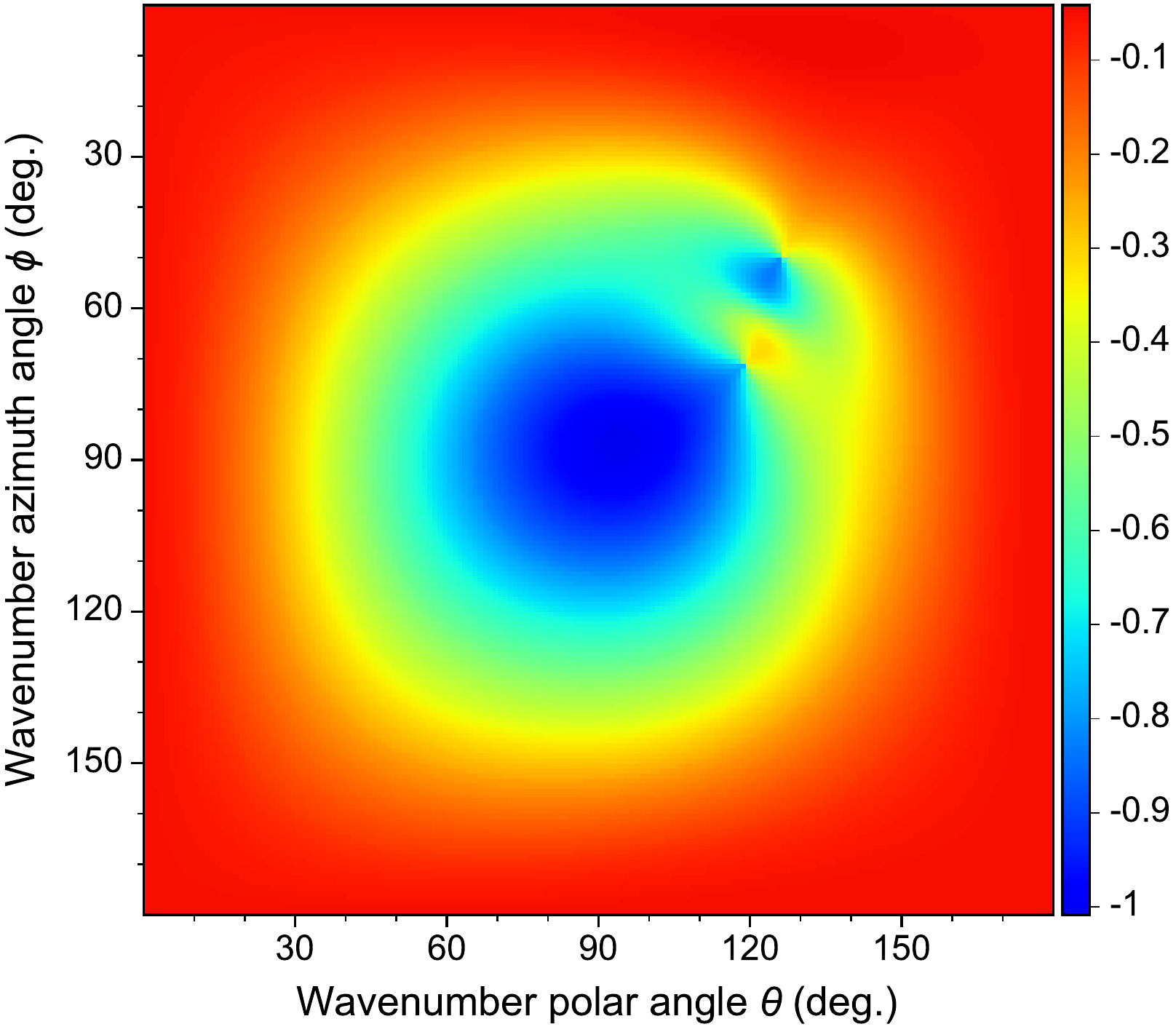}}
\subfigure[]{\includegraphics[width=0.3\textwidth]{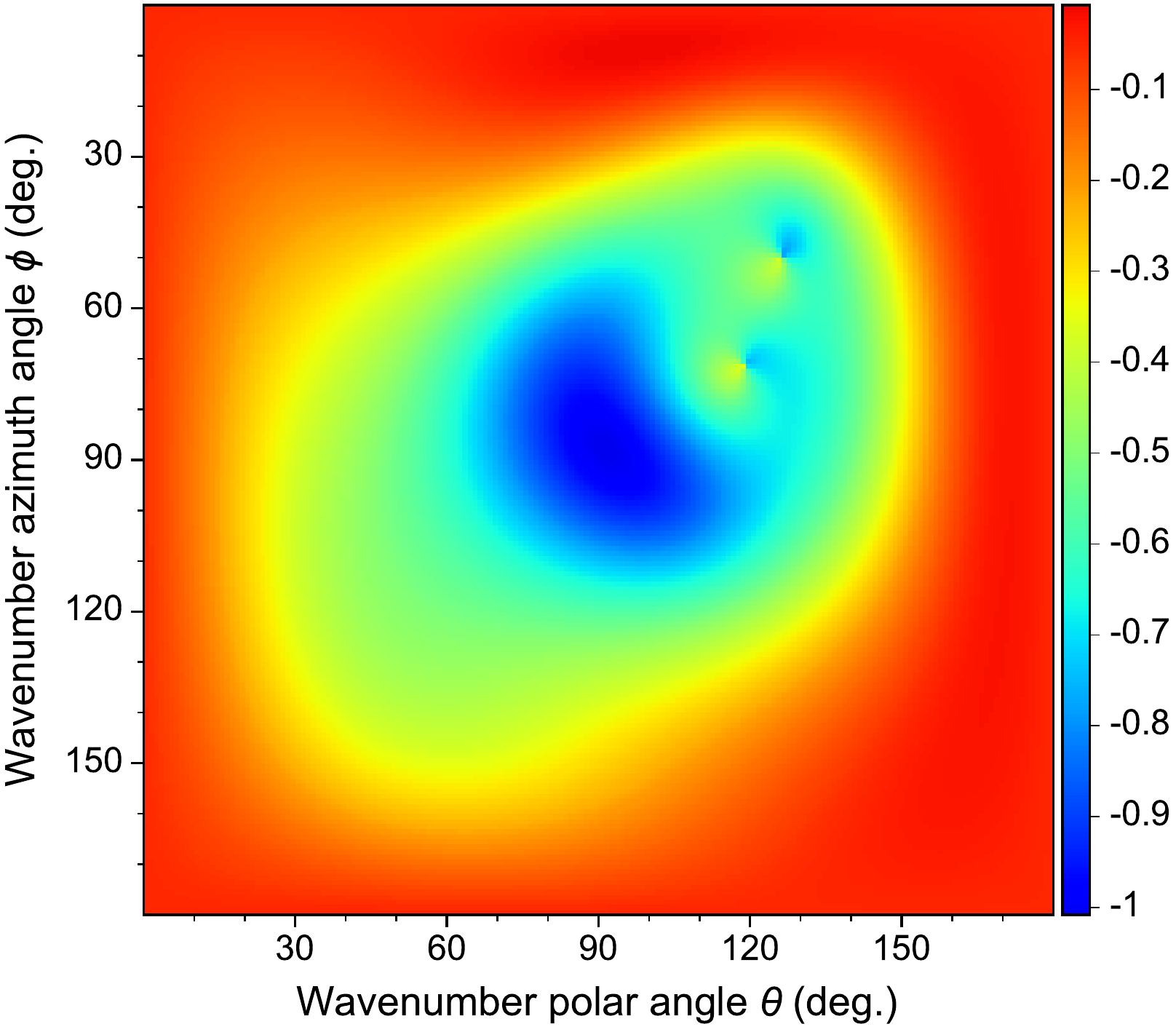}}
\subfigure[]{\includegraphics[width=0.3\textwidth]{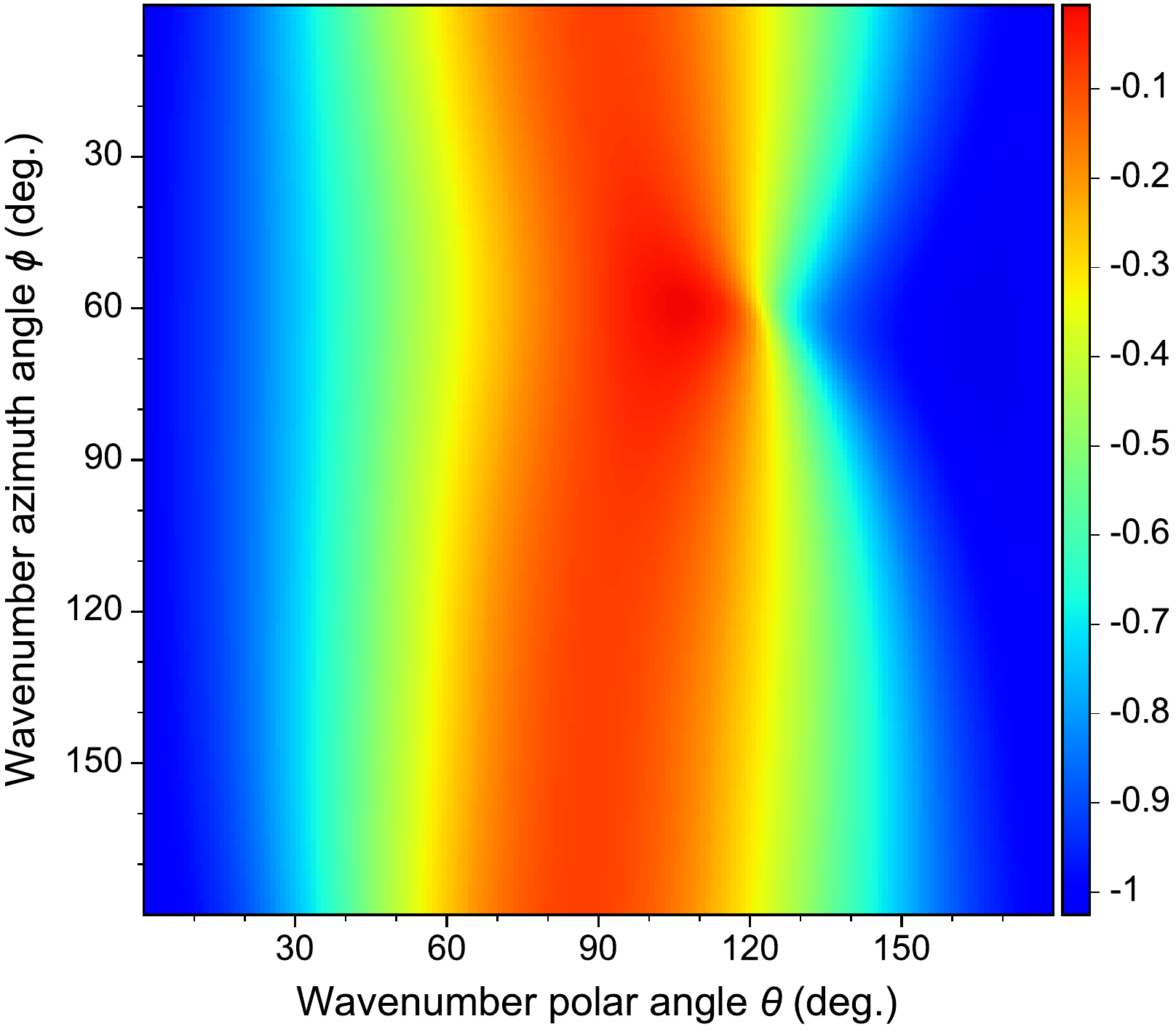}}
\subfigure[]{\includegraphics[width=0.3\textwidth]{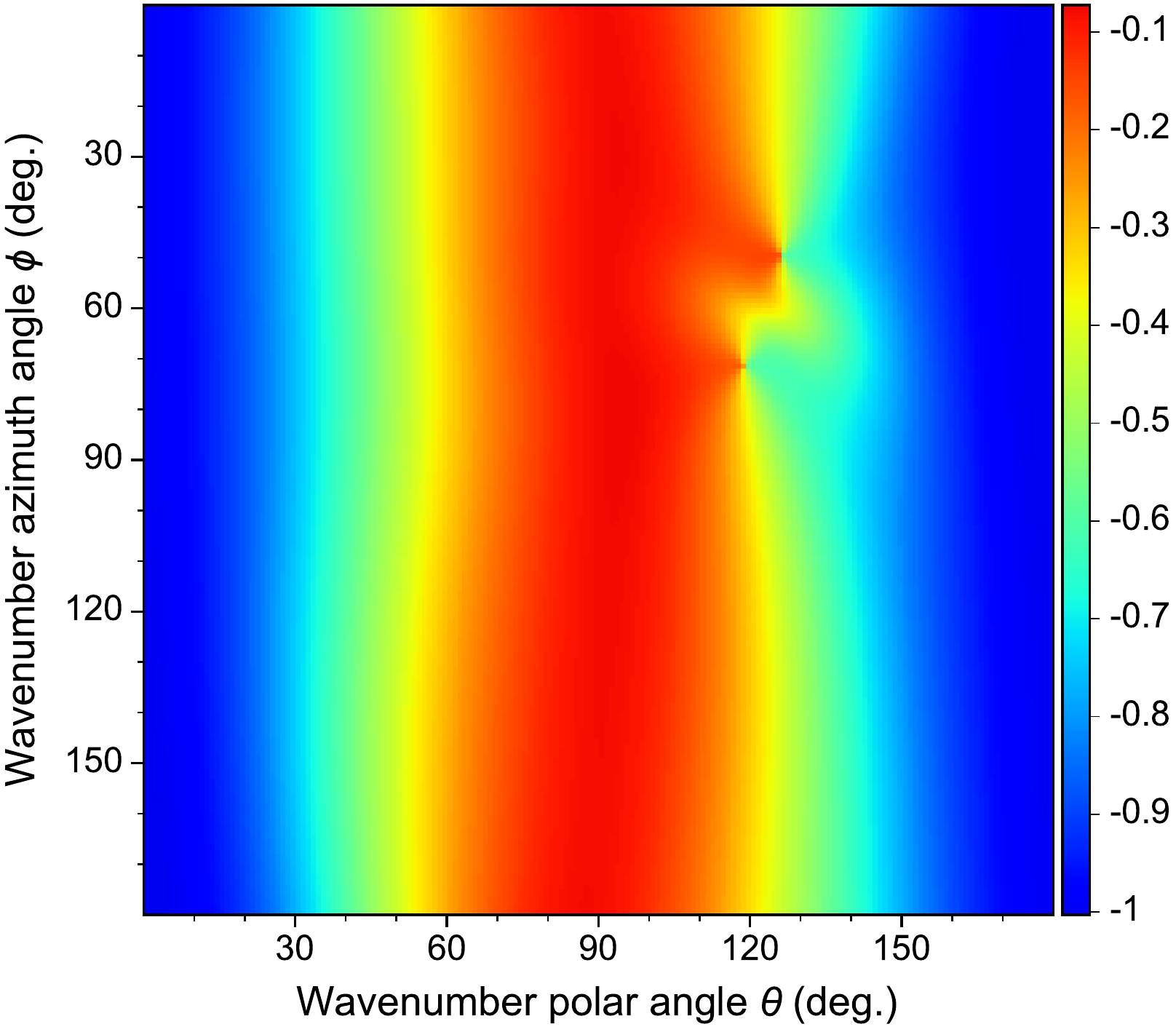}}
\subfigure[]{\includegraphics[width=0.3\textwidth]{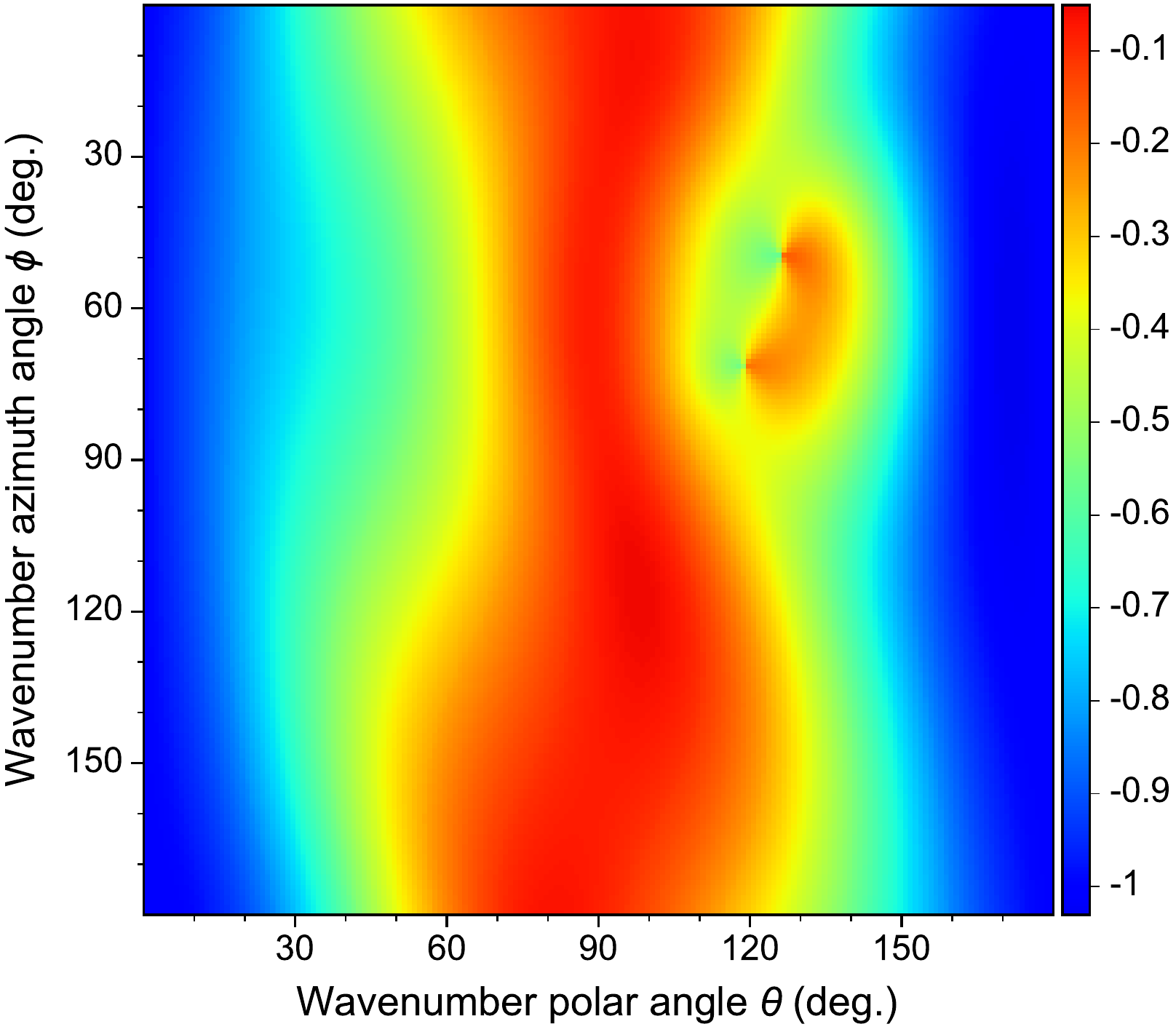}}
\caption{Eigenvalue derivatives of $\tilde{\bfA}$ of MPML in the (a)-(c) $x_1$-, (d)-(f) $x_2$-, and (g)-(i) $x_3$-directions under calculated optimal damping ratios in eq.~\eqref{eq:tti_3d_ratio} for the 3D quasi-TTI medium with elasticity matrix \eqref{eq:tti_3d}. (a), (d) and (g) represent qP-wave, (b), (e) and (h) represent qS1-wave, (c), (f) and (i) represent qS2-wave.}
\label{fig:tti_3d_deriv}
\end{figure}

Similar to the 3D quasi-VTI case, we obtain the following optimal damping 
ratios using Algorithm~\ref{alg:damping_ratio_3}:
\begin{equation}
\xi_1=0.089,\qquad\xi_2=0.051, \qquad \xi_3=0.080.
\label{eq:tti_3d_ratio}
\end{equation}
The eigenvalue derivatives under this set of damping ratios for three 
axis directions are shown in Fig.~\ref{fig:tti_3d_deriv}. Obviously, for the quasi-TTI medium where the symmetric axes 
are not aligned with coordinate axes, the 
eigenvalue derivatives of all three wave modes along any coordinate axis 
is no longer symmetric about any $\theta$ or $\phi$ lines. Therefore, it 
is necessary to use the entire range of wavenumber polar angle $\theta$ and azimuth angle $\phi$, i.e., 
$(0,\pi]\times (0,\pi]$, to determine the optimal damping ratios.

\begin{figure}
\centering
\includegraphics[width=0.65\textwidth]{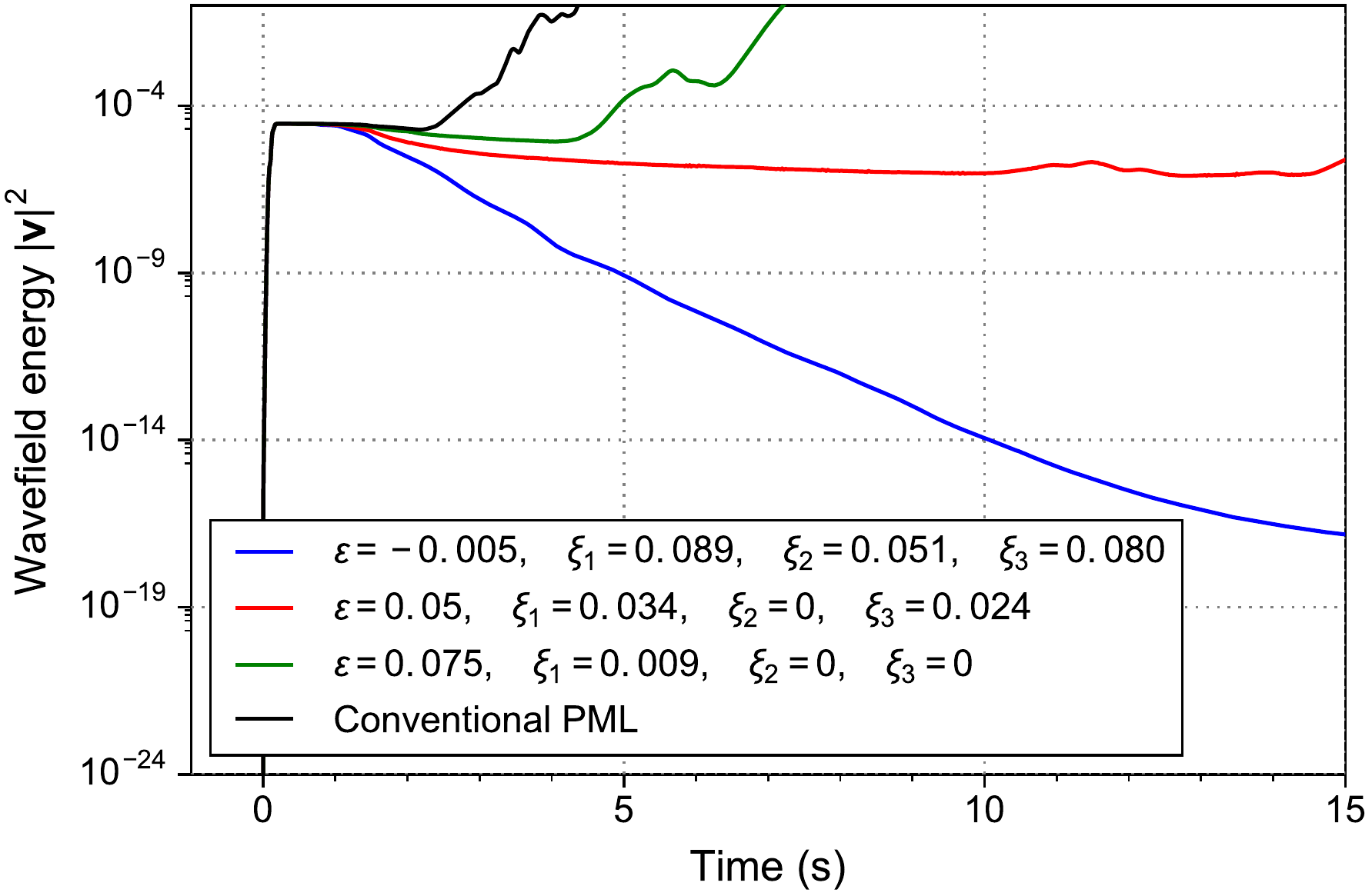}
\caption{Wavefield energy decay curve in the 3D quasi-TTI medium with elasticity matrix \eqref{eq:tti_3d}.}
\label{fig:tti_3d_energy}
\end{figure}

Figure~\ref{fig:tti_3d_energy} depicts the wavefield energy decays 
under the optimal damping ratios in eq.~\eqref{eq:tti_3d_ratio} and damping ratios with thresholds $\epsilon=0.05$ and $\epsilon=0.075$.  
For the case where $\epsilon=0.05$, the wavefield energy does not diverge immediately after maximum 
energy value occurred. Instead, the curve indicates a very slow energy decay after 
about 1~s. In contrast, the optimal MPML with threshold 
$\epsilon=-0.005$ shows a ``normal'' energy decay.  Therefore, although 
the MPML with $\epsilon=0.05$ does not show energy divergence within 
15~s, it fails to effectively absorb the outgoing wavefield, and we consider this as a ``quasi-divergence.'' Meanwhile, the MPML with $\epsilon=0.075$ shows an energy divergence after about 4~s. These results further demonstrate that the behavior of MPML with a positive eigenvalue derivative threshold is different and unpredictable for different kinds of anisotropic media.  
Figure~\ref{fig:tti_3d_energy} also shows that the conventional PML gives an 
unstable result.

\begin{figure}
\centering
\subfigure[]{\includegraphics[width=0.45\textwidth]{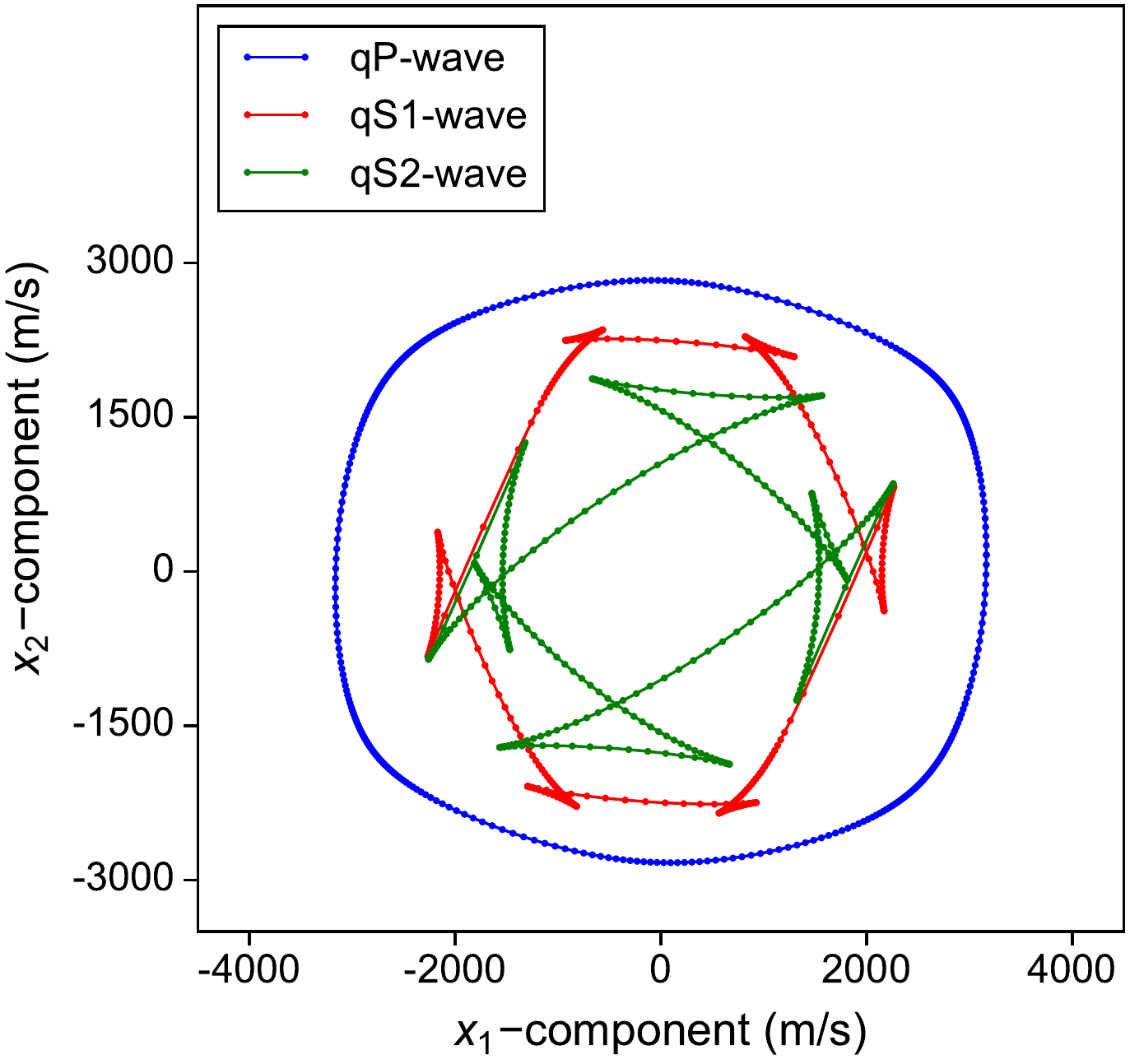}}
\subfigure[]{\includegraphics[width=0.45\textwidth]{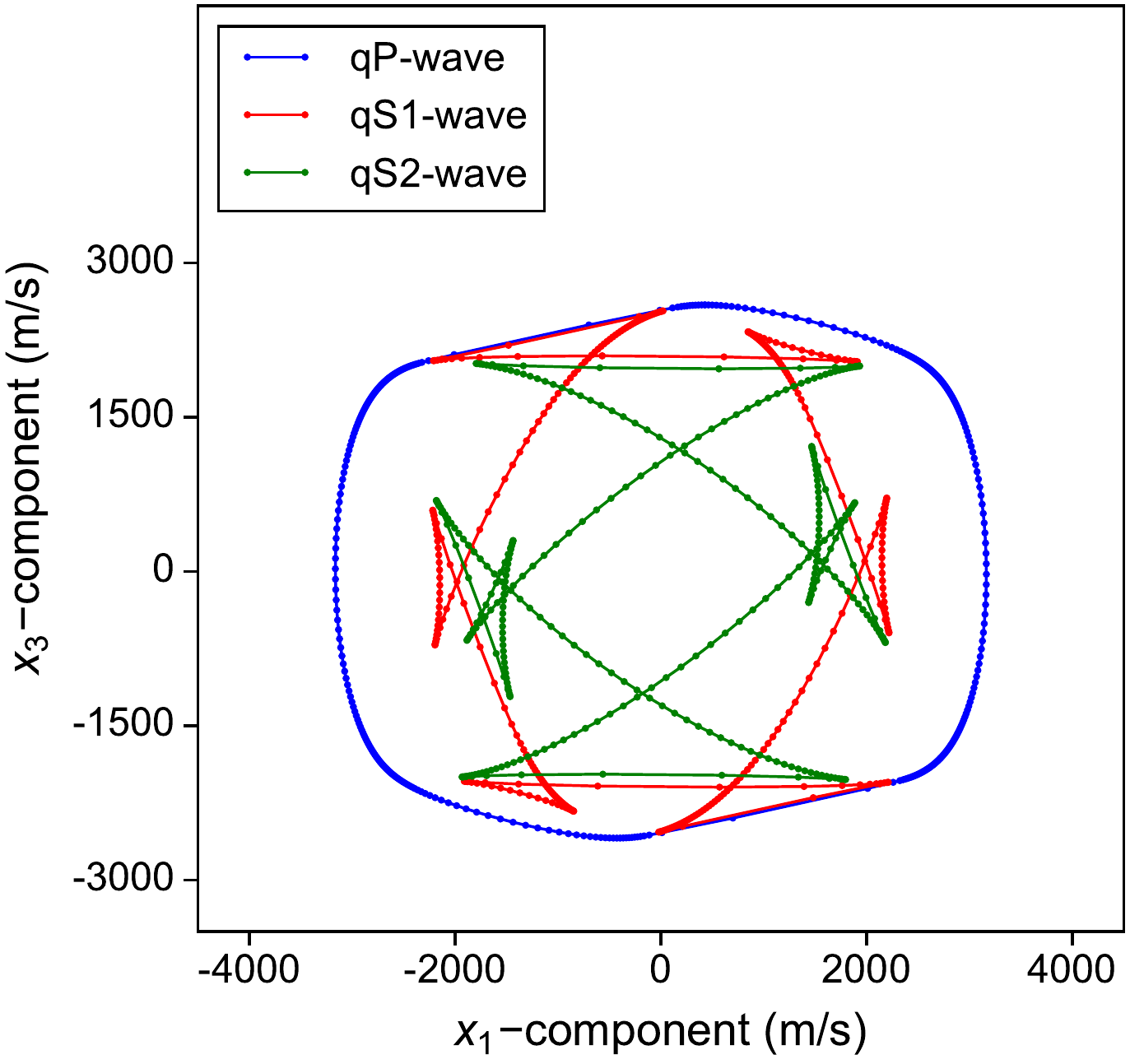}}
\subfigure[]{\includegraphics[width=0.45\textwidth]{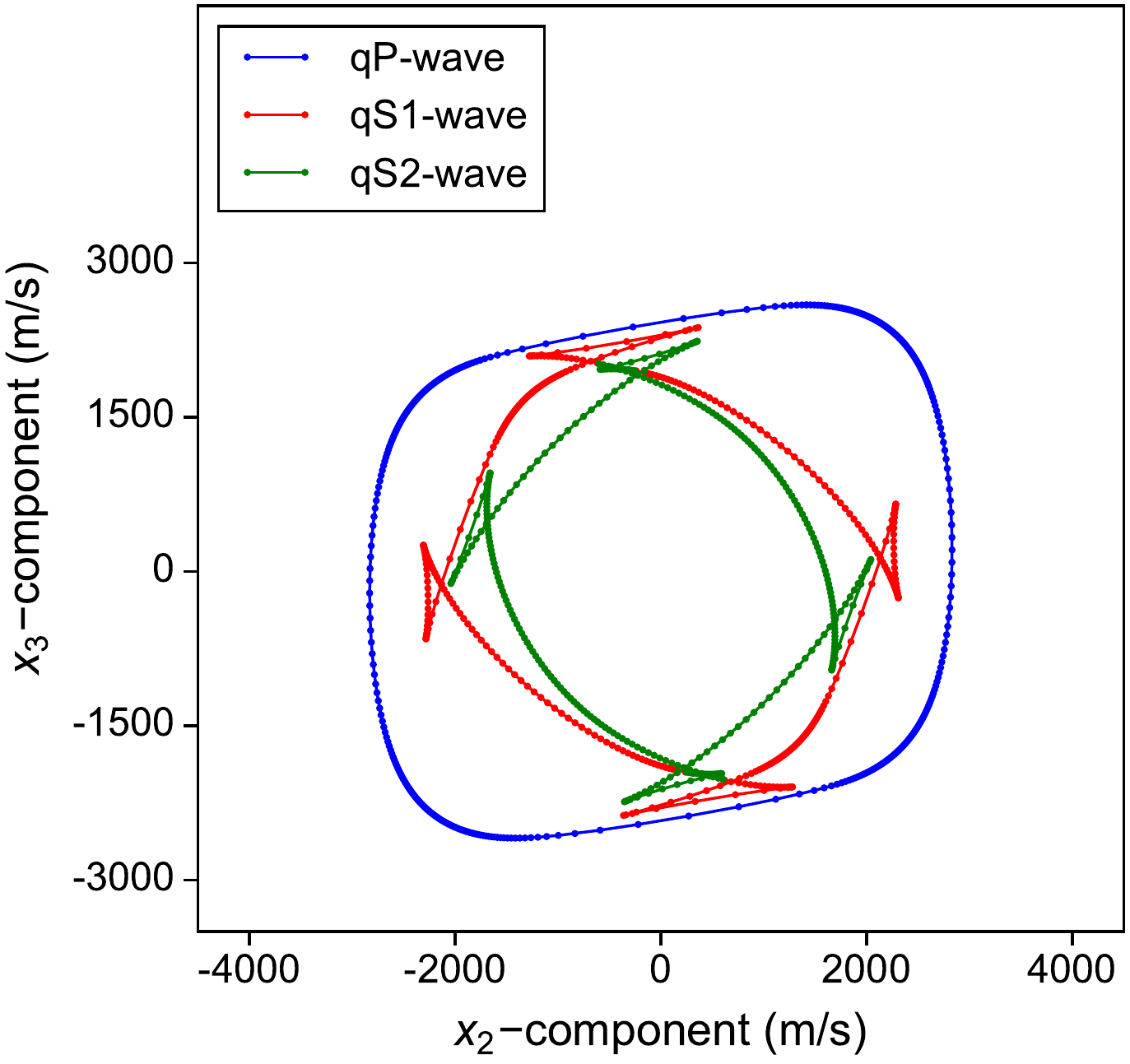}}
\caption{Wavefront curves in the 3D triclinic anisotropic medium with elasticity matrix \eqref{eq:tti_3d} on the (a) $x_1x_2$ (b) $x_1x_3$ and (c) $x_2x_3$ axis plane. qS1 and qS2 represents the two qS-waves. }
\label{fig:tri_3d_curve}
\end{figure}

Our last 3D numerical example is based on a triclinic anisotropic medium 
represented by
\begin{equation}
\bfC =\left[\begin{array}{cccccc}
10   & 3.5  &  2.5  & -5  & 0.1  &  0.3 \\
&8   & 1.5   & 0.2   &-0.1   &-0.15 \\
&&6   & 1   & 0.4   & 0.24 \\
&&&5  &  0.35   & 0.525 \\
&&&&4  & -1 \\
&&&&&3
\end{array}\right],
\label{eq:tri_3d}
\end{equation}
with unit GPa. The wavefront curves on three axis planes are shown in Fig.~\ref{fig:tri_3d_curve}.

\begin{figure}
\centering
\subfigure[]{\includegraphics[width=0.3\textwidth]{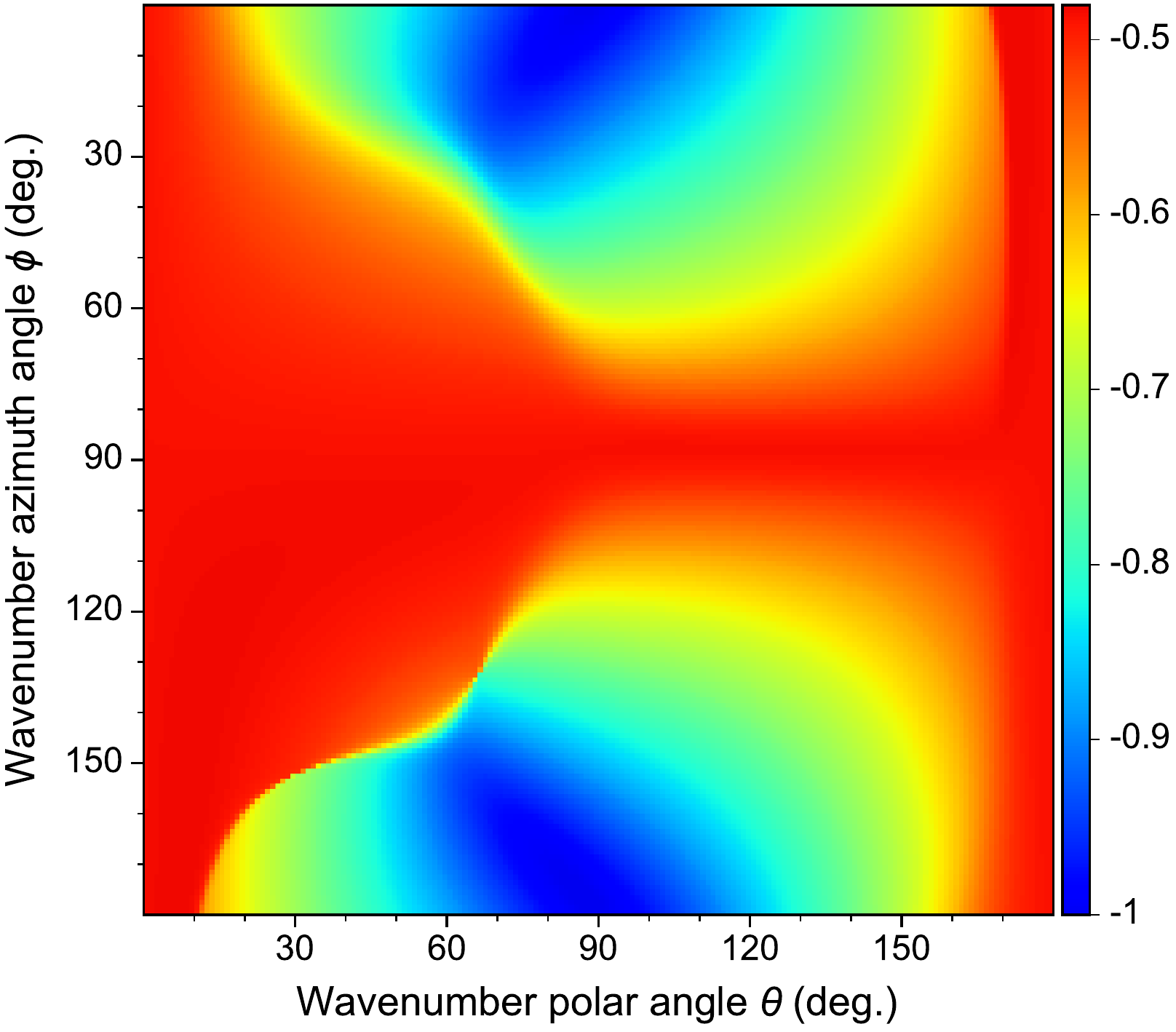}}
\subfigure[]{\includegraphics[width=0.3\textwidth]{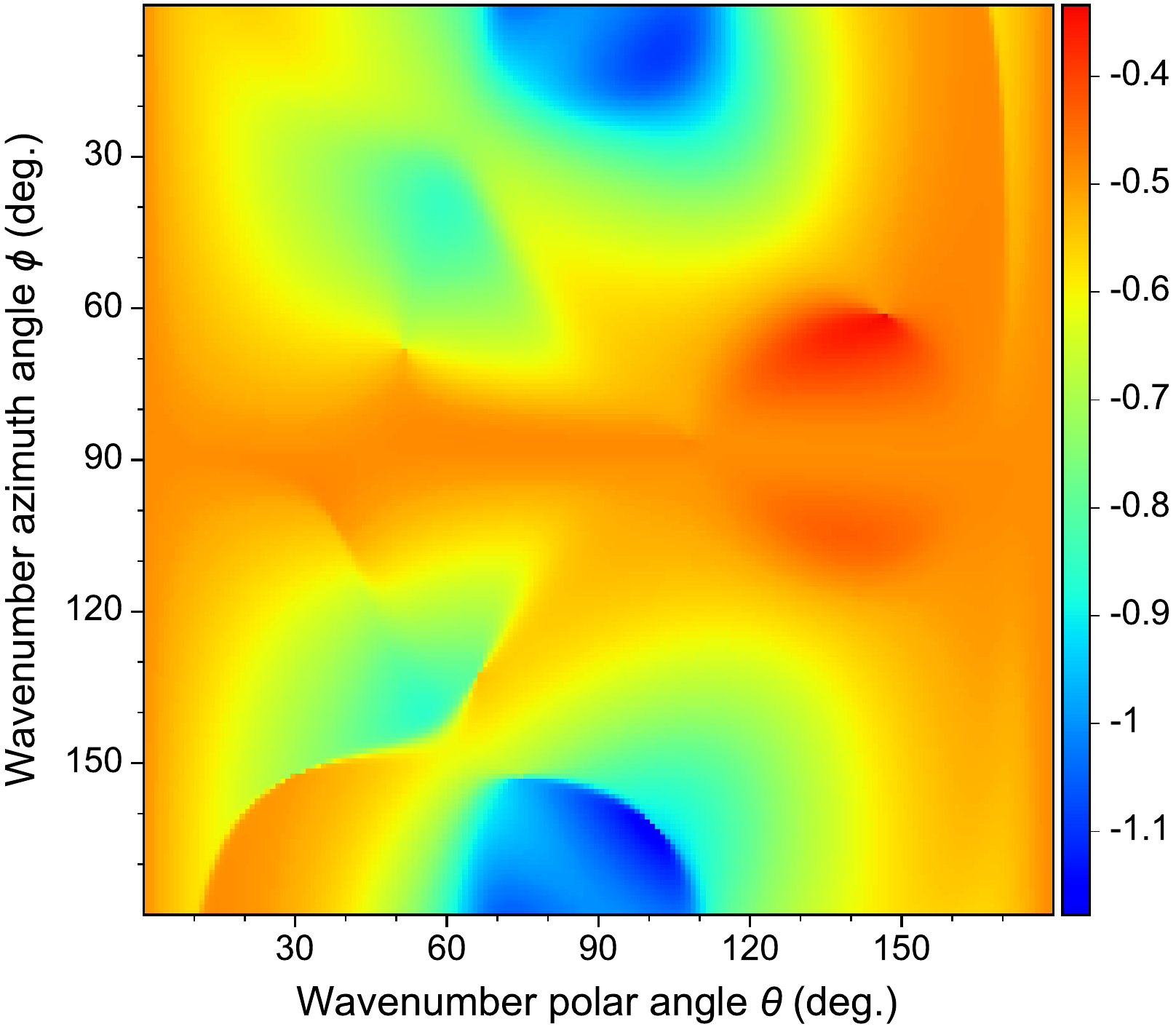}}
\subfigure[]{\includegraphics[width=0.3\textwidth]{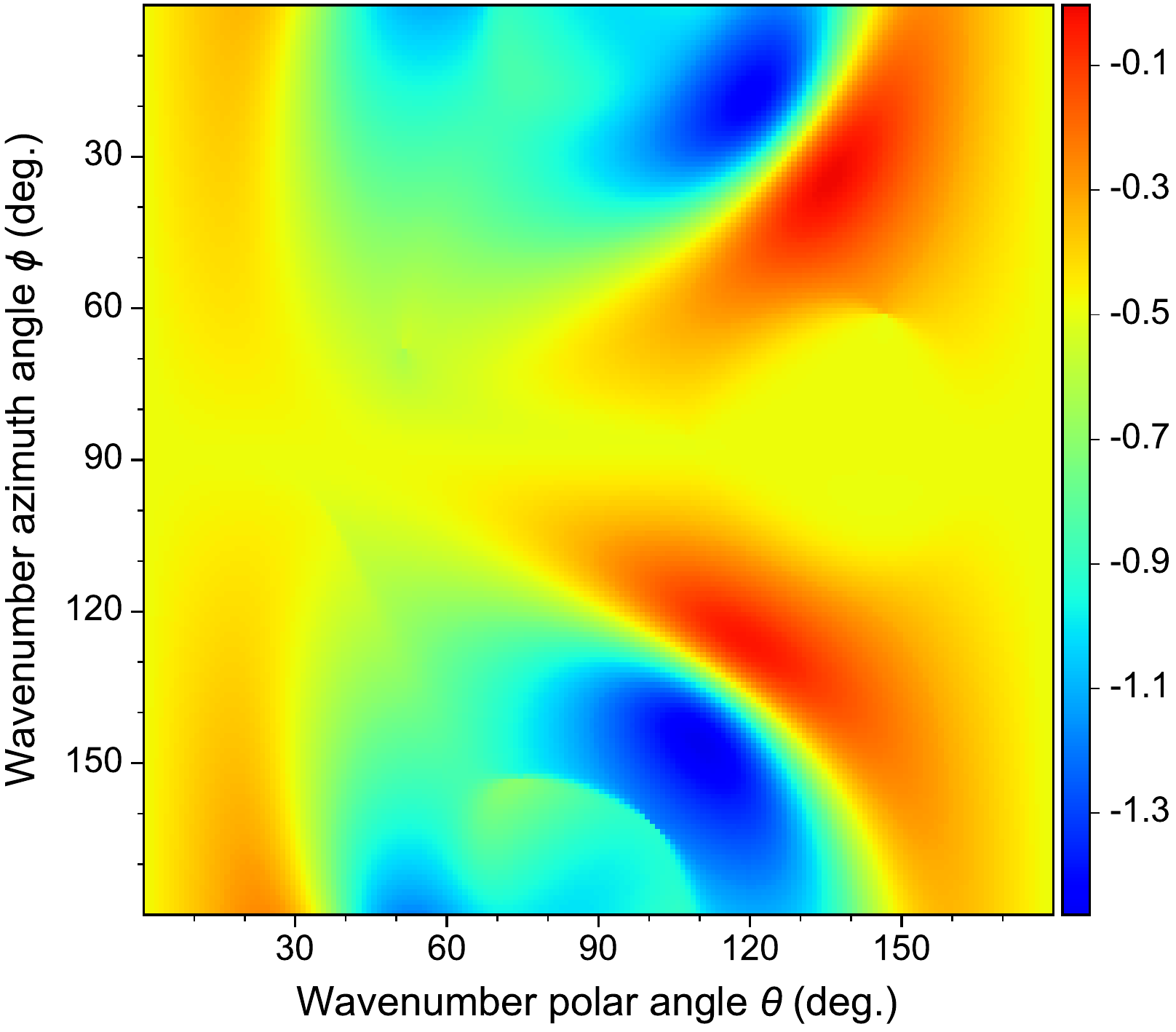}}
\subfigure[]{\includegraphics[width=0.3\textwidth]{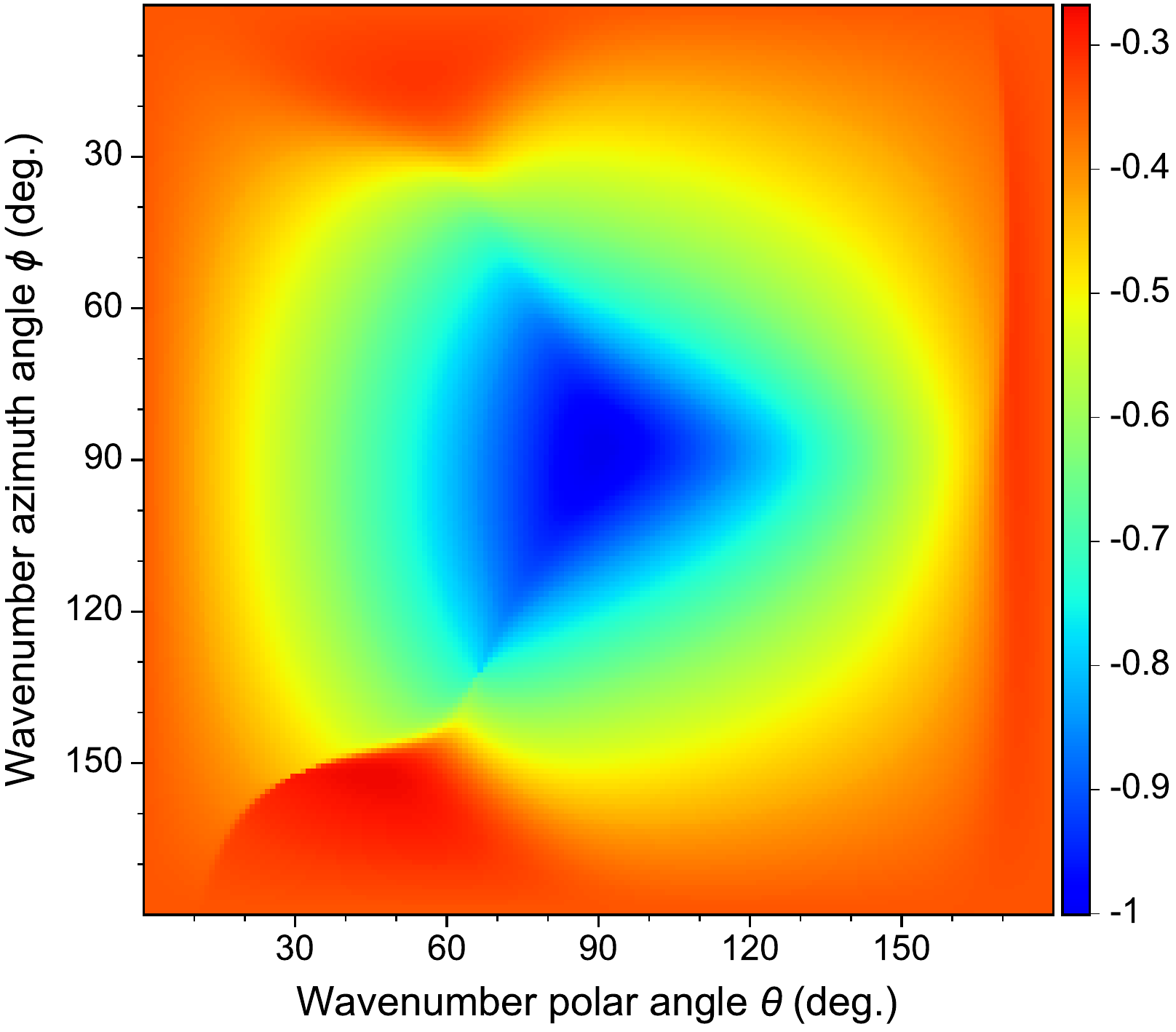}}
\subfigure[]{\includegraphics[width=0.3\textwidth]{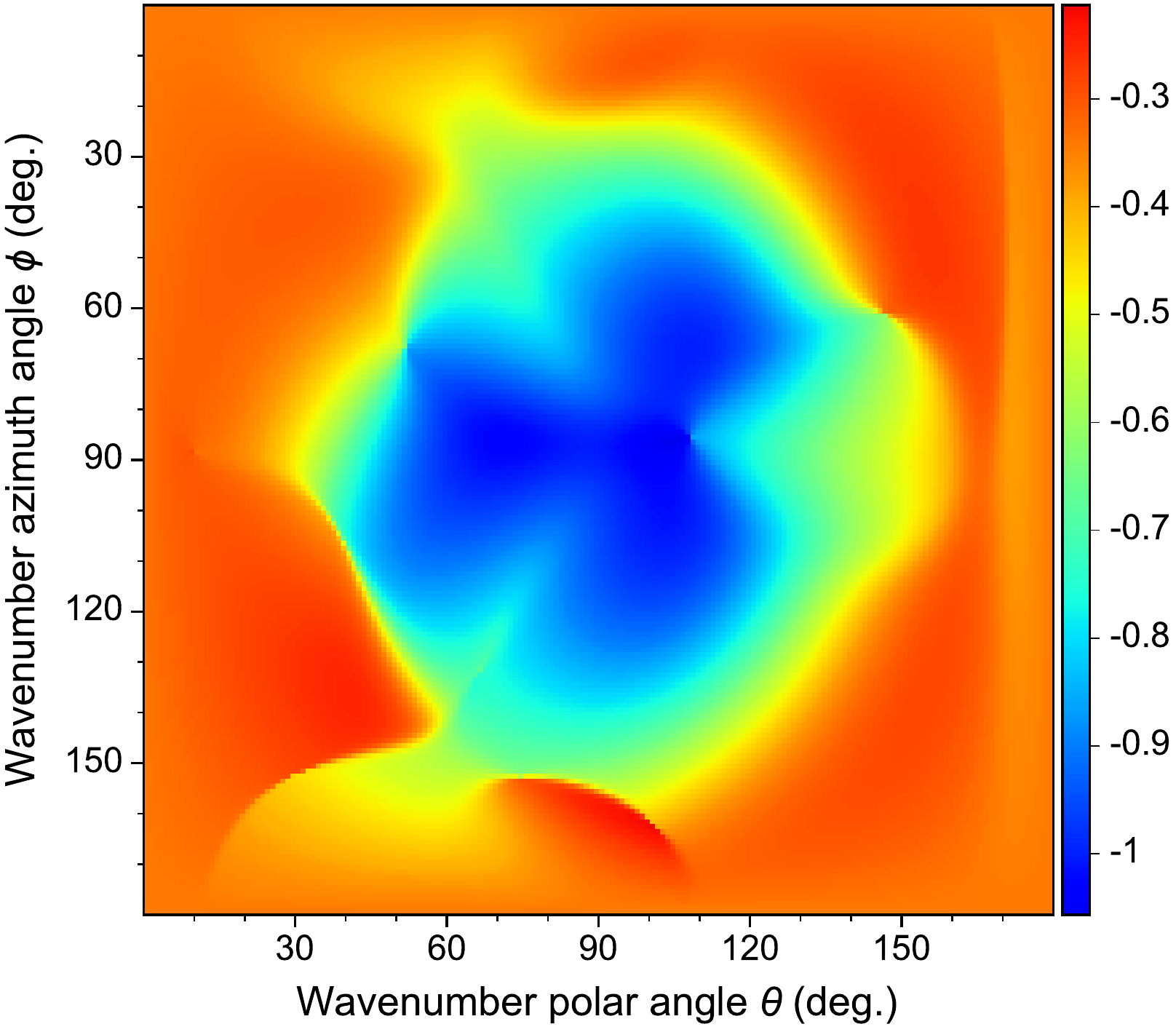}}
\subfigure[]{\includegraphics[width=0.3\textwidth]{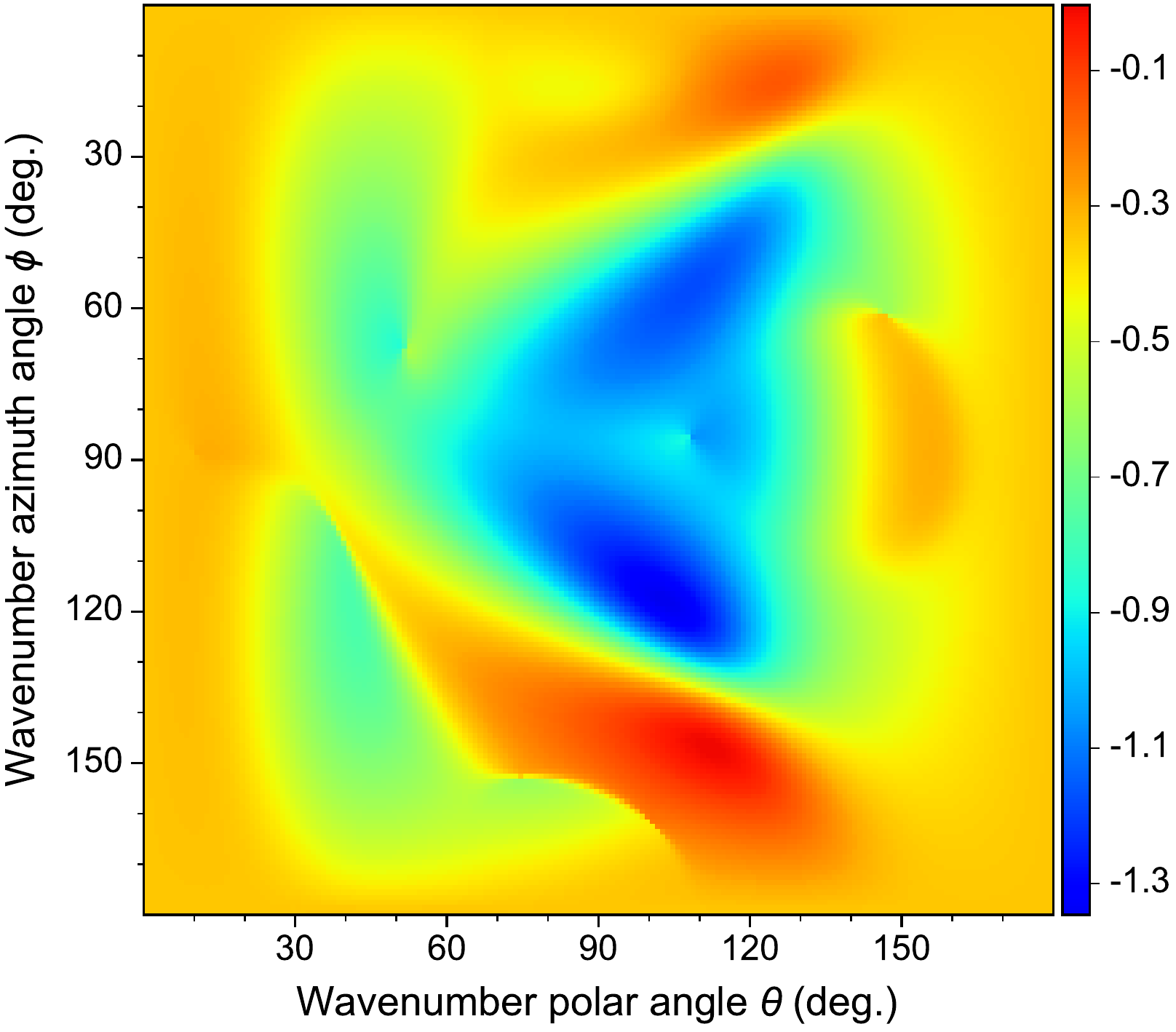}}
\subfigure[]{\includegraphics[width=0.3\textwidth]{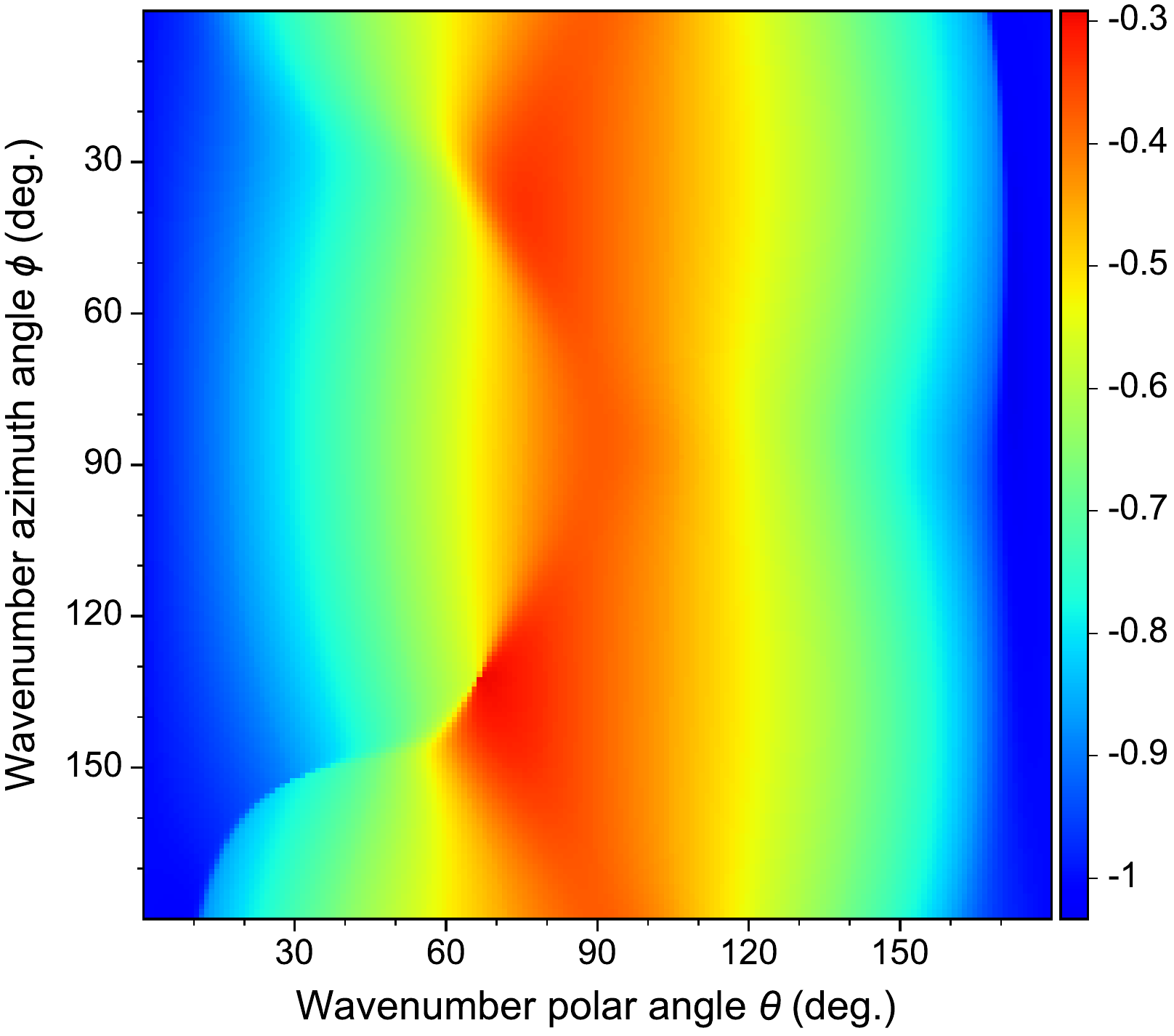}}
\subfigure[]{\includegraphics[width=0.3\textwidth]{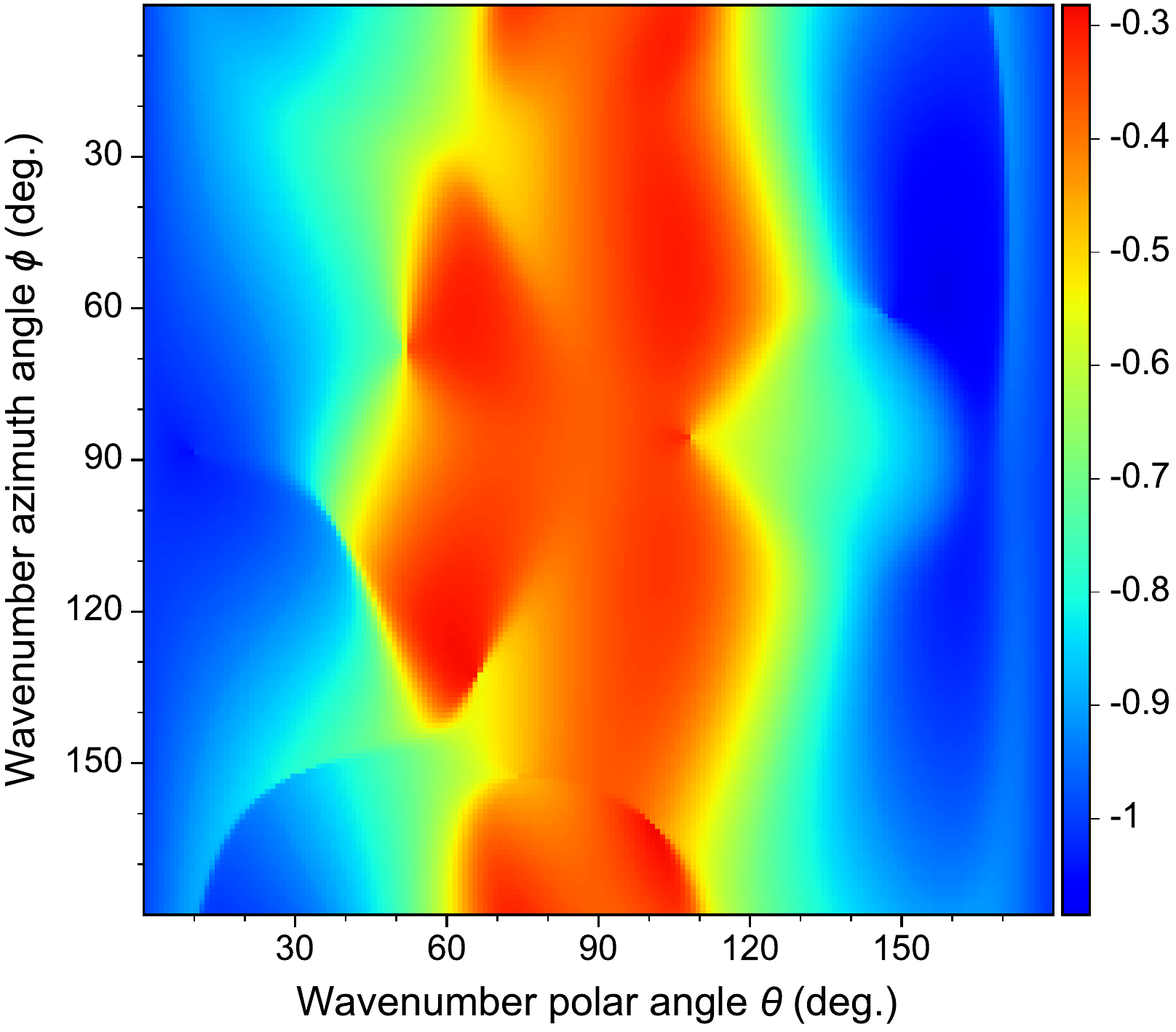}}
\subfigure[]{\includegraphics[width=0.3\textwidth]{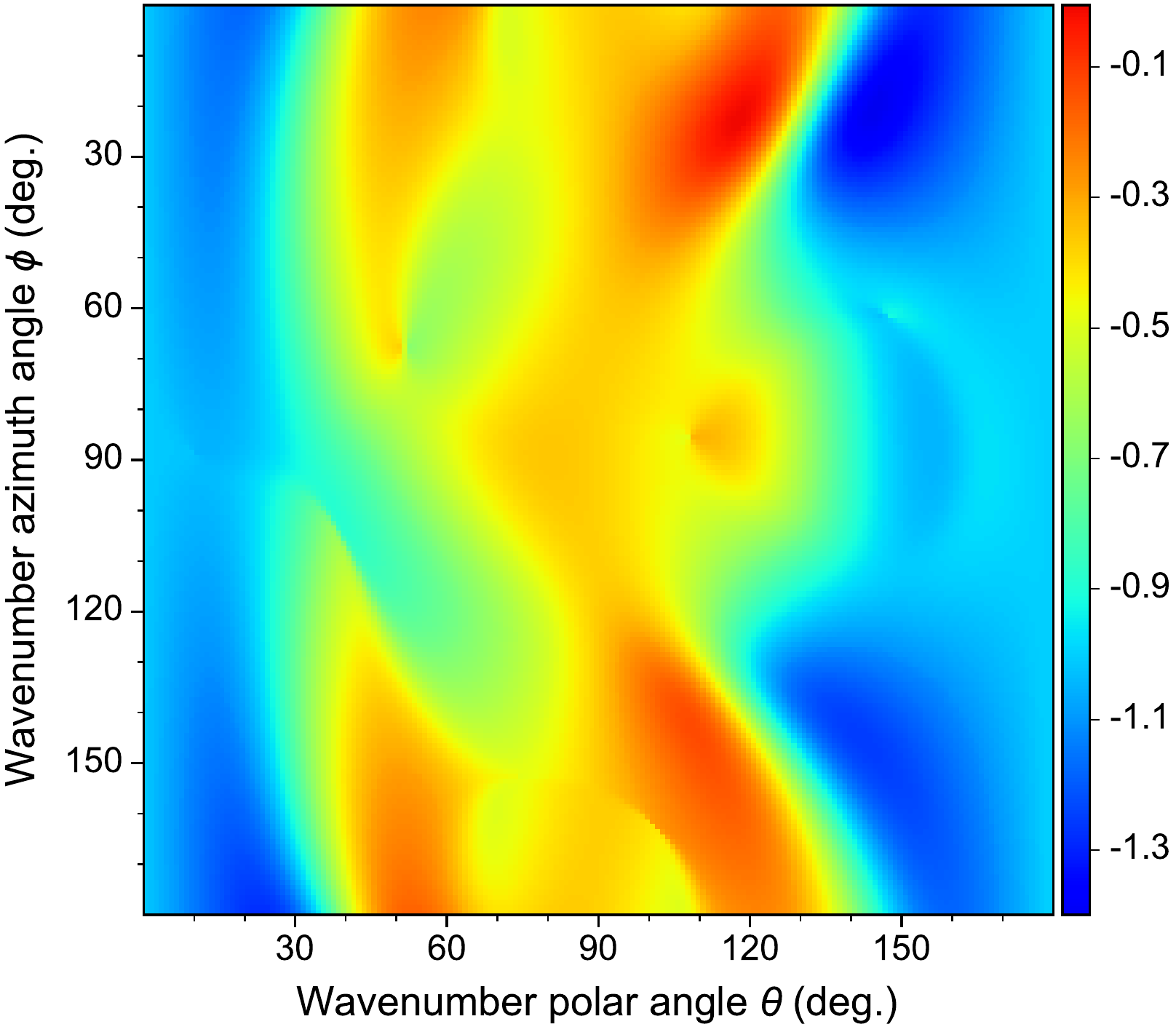}}
\caption{Eigenvalue derivatives of $\tilde{\bfA}$ of MPML in the (a)-(c) $x_1$-, (d)-(f) $x_2$-, and (g)-(i) $x_3$-directions under calculated optimal damping ratios in eq.~\eqref{eq:tri_3d_ratio} for the 3D triclinic anisotropic medium with elasticity matrix \eqref{eq:tri_3d}. (a), (d) and (g) represent qP-wave, (b), (e) and (h) represent qS1-wave, (c), (f) and (i) represent qS2-wave.}
\label{fig:tri_3d_deriv}
\end{figure}

We solve for the optimal damping ratios for this triclinic anisotropic 
medium using Algorithm~\ref{alg:damping_ratio_3}, and obtain the 
following optimal damping ratios with $\epsilon=-0.005$:
\begin{equation}
\xi_1=0.487, \qquad \xi_2=0.345,\qquad \xi_3=0.374.
\label{eq:tri_3d_ratio}
\end{equation}
The damping ratios for this anisotropic medium are unexpectedly very 
large compared with those for the heretofore 2D and 3D examples. We seek 
the reasons of these large damping ratios from the eigenvalue derivatives 
shown in Fig.~\ref{fig:tri_3d_deriv}, and find that it is the qS2-wave 
that leads to such large damping ratios to achieve a stable MPML.  In 
fact, for the damping ratios in eq.~\eqref{eq:tri_3d_ratio}, the 
corresponding eigenvalue derivatives of qP- and qS1-waves are far smaller 
than zero, yet the eigenvalue derivative of qS2-wave merely smaller than 
zero ($-0.005$  under our threshold setting), leading to a set of 
relatively large damping ratios for this 3D anisotropic medium. 

\begin{figure}
\centering
\includegraphics[width=0.65\textwidth]{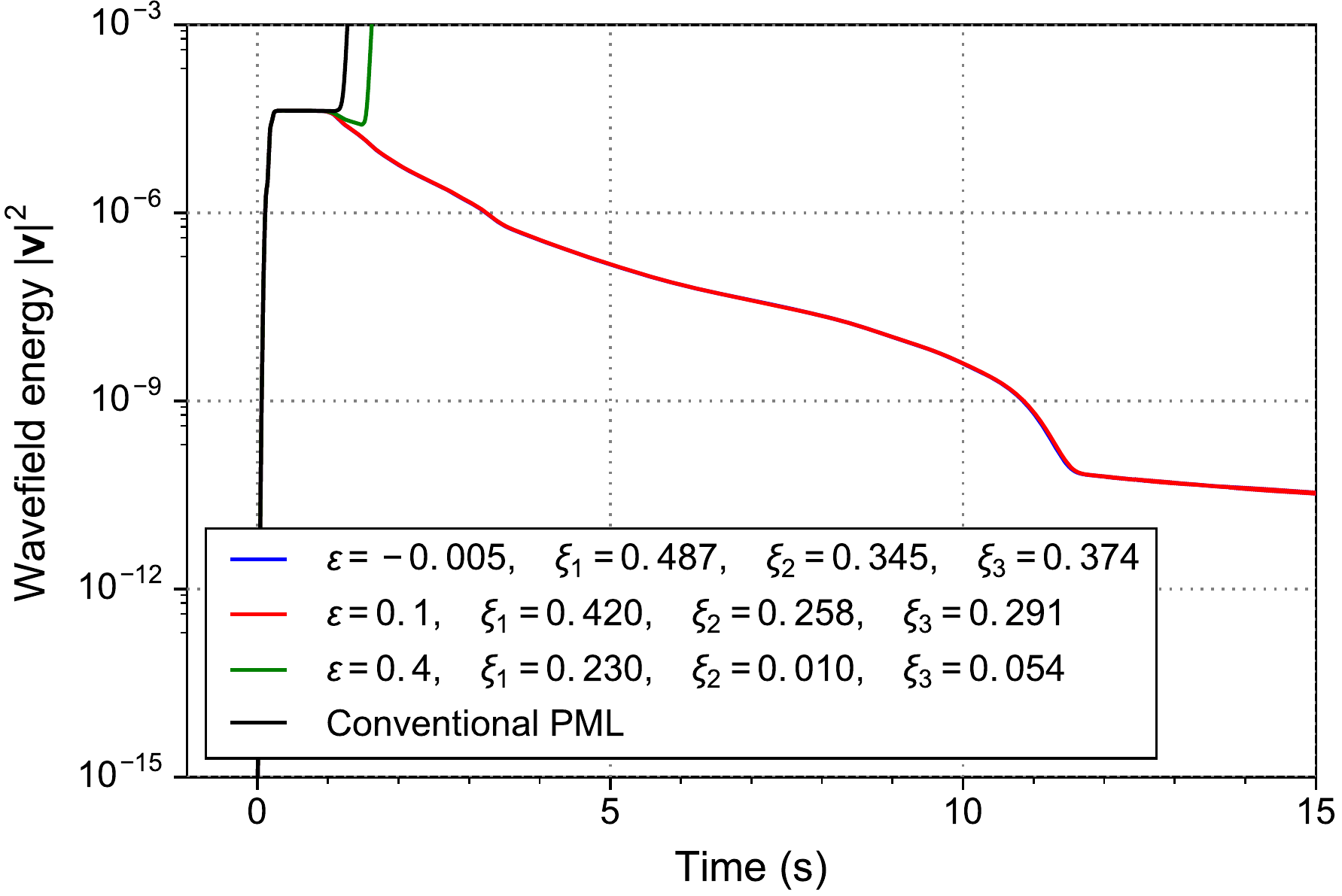}
\caption{Wavefield energy decay curve in the 3D triclinic anisotropic 
	medium with elasticity matrix \eqref{eq:tri_3d}. 
	The blue curve ($\epsilon=-0.005$) and red curve ($\epsilon=0.1$) are almost identical. }
\label{fig:tri_3d_energy}
\end{figure}

Our calculated optimal damping ratios result in a stable MPML, as 
indicated by the corresponding energy decay curve shown in 
Fig.~\ref{fig:tri_3d_energy}. The wavefield energy decay curve with 
a threshold of $\epsilon=0.1$ displayed in Fig.~\ref{fig:tri_3d_energy} 
is surprisingly almost identical with that of $\epsilon=-0.005$.  When 
using a threshold $\epsilon=0.4$, MPML become unstable, indicating that 
the positive threshold 0.4 is too large to make MPML stable.  This verifies 
again that, although a positive threshold might result in stable MPML, we 
should use a negative threshold to ensure a stable MPML for general  
anisotropic media. This is consistent with the stability condition 
described in \cite{Fajardo-Papa_2008}, and is perhaps the only practical 
method to stabilize PML using nonzero damping ratios. 

\section{Conclusions}

A definite analytic method for determining the 
optimal damping ratios of multi-axis perfectly 
matched layers (MPML) is generally impossible for 3D general anisotropic media 
with possible all nonzero elasticity parameters. 
We have developed a new method to efficiently determine the optimal 
damping ratios of MPML
for absorbing unwanted, outgoing propagating waves in 2D and 3D general 
anisotropic media. This numerical approach is very straightforward using the left and right 
eigenvectors of the damped system coefficient matrix.  We have used six 
numerical modeling examples of elastic-wave propagation in 2D and 3D 
anisotropic media to demonstrate that our new algorithm can effectively 
and correctly provide the optimal MPML damping ratios for even very 
complex, general anisotropic media.

\section{Acknowledgments}

This work was supported by U.S.\ Department of Energy through contract
DE-AC52-06NA25396 to Los Alamos National Laboratory (LANL). The 
computation was performed using super-computers of LANL's 
Institutional Computing Program.

\bibliographystyle{elsarticle-harv} 
\bibliography{refs}

\end{document}